\documentclass[a4paper,fleqn,usenatbib]{mnras}

\usepackage{newtxtext,newtxmath}

\usepackage[T1]{fontenc}
\usepackage{ae,aecompl}

\usepackage{graphicx}	
\usepackage{amsmath}	
\usepackage{amssymb}	

\usepackage{multirow}
\usepackage{booktabs,caption}
\usepackage[flushleft]{threeparttable}
\usepackage{lscape}  
\usepackage{adjustbox}

\title[Comparison between light curves of SN~II and SN~IIb]{Comparison of the optical light curves of hydrogen-rich and hydrogen-poor type II supernovae}

\author[P. J. Pessi et al.]{
P. J. Pessi$^{1,2}$\thanks{E-mail: pjpessi@fcaglp.unlp.edu.ar (P.J.P.)},
G. Folatelli$^{1,2,3}$,
J. P. Anderson$^{5}$,
M. Bersten$^{1,2,3}$,
C. Burns$^{6}$,
\newauthor
C. Contreras$^{7,8}$,
S. Davis$^{9}$,
B. Englert$^{2,4}$,
M. Hamuy$^{10}$,
E.Y. Hsiao$^{9}$,
L. Martinez$^{1,2}$,
\newauthor
N. Morrell$^{7}$,
M. M. Phillips$^{7}$,
N. Suntzeff$^{11}$
and M. D. Stritzinger$^{12}$
\\
$^{1}$Instituto de Astrof\'isica de La Plata (IALP), CONICET, Paseo del bosque S/N, 1900, Argentina\\
$^{2}$Facultad de Ciencias Astron\'omicas y Geof\'isicas (FCAG), Universidad Nacional de La Plata (UNLP), Paseo del bosque S/N, 1900, Argentina\\
$^{3}$Kavli Institute for the Physics and Mathematics of the Universe, Todai Institutes for Advanced Study,\\
University of Tokyo, 5-1-5 Kashiwanoha, Kashiwa, Chiba 277-8583, Japan.\\
$^{4}$Agencia Nacional de Promoci\'on Cient\'ifica y Tecnol\'ogica (ANPCyT), Godoy Cruz 2370, C142FQD, Buenos Aires, Argentina.\\
$^{5}$European Southern Observatory, Alonso de C\'ordova 3107, Casilla 19, Santiago, Chile\\
$^{6}$Observatories of the Carnegie Institution for Science, 813 Santa Barbara St, Pasadena, CA 91101, USA\\
$^{7}$Las Campanas Observatory, Carnegie Observatories, Casilla 601, La Serena, Chile\\
$^{8}$Space Telescope Science Institute, 3700 San Martin Drive, Baltimore, MD 21218, USA\\
$^{9}$Department of Physics, Florida State University, 77 Chieftan Way, Tallahassee, FL 32306, USA\\
$^{10}$Departamento de Astronom\'ia, Universidad de Chile, Casilla 36-D, Santiago, Chile\\
$^{11}$George P. and Cynthia Woods Mitchell Institute for Fundamental Physics and Astronomy, Department of Physics and Astronomy,\\
Texas A\&M University, 4242 TAMU, College Station, TX 77843, USA\\
$^{12}$Department of Physics and Astronomy, Aarhus University, Ny Munkegade 120, DK-8000 Aarhus C, Denmark
}

\date{Accepted 2019 July 2. Received 2019 July 1; in original form 2019 February 6}

\pubyear{2018}

\begin{document}
\def \msun{\ifmmode{{\rm\ M}_\odot}\else{${\rm\ M}_\odot$}\fi}
\label{firstpage}
\pagerange{\pageref{firstpage}--\pageref{lastpage}}
\maketitle

\begin{abstract}
  Type II supernovae (SNe~II) show strong hydrogen features in their spectra throughout their whole evolution while type IIb supernovae (SNe~IIb) spectra evolve from dominant hydrogen lines at early times to increasingly strong helium features later on. However, it is currently unclear whether the progenitors of these supernova (SN) types form a continuum in pre-SN hydrogen mass or whether they are physically distinct. SN light-curve morphology directly relates to progenitor and explosion properties such as the amount of hydrogen in the envelope, the pre-SN radius, the explosion energy and the synthesized mass of radioactive material. In this work we study the morphology of the optical-wavelength light curves of hydrogen-rich SNe~II and hydrogen-poor SNe~IIb to test whether an observational continuum exists between the two. Using a sample of 95 SNe (73 SNe~II and 22 SNe~IIb), we define a range of key observational parameters and present a comparative analysis between both types. We find a lack of events that bridge the observed properties of SNe~II and SNe~IIb. Light curve parameters such as rise times and post-maximum decline rates and curvatures clearly separate both SN types and we therefore conclude that there is no continuum, with the two SN types forming two observationally distinct families. In the $V$-band a rise time of 17 days (SNe~II lower, SNe~IIb higher), and a magnitude difference between 30 and 40 days post explosion of 0.4 mag (SNe~II lower, SNe~IIb higher) serve as approximate thresholds to differentiate both types.
\end{abstract}
\begin{keywords}
supernovae: general -- supernovae: individual (SN~2001fa, SN~2004ff, SN~2006Y, SN~2007fz, SN~2008M, SN~2013ai, SN~2013fs)
\end{keywords}


\section{Introduction}
\label{sec:intro}
It is believed that stars with initial masses $>$ 8\msun\  end their lives exploding as core collapse SNe (CCSNe). CCSNe can be divided into two main groups based on their spectral properties. Those SNe that show hydrogen in their spectra are classified as Type II supernovae (SNe~II) while those that do not are of Type I (\citealt{1941PASP...53..224M}, see \citealt{2017hsn..book..195G} and references therein for a review of spectroscopic SNe classification). Subsequent division of SNe~II into ``plateau'' (IIP) and ``linear'' (IIL) was introduced based on the shape of the light curve (\citealt{1979A&A....72..287B}), however recent works have found a continuum of observed properties between the two (\citealt{2014ApJ...786...67A}, \citealt{2015ApJ...799..208S}, \citealt{2016AJ....151...33G}, \citealt{2016ApJ...828..111R}). Therefore, we refer to these two historical groups together as ``hydrogen-rich SNe~II'' for the remainder of this work. A number of further sub-classifications exists, such as those SNe~II showing similar properties to SN~1987A (with particularly long-rising light curves, see e.g. \citealt{2012A&A...537A.141P}, \citealt{2016A&A...588A...5T} and references therein), and those displaying narrow emission features in their spectra, the Type IIn (\citealt{1990MNRAS.244..269S}). However, both of these groups show characteristics sufficiently distinct from hydrogen-rich SNe~II (as defined above) that we do not consider them in this work (see \citealt{2017hsn..book..239A} and references therein for additional discussion on hydrogen-rich CCSNe). Type I CCSNe were also separated into Types Ib and Ic based on the presence or absence of helium lines in their spectra (see \citealt{1997ARA&A..35..309F}, \citealt{2017hsn..book..195G} and references therein). Due to the lack of hydrogen these latter events are also referred to as stripped-envelope SNe (SE SNe) in reference to their progenitors having lost most, if not all, of their hydrogen envelopes before explosion. SNe~Ib and SNe~Ic are not studied in the present work.

An intermediate group named Type IIb was identified that undergoes a transition from a hydrogen-rich (similar to SNe~II) to a hydrogen-poor (similar to SNe~Ib) spectrum (\citealt{1993ApJ...415L.103F}). This transition is attributed to a small amount of hydrogen in their progenitor envelopes, and we thus refer to this group simply as ``SNe~IIb''\footnote{There are a number of Type Ib SNe that also show evidence of hydrogen in their spectra - e.g.: {SN~1996N (\citealt{1998A&A...337..207S}, \citealt{2002ApJ...566.1005B}), SN~1999ex (\citealt{2002AJ....124..417H}, \citealt{2006PASP..118..791B}), SN~2005bf (\citealt{2006ApJ...641.1039F}, \citealt{2006PASP..118..791B}, \citealt{2007PASP..119..135P}), SN~2007Y (\citealt{2009ApJ...696..713S})} - and could also be considered to be hydrogen-poor SE-SNe. These events are not included in this study.} here. Their characteristic bell-shaped light curve, in addition to their overall similarity to SNe~Ib and SNe~Ic, has led the community to associate SNe~IIb with SE~SNe (e.g., \citealt{1996ApJ...459..547C}, \citealt{2001AJ....121.1648M}) and model the main peak of SN~IIb light curves as powered by radioactive decay of $^{56}$Ni (e.g. \citealt{1994ApJ...420..341S}, \citealt{1994ApJ...429..300W}, \citealt{2012ApJ...757...31B}). However, there are few studies that focus on a quantitative comparison of light-curve morphologies between SNe~II and SNe~IIb (c.f.: \citealt{2012ApJ...756L..30A}, \citealt{2014MNRAS.445..554F}, \citealt{2018PASA...35...49E}). 

An outstanding question in the study of CCSNe observations and their implications for our understanding of progenitor systems, is whether in spite of the spectroscopic distinction at late times, the early spectral similarities of SNe~II and SNe~IIb imply a shared common origin. Theoretical works have suggested that the majority (if not all) SNe~IIb could come from interacting binaries (see e.g. \citealt{1992ApJ...391..246P}, \citealt{1995PhR...256..173N}, \citealt{2013ApJ...762...74B}, \citealt{2017ApJ...840...10Y}, \citealt{2019MNRAS.486.4451G}). Although the bulk of hydrogen-rich SNe can be satisfactorily explained by the single-progenitor scenario, in principle some SNe II may also come from binary systems. For isolated stars the degree of envelope stripping should be chiefly dictated by the initial progenitor mass (with some additional dependence on progenitor metallicity, see e.g. \citealt{2003fthp.conf....3H}). However, the situation is complicated by the effect of mass transfer in binary progenitors. Investigating whether SNe~II and SNe~IIb arise from similar or distinct progenitor scenarios can shed light on the importance of these competing mass-loss mechanisms.

It has been argued that the diversity in the light curves of hydrogen-rich SNe~II is driven by the mass of the hydrogen envelope retained by the progenitor at the epoch of explosion. This follows early theoretical modelling of SNe~II (\citealt{1993ApJ...414..712P}) claiming that faster declining SNe~II (that also show shorter-duration plateau phases, see e.g. \citealt{1983Ap&SS..89...89L}, \citealt{2004ApJ...617.1233Y} for theoretical evidence and \citealt{2014ApJ...786...67A}, \citealt{2017ApJ...850...90G} for observational evidence of this phenomenon) arise from progenitors with lower mass envelopes. Since the hydrogen lines of SN~IIb spectra disappear with time, it is believed that their progenitors retain an even smaller fraction of their hydrogen. If there were indeed a continuum of hydrogen envelope mass between the progenitors of SNe~II and SNe~IIb then we may expect to see a continuum in their observable light curve properties, similar to what is seen between fast and slow declining hydrogen-rich SNe~II.

\cite{2012ApJ...756L..30A} studied the light curve behaviour of a sample of 15 SNe~IIb and SNe~II in the $R$-band and argued that they separate into three distinct groups according to light curve shape: IIP, IIL, and IIb. \cite{2014MNRAS.445..554F} studied 11 fast-declining SNe~II (IIL) and found that they seem to be photometrically related to SNe~IIb. 
Here, our study follows those previous works, using a larger sample and the addition of a range of observational parameters defined consistently across all SN types. We present an analysis of 95 SNe (73 hydrogen-rich SNe~II and 22 SNe~IIb) with the aim of determining whether there is a clear distinction in the shape of the light curves between these SNe subtypes, or whether a continuum is observed. 

The paper is organized as follows: in Section \ref{snsamples} we present the data sample. In Section \ref{sec:lc} we describe our light curve analysis method and explain how we obtain explosion and maximum-light epochs. Section \ref{param} introduces the parameters we have used to compare light curve properties. Results are provided in Section \ref{results}. Section \ref{sec:discussion} presents a discussion of deviating cases and possible implications for progenitor properties. Finally in Section \ref{sec:conc} we present our conclusions.
In addition, an appendix is included where figures not included in the main body of the manuscript are presented along with tables containing information on each SN considered and the numeric values of the defined comparison parameters.

\section{Supernova samples}
\label{snsamples}
Our sample includes 73 SNe~II (see Table \ref{tab:snII}), with 64 taken from the Carnegie Type II Supernova Program (CATS; \citealt{2016AJ....151...33G}; PI: Hamuy, 2002$-$2003) and the Carnegie Supernova Project (CSP-I; \citealt{2006PASP..118....2H}; PIs: Phillips and Hamuy, 2004$-$2009) samples from the work of  \citet{2014ApJ...786...67A}. This sample was selected as it originated from a well defined dataset that was observed, reduced and calibrated in a consistent manner. Additionally, nine objects having well-defined light curve maxima were included from the literature. All the objects with well defined light curve maxima were then used as templates to obtain the date of maximum light for the rest of the sample. This was required because such dates were generally poorly constrained in \cite{2014ApJ...786...67A},  but are necessary to obtain rise times (see Sec. \ref{RT}): an important parameter differentiating SNe~II from SNe~IIb.
In the case of SNe~IIb, no sufficiently large sample exists from a single follow-up program. Therefore, we collate all SNe~IIb from the literature (see Table \ref{tab:snIIb}) that fit the following selection criteria (that also apply to the SNe~II):
\begin{itemize}
\item Good sampling. At least nine data points in our primary photometric band ($V$), in order to have a good approximation of the intrinsic shape of the light curve. 
\item Existing photometric data points around maximum light for SNe~IIb and prior to $\sim$25 days after explosion for SNe~II to have a dense enough coverage of the early part of the light curve. This means that our SN~II light curves do not necessarily cover the epoch of maximum light.
\item When the explosion date is not well constrained by prediscovery non-detections, at least three available spectra in order to be able to estimate the explosion dates through spectral matching (see Section \ref{exp}).
\end{itemize}
The parameters defined in this work (see Section \ref{param}) are only constrained by photometric data points up to $\sim$60 days after explosion. The reasons to adopt this limit are: a) to enable a characterization of the light curve properties around maximum light and the subsequent initial decline; and b) the light-curve coverage is more uniform among the SN sample than at later times. 
We choose $V$ as our primary photometric band as historically this is the band in which most SNe were observed with the best cadence. $B$- and $r$-band were also considered in order to study how the light curves behave at shorter and longer wavelengths, respectively.
K-corrections have not been performed to the SN photometry. However, given the redshift range considered here (z < 0.08), such corrections are too small to affect our analysis. Also, all time intervals are given in the SN rest frame.

\section{Light curves}
\label{sec:lc}
\begin{figure}
  \includegraphics[width=\columnwidth]{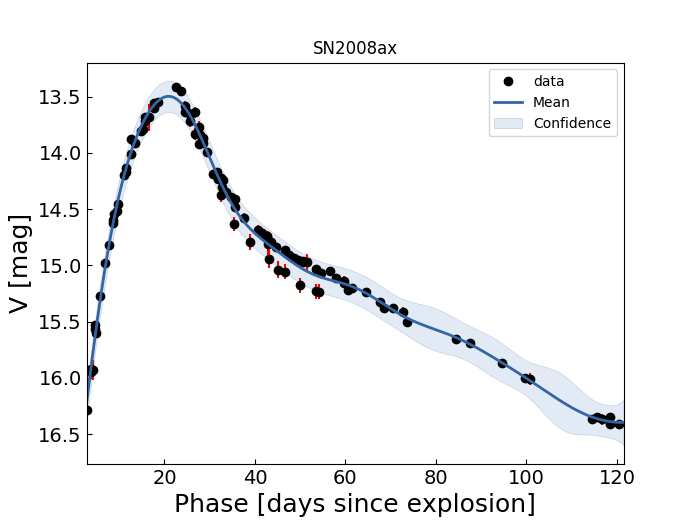}
    \caption{Resampled light curve of SN~2008ax. The black dots are the original photometric data points (obtained from different sources, see Table \ref{tab:snIIb}, hence the observable offset). Some error bars are smaller than the dot size. The mean of the resampling is presented as a blue continuous line and the confidence interval is presented in light blue.}
    \label{GPexample}
\end{figure}

To be able to compare the light curves and define the parameters described in Section \ref{param} in a consistent manner, a uniform sampling of the light curve is needed. Hence, we resample the photometry via the Gaussian process method (see \citealt{2006gpml.book.....R} for details on Gaussian processes) using a self-made code that implements the Python library \citet{gpy2014}, using a squared-exponential kernel\footnote{The kernel takes as input two hyperparameters, lengthscale and variance. Our first approximation for these values were the mean separation of the photometric points and the standard deviation of said separation respectively. In three cases (SN~2012aw, SN~2013ej and SN~2013df) a different value (selected by trial and error) was used in order to obtain a better fit.} plus a kernel to account for the errors in the measured magnitudes. After running the code we not only obtain a regular sampling but also an uncertainty or confidence interval of said sampling. An example of the Gaussian process fit is presented in Fig. \ref{GPexample}.

The Gaussian process works well when the photometry is well sampled. However, when the data points are not evenly sampled, the method tends to introduce spurious wiggles. At the same time, the confidence interval increases to accommodate to the lack of data. In our sample this effect is only considerable for the late part of the light curve, which is not considered in the analysis.

\subsection{Explosion epoch estimation}
\label{exp}

CSP-I and CATS SN~II dates of explosion are those published in \citet{2017ApJ...850...89G}, which are a combination of explosion epochs from spectral matching and prediscovery non-detections. 
For the SNe~II from other sources, we derive explosion epochs in the same way as \citet{2017ApJ...850...89G}. Whenever the interval between the last non-detection and the first detection is shorter than 20 days, the mid point of the interval is adopted as the explosion date and its uncertainty is set as half the interval length. When the previous requirement is not met, we use the spectral matching tool SNID (Supernova Identification software, \citealt{2007ApJ...666.1024B}) to derive an explosion date. In order to do this, we have added the spectra from \citet{2017ApJ...850...89G} as SNID templates.

In the case of SNe~IIb we create additional SNID templates using those events where the explosion epoch is well constrained from the discovery data (i.e., an interval shorter than 20 days between last non-detection and discovery). For the rest of the SNe~IIb (without constraining non detections) we run SNID on each of their spectra. From each SNID run we select the spectral matches that provide an estimated age from maximum light that is within five days from the known age based on the light-curve fit. Every spectral match provides an age from explosion ($t_\mathrm{exp\_match}$) for the given spectrum. These $t_\mathrm{exp\_match}$ have an associated uncertainty that is given by that of the date of explosion of the SN template. We fix a minimum uncertainty of five days to account for the typical uncertainty in the spectral matching procedure, that is mainly determined by the size and diversity of the available template sample. Using the observation epoch ($t_\mathrm{obs}$) of the fitted spectrum we can compute a date of explosion for each match as $t_\mathrm{obs} - t_\mathrm{exp\_match}$. Finally, we compute the explosion date for the given SN~IIb as the weighted average of the results from all selected SNID matches.

To test the validity of this procedure we remove each SN~IIb template from the SNID library and calculate its explosion epoch as above. The results are presented in Table \ref{tab:compIIbsnid}. The average difference between the explosion epoch calculated with SNID and the one calculated from non-detections is of $-0.65 \pm 5.0$ days, which validates our procedure. 

\begin{table}
  \centering
  \caption{Comparison between the explosion epochs obtained from non-detection constraints  and those calculated using SNID for SNe~IIb.}
  \label{tab:compIIbsnid}
  \begin{tabular}{lccccr} 
    \hline
    \multicolumn{2}{c}{}SNID &
    \multicolumn{2}{c}{}Observed\\\cmidrule(r){2-3}\cmidrule(l){4-5}
    SN & Exp[MJD] &  e\_Exp & Exp[MJD] & e\_Exp & diff  \\
    \hline
    1993J   & 49074.4 &   5.79 & 49071.7 & 2.60 &  2.7  \\  
    2004ex  & 53286.9 &   7.68 & 53288.8 & 0.50 & -1.9  \\  
    2004ff  & 53291.4 &   3.25 & 53302.9 & 3.49 &-11.5  \\  
    2006el  & 53964.0 &   7.88 & 53968.8 & 3.48 & -4.8  \\  
    2006T   & 53760.5 &   7.69 & 53759.0 & 7.02 &  1.5  \\  
    2008aq  & 54508.0 &   9.51 & 54514.9 & 8.48 & -6.9  \\  
    2008ax  & 54529.4 &   6.85 & 54528.3 & 0.13 &  1.1  \\  
    2009K   & 54850.2 &   9.12 & 54843.6 & 1.49 &  6.6  \\  
    2011dh  & 55712.1 &   8.19 & 55712.4 & 0.52 & -0.3  \\  
    2013df  & 56448.2 &   5.04 & 56443.9 & 6.93 &  4.3  \\   
    2016gkg & 57653.2 &  10.07 & 57651.2 & 0.06 &  2.0  \\  
     \hline
  \end{tabular}
\end{table}

\subsection{Maximum light epoch estimation}
\label{sec:ml}
In the case of SNe~IIb, we consider only the main maximum (thought to be powered by radioactive decay), and not that which occurs earlier during the cooling phase (see e.g. \citealt{2012ApJ...757...31B}). The date of the main maximum  is taken from the light curve through the Gaussian process resampling. To determine an error on the epoch of maximum light we adopted the mean separation in time between subsequent data points in the range from 15 days before to 15 days after maximum. This takes into account the uncertainty introduced by the cadence of the observations.

In the case of hydrogen-rich SNe~II, when possible, the date of maximum is obtained directly from the resampled light curve, in the same way as described for SNe~IIb. However, most of the objects in the sample of \cite{2014ApJ...786...67A} do not have a well-observed maximum. In order to estimate the light curve maxima for those objects we searched the literature for SNe~II where the epoch of maximum light was observed. The spectra of these SNe were then used to produce additional SNID templates. SNID was then run for the rest of the sample. If the difference between the age of explosion of the sample spectrum (calculated in the previous section) and the age of explosion of the template is less than five days, the template is considered a good match and it is used to estimate the epoch of maximum light. Again we cross-validate this procedure by removing each SN~II used as a template, running that template through our steps, and comparing the values estimated using SNID with those obtained directly from the re-sampled light curve. To further test this method, the same was done with SNe~IIb. The results are presented in Table \ref{tab:compIIsnid}. Average differences on the maximum light for SNe~II is $-0.93 \pm 4.6$ days, again validating our procedure. 

\begin{table}
  \centering
  \caption{Comparison between the date of maximum light in the $V$-band obtained directly from SN photometry and that obtained from SNID for SNe~II (top) and SNe~IIb (bottom).}
  \label{tab:compIIsnid}
  \begin{tabular}{lccccr} 
    \hline
    \multicolumn{2}{c}{}SNID &
    \multicolumn{2}{c}{}Observed\\\cmidrule(r){2-3}\cmidrule(l){4-5}
    SNII & Max[MJD] & e\_Max & Max[MJD] & e\_Max & diff \\
    \hline
    1999gi & 51529.0 & 3.76 &  51529.2 & 2.20 &  -0.2  \\
    2002gd & 52564.5 & 4.84 &  52562.6 & 2.31 &   1.9  \\
    2004er & 53284.7 & 5.23 &  53287.6 & 1.78 &  -2.9  \\
    2007U  & 54146.1 & 3.97 &  54144.9 & 3.32 &   1.2  \\
    2009ib & 55054.4 & 4.52 &  55057.1 & 1.63 &  -2.7  \\
    2012aw & 56015.7 & 4.72 &  56012.5 & 0.61 &   3.2  \\
    2013ai & 56352.4 & 3.39 &  56365.4 & 1.00 &  -13.0 \\
    2013by & 56413.7 & 4.37 &  56414.7 & 1.10 &  -1.0  \\
    2013ej & 56509.2 & 4.58 &  56510.4 & 0.91 &  -1.2  \\
    2013fs & 56584.3 & 4.46 &  56578.1 & 2.01 &   6.2  \\
    2014cx & 56914.5 & 2.88 &  56917.5 & 0.77 &  -3.0  \\
    2014G  & 56682.0 & 4.34 &  56681.7 & 0.61 &   0.3  \\
    \hline
    SNIIb & Max[MJD] & e\_Max & Max[MJD] & e\_Max & diff \\
    \hline
    1993J  & 49093.6 & 3.6 & 49095.0 & 0.11 & 1.4  \\
    1996cb & 50450.6 & 3.1 & 50450.4 & 1.84 & 0.2  \\
    2003bg & 52718.8 & 3.3 & 52717.6 & 10.3 & 1.2  \\
    2004ex & 53311.3 & 3.0 & 53307.2 & 1.81 & 4.1  \\
    2004ff & 53323.3 & 3.0 & 53313.1 & 1.88 & 10.2 \\
    2005Q  & 53408.1 & 4.0 & 53406.4 & 1.50 & 1.7  \\
    2006ba & 53825.5 & 3.2 & 53823.1 & 2.49 & 2.4  \\
    2006el & 53983.9 & 3.2 & 53983.8 & 1.60 & 0.1  \\
    2006T  & 53780.5 & 4.2 & 53780.3 & 1.74 & 0.2  \\
    2008aq & 54537.0 & 3.3 & 54532.0 & 1.35 & 5.7  \\
    2008ax & 54548.3 & 5.1 & 54549.6 & 0.52 & -1.3 \\
    2008bo & 54569.0 & 3.1 & 54567.6 & 1.21 & 1.4  \\
    2009mg & 55190.6 & 3.3 & 55188.5 & 2.04 & 2.1  \\
    2009K  & 54864.5 & 4.7 & 54869.4 & 2.37 & -4.9 \\
    2009Z  & 54881.4 & 2.7 & 54877.7 & 2.37 & 3.7  \\
    2010jr & -       & -   & 55531.7 & 1.60 & -    \\     
    2011dh & 55733.3 & 6.5 & 55732.1 & 0.88 & 1.2  \\
    2011ei & 55785.1 & 3.0 & 55784.4 & 0.91 & 0.7  \\
    2011fu & 55847.0 & 4.5 & 55846.0 & 0.49 & 1.0   \\
    2011hs & 55889.8 & 3.5 & 55888.2 & 0.75 & 1.6  \\
    2013ak & 56370.4 & 3.6 & 56368.8 & 1.79 & 1.6  \\
    2013df & 56466.3 & 3.2 & 56469.8 & 0.38 & -3.5 \\
    2016gkg & 57671.9 & 4.3 & 57671.7 & 1.92 & 0.2  \\
    \hline
  \end{tabular}
\end{table}

Both SNe~II and SNe~IIb have fewer objects with well-sampled light curve peaks in the $B$- and $r$-band. For those that do have a well-sampled peak, the maximum date is obtained directly from the photometry. For these same SNe we then calculate the weighted average of the difference of JD at peak for $V$ and $B$ together with $V$ and $r$, using as weights the errors on the determination of the date of maximum light. Results are presented in Table \ref{tab:difjdmax}. These average differences are then used to estimate the date of maximum in $B$ and $r$ for the rest of the objects as the difference between the SN maximum epoch in $V$ and the weighted average of the corresponding band.

\begin{table}
  \centering
  \caption{Weighted average and error of $B_\mathrm{max}\,-\,V_\mathrm{max}$ and $V_\mathrm{max}\,-\,r_\mathrm{max}$, considering only those SNe with data points before light curve peak.}
  \label{tab:difjdmax}
  \begin{tabular}{lcccr} 
    \hline
    Type & $\#$SNe & Band  &  Weighted average & error \\
    \hline    
    \multirow{2}{*}{}IIb & 20 & $B_\mathrm{max}\,-\,V_\mathrm{max}$  &   -1.32  & 0.40 \\ 
    & 10 & $V_\mathrm{max}\,-\,r_\mathrm{max}$  &  -0.56  & 0.75 \\
    \hline
    \multirow{2}{*}{}II & 8& $B_\mathrm{max}\,-\,V_\mathrm{max}$  &   -5.03  & 2.14 \\ 
    & 8 & $V_\mathrm{max}\,-\,r_\mathrm{max}$  &  -3.04  & 2.66 \\
    \hline
  \end{tabular}
\end{table}

\section{Resampled SN light curves and observational parameters}
\label{param}

In this section we first present the resampled $V$-band hydrogen-rich SN~II and SN~IIb light curves and discuss their morphology. A number of observational parameters are then defined. These are used to compare the two SN types quantitatively in Section \ref{results}.

\begin{figure}
  \includegraphics[width=\columnwidth]{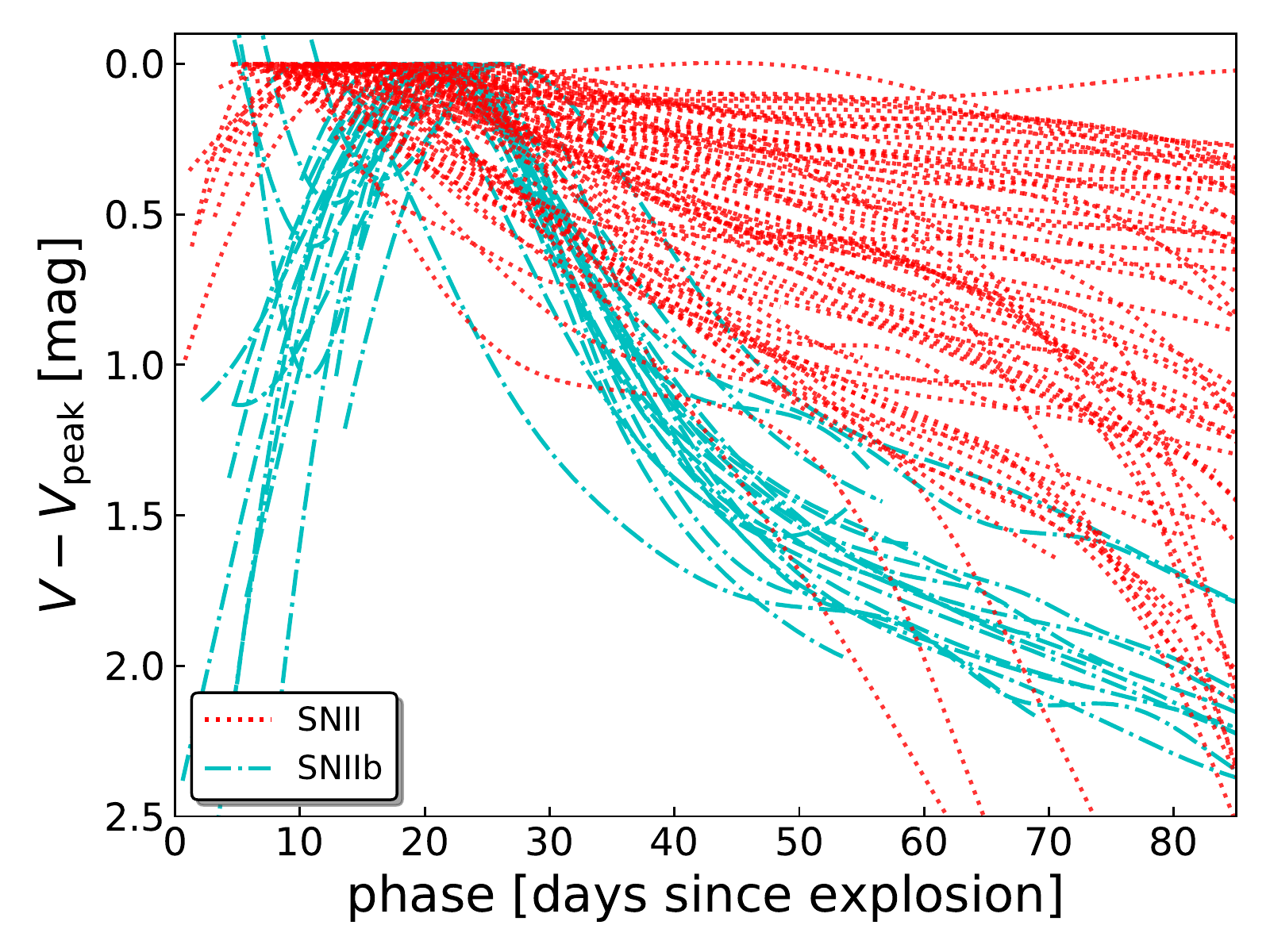}
  \includegraphics[width=\columnwidth]{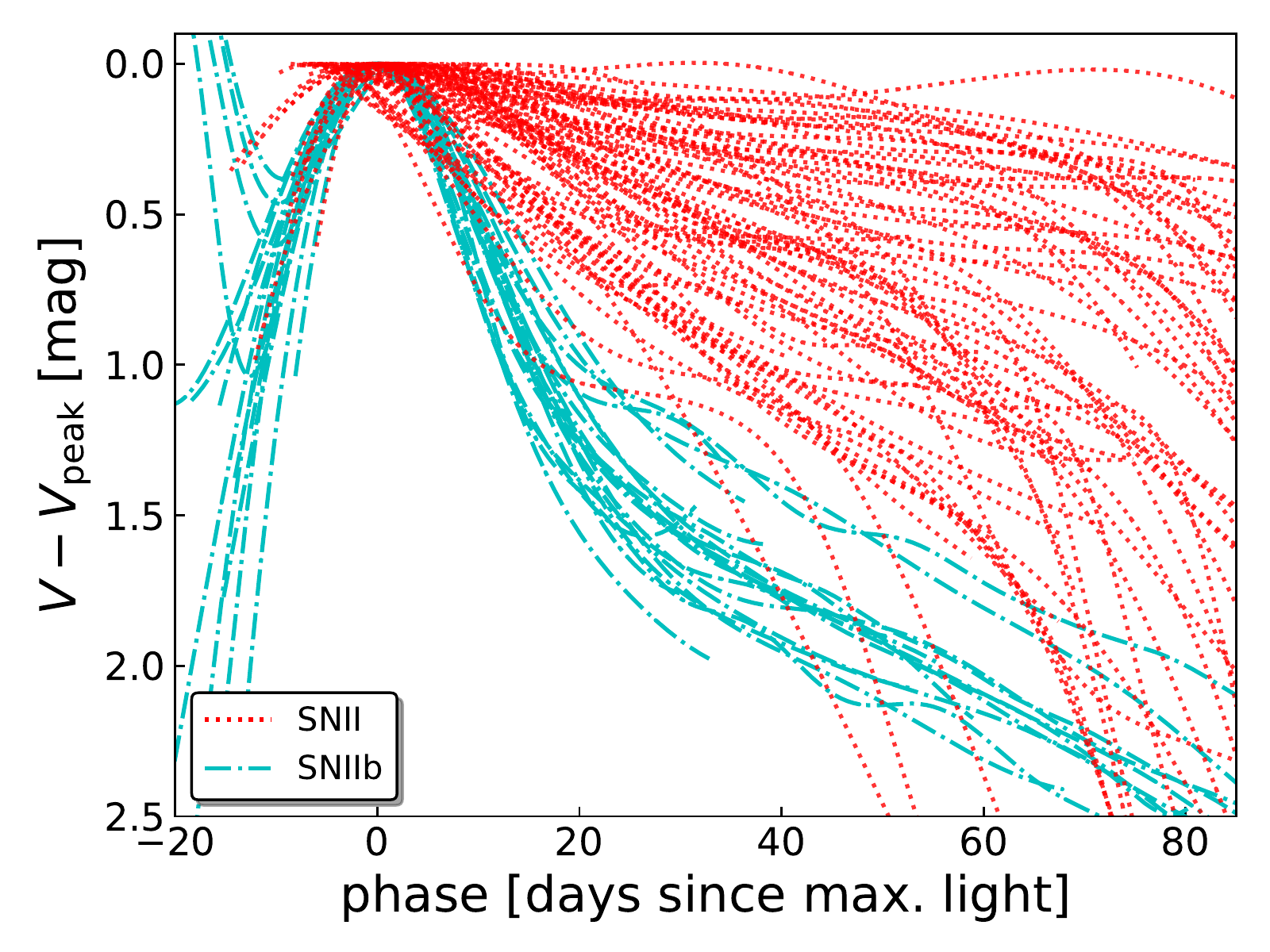}
    \caption{Resampled light curves normalized with respect to the $V$-band peak magnitude. SNe~II are plotted in red dotted lines while SNe~IIb are plotted in cyan dash-dot lines. Top panel: phase with respect to explosion epoch. Bottom panel: phase with respect to date of maximum light. Similar plots are presented for $B$- and $r$-band in the Appendix \ref{apend}.}
    \label{fig:norm}
\end{figure}

In Fig. \ref{fig:norm} we present the resampled light curves normalized to the peak magnitude in $V$-band (if peak magnitude is not observed, the magnitude of the first photometric data point is used), both with respect to explosion date and to epoch of maximum light. Similar plots can be seen for $B$ and $r$ bands in Fig. \ref{fig:normB} and Fig. \ref{fig:normr} respectively.
The different behaviour of SNe~II and SNe~IIb can be seen in a qualitative, visual manner in these figures: SNe~IIb take a longer time to reach maximum light (see top panel of Fig. \ref{fig:norm}) and decline more quickly post maximum (most easily seen in the bottom panel of Fig. \ref{fig:norm}) than SNe~II. SNe~IIb show a change in curvature following the first decline after peak, commonly thought to be the radioactive decay tail. This behaviour is not observed in SNe~II; even though some of them show a change in the curvature, they generally seem to have a small ``plateau'' before continuing to decline.

\begin{figure}
  \includegraphics[width=\columnwidth]{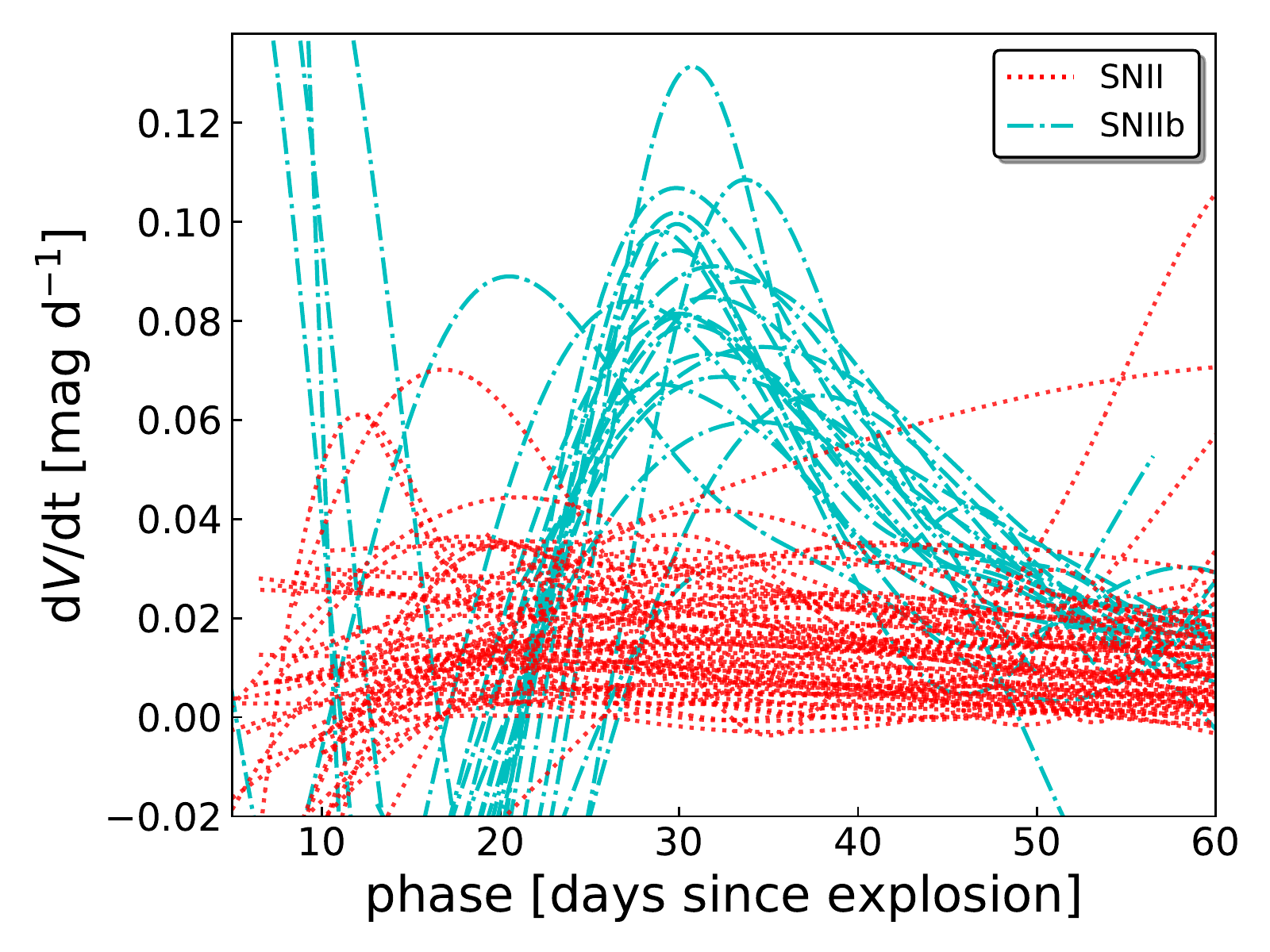}
  \includegraphics[width=\columnwidth]{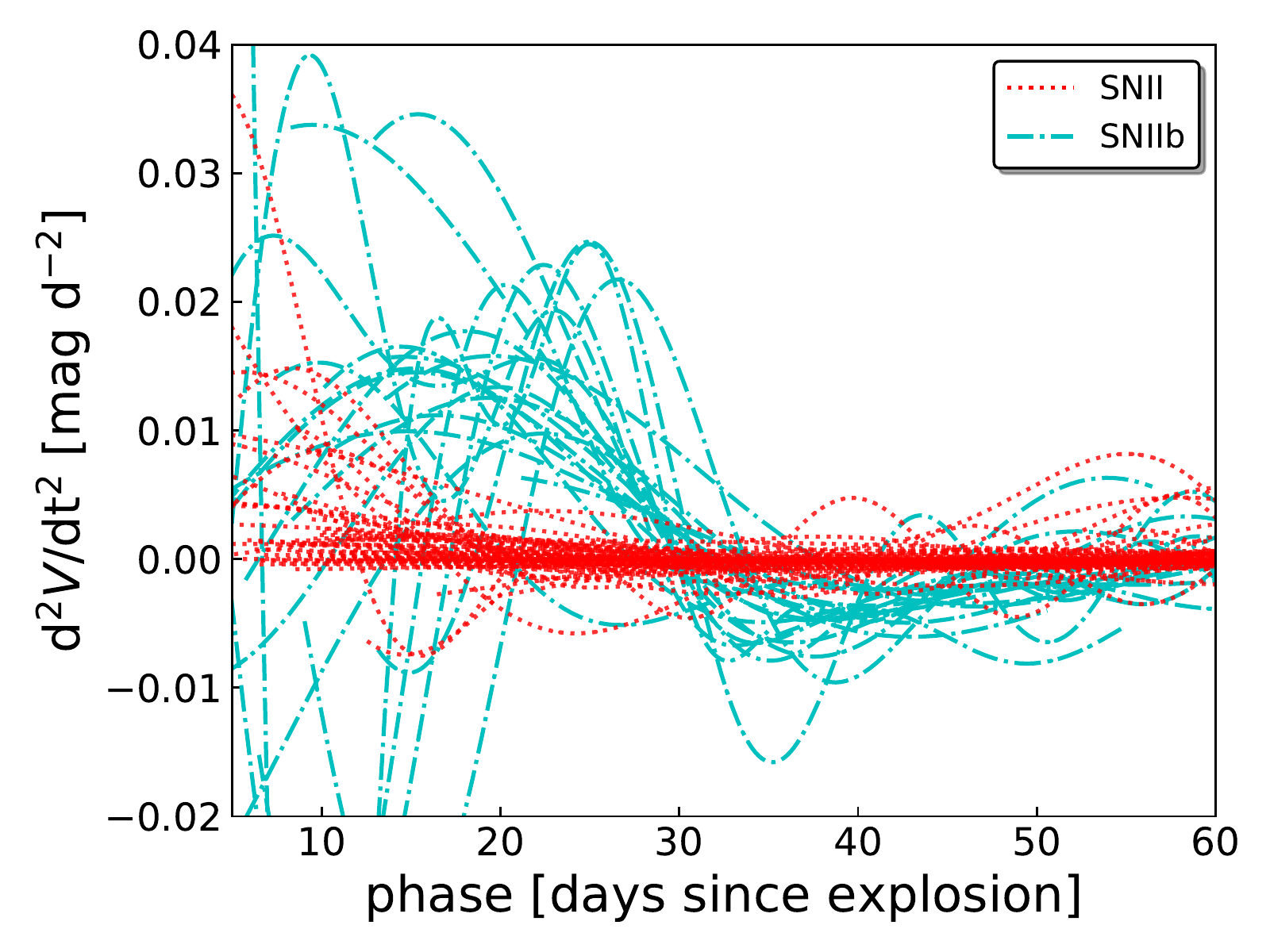}
    \caption{Derivatives of the $V$-band light curves with respect to explosion. SNe~II are plotted in red dotted lines while SNe~IIb are plotted in cyan dash-dot lines. Top panel: first derivative. Bottom panel: second derivative. Similar trends are observed in the $B$- and $r$-band (see Appendix \ref{apend}).}
    \label{fig:der}
\end{figure}

To better study the behaviour of the light curves, the continuous resampling was used to compute their time derivatives, which are presented for $V$ band in Fig. \ref{fig:der}. Similar plots can be seen for $B$ and $r$ bands in Fig. \ref{fig:derB} and Fig. \ref{fig:derr} respectively. Finite differences were performed in order to obtain the first and second derivatives of the light curve. Both derivatives of SN~II are closer to zero than SN~IIb during the first 50 days after explosion, suggesting that their light curve shapes change more slowly than those of SNe~IIb during this period of time. Similarly to Fig. \ref{fig:norm}, Fig. \ref{fig:der} shows that SNe~II and SNe~IIb separate qualitatively in their derived light curves.

The parameters we define in order to quantitatively compare the light curves are described below:
\begin{itemize}
\item {\bf $t_{\mathrm{rise}}$:} time to maximum light or rise time, considered to be the elapsed time between the date of explosion and the date of maximum light.
\item {\bf $\Delta m_{40-30}$:} magnitude difference between phases 30 and 40 days post explosion, which accounts for the decline rate of the light curve. The phases are chosen in order to capture the post-maximum behaviour of both SN types. This parameter is similar to $\mathrm{dm}_{1}$ defined below in the sense that it measures a slope of the light curve. We include it here because it is easier to visualize and to measure when data are sparse.
\item {\bf $\mathrm{dm}_{1}$:} earliest maximum of the first light-curve derivative after maximum light. This parameter captures the maximum rate at which the light curve declines after peak.
\item {\bf $\mathrm{dm}_{2}$:} earliest minimum of the second light-curve derivative after maximum light. This parameter gives an idea of how much the curvature of the light curve changes in the region where it changes the most.
\end{itemize}
To be sure that $\mathrm{dm}_{1}$ and $\mathrm{dm}_{2}$ describe the early part of the light curve we add the constraint that $\mathrm{dm}_{1}$ should occur before 80 days (to exclude the decline after the ``plateau'' of the slow declining SN~II) and  $\mathrm{dm}_{2}$ should occur after $\mathrm{dm}_{1}$ (to exclude the change in curvature that happens before the main peak of the SN~IIb). Some SN~II light curves appear nearly as straight lines. In such cases, we do not provide values of $\mathrm{dm}_{1}$ and $\mathrm{dm}_{2}$ since the derivative of a straight line is a constant.

\section{Results}
\label{results}
\subsection{Rise times}
\label{RT}

\begin{figure}
  \includegraphics[width=\columnwidth]{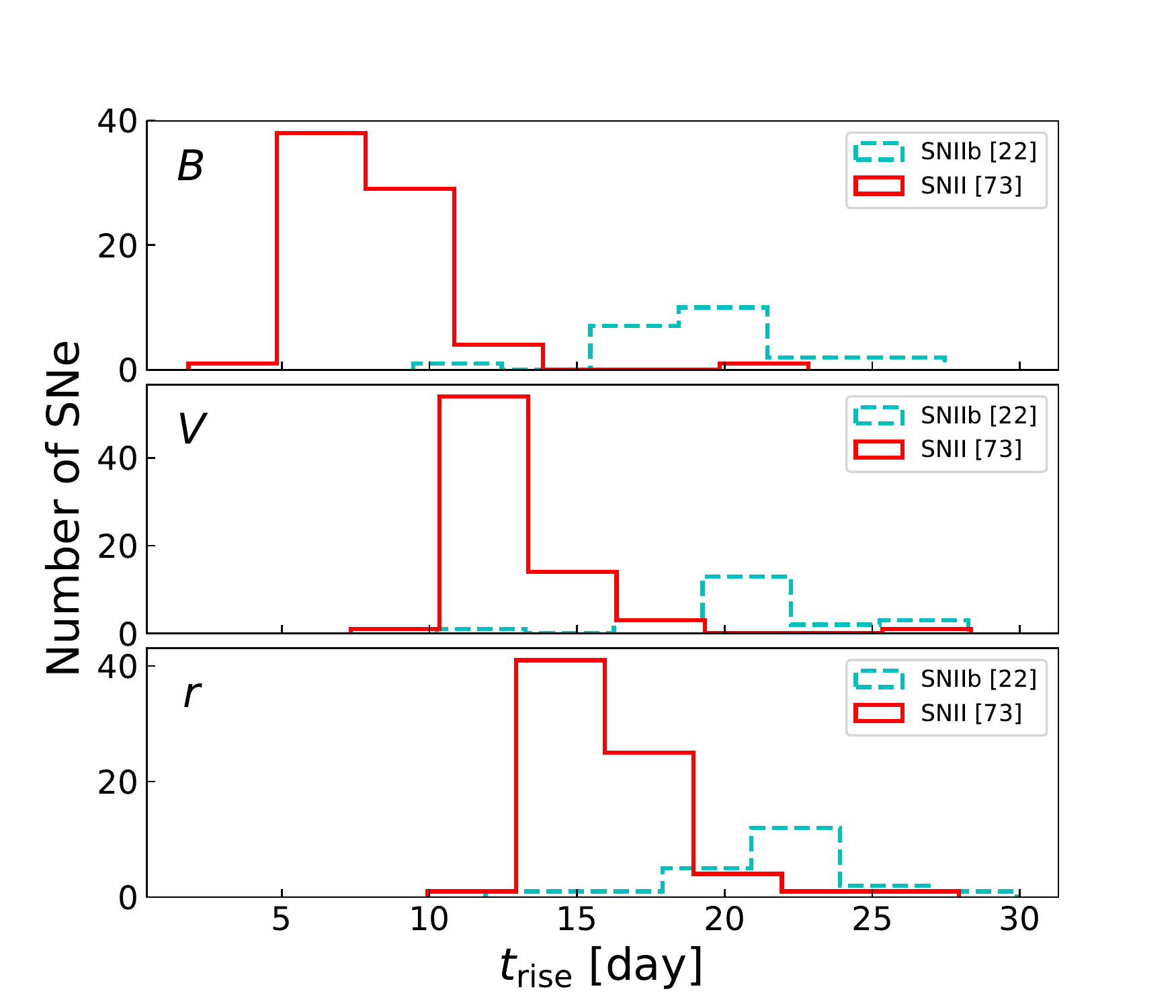}
    \caption{$t_{\mathrm{rise}}$ distributions. $B$-, $V$- and $r$-band from top to bottom. The total number of objects is specified in brackets for each band. SNe~II are plotted in red solid lines and SNe~IIb in cyan dashed lines.}
    \label{fig:risetime}
\end{figure}

In Fig. \ref{fig:risetime} we present histograms of $t_{\mathrm{rise}}$ in the three bands considered in this work. We note that while the mean values of the distributions change considerably between bands for SNe~II (larger values for longer wavelengths), for SNe~IIb the distribution means appear nearly constant (see Table \ref{tab:meanrisetime}). 
This is expected to happen because, on the one hand, the light-curve maximum of SNe~II is attributed to post-shock cooling emission. As the ejecta cools down, the emission peak moves to longer wavelengths and therefore the light-curve maximum is later in redder bands. On the other hand, the light-curve maximum of SNe~IIb is thought to be regulated by the diffusion time-scale of $^{56}$Ni decay radiation, which is nearly constant in the optical. An Ashman's D test (\citealt{1994AJ....108.2348A}) is performed in order to test if there is a possible bimodal behaviour between the distributions of the different parameters, this is accounted for by the D factor. A D value $\geq 2$ suggests bimodality. In this case we obtain $D_{B} \, = \, 3.9$, $D_{V} \, = \, 2.9$ and $D_{r} \, = \, 1.9$, which we interpret as a clear distinction between the $t_{\mathrm{rise}}$ values of both SN types in all three bands. Since the variables under study do not strictly come from Gaussian distributed parent populations, we also performed a Silverman's test (\citealt{10.2307/2985156}) after which two modes can be observed, supporting a distinction between SNe~II and SNe~IIb.

The weighted average of $t_{\mathrm{rise}}$ is calculated for the three bands (see Table \ref{tab:meanrisetime}), using as weights the errors on $t_{\mathrm{rise}}$. We compared these values with those obtained in previous works.
\begin{itemize}
\item \citet{2016MNRAS.459.3939V} present for SNe~II in their Figure 16 a distribution of $t_{\mathrm{rise}}$ vs. $V$-band magnitude that goes from $\sim$0 days to $\sim$20 days. The mean value of $t_{\mathrm{rise}}$ is $\sim$10 days which is consistent with our result;
\item \citet{2016ApJ...820...33R} present the $t_{\mathrm{rise}}$ of 57 SNe~II in the $R$-band (that can be considered similar to $r$-band). From their Table 3 we calculate an average $t_{\mathrm{rise}}$  of 7.74 $\pm$ 2.9 days. Subsequently, \citet{2016ApJ...828..111R} using a similar sample to \citet{2016ApJ...820...33R} found three groups: fast rise/slow decline (II-FS), fast rise/fast decline (II-FF) and slowly rising (II-S). The first two groups with $t_{\mathrm{rise}}$ $\sim$ 6.5 days and the last one with  $t_{\mathrm{rise}}$ $\sim$ 10 days. Furthermore \citet{2015MNRAS.451.2212G} find for SNe~II a $t_{\mathrm{rise}}$ of 7.5 $\pm$ 0.3 days in the $g$-band (that can be compared to the $V$-band). The distributions of these studies are systematically smaller than what we derive for our sample. The studies mentioned before used photometry from surveys with higher cadence than those producing the majority of our sample data, which could be the origin of these differences.
\item \citet{2015A&A...582A...3G} present $t_{\mathrm{rise}}$ of a sample of SNe~II in the $r'/R$-band. From their Table 4 we calculate an average of $10.14 \pm 4.42$ days which is consistent within uncertainties with our values;
\item \citet{2016MNRAS.458.2973P}  present in their Table 7 a list of literature bolometric  $t_{\mathrm{rise}}$ values for SE~SNe. With the values corresponding to SNe~IIb we calculate an average of $15.89 \pm 4.8$ days. Considering that we find that $B$-, $V$- and $r$-band light curves for SN IIb reach their maximum at very similar epochs, we can assume that the bolometric light curves also reach their maxima at a similar time. Their bolometric $t_{\mathrm{rise}}$ is smaller than what we find here, although still compatible within the uncertainties;
\item \citet{2015A&A...574A..60T}  present in their Table 4 values for SNe~IIb $t_{\mathrm{rise}}$ in the $R/r$-band. The resulting average rise time is $23.07 \pm 1.96$ days which is consistent with our results;
\item A subgroup of our SNe~IIb sample is part of the work of \citet{2018A&A...609A.136T}. They present the explosion epoch and the epoch of maximum light in different bands for each object. From this, one can infer the $t_{\mathrm{rise}}$. We calculate an average $t_{\mathrm{rise}}$ for $B$-band of $18.97 \pm 2.73$ days, for $V$-band an average of $20.84 \pm 2.94$ days and for $r$-band an average of $21.78 \pm 3.32$ days which closely match our results.  
\end{itemize}

In summary, the $t_{\mathrm{rise}}$ of SNe~IIb are consistently longer than for SNe~II, both in the current work and in the literature. Note that the computed $t_{\mathrm{rise}}$ average is bigger for our SN~II sample than for other higher cadence samples (see above). While this is a caveat to our estimated $t_{\mathrm{rise}}$ distributions, assuming that our values are slightly overestimated only strengthens our conclusion that SNe~IIb have longer $t_{\mathrm{rise}}$ than SNe~II.

\begin{table}
  \centering
  \caption{Weighted average and $rms$ of $t_{\mathrm{rise}}$ for every band.}
  \label{tab:meanrisetime}
  \begin{tabular}{lcr} 
    \hline
    Type &  Band &  Weighted average   \\
    \hline    
    \multirow{3}{*}{}IIb &$B$ &  19.0 $\pm$ 1.8 \\
    & $V$ &  20.9 $\pm$ 1.6\\ 
    &$r$  & 21.3 $\pm$ 2.1 \\
    \hline
    \multirow{3}{*}{}II &$B$  &   8.3 $\pm$ 2.0\\
    & $V$  &  12.8 $\pm$ 2.4 \\ 
    &$r$  &  16.0 $\pm$ 3.6 \\
    \hline
  \end{tabular}
\end{table}

\subsection{Magnitude difference between 40 and 30 days post-explosion}

\begin{figure}
  \includegraphics[width=\columnwidth]{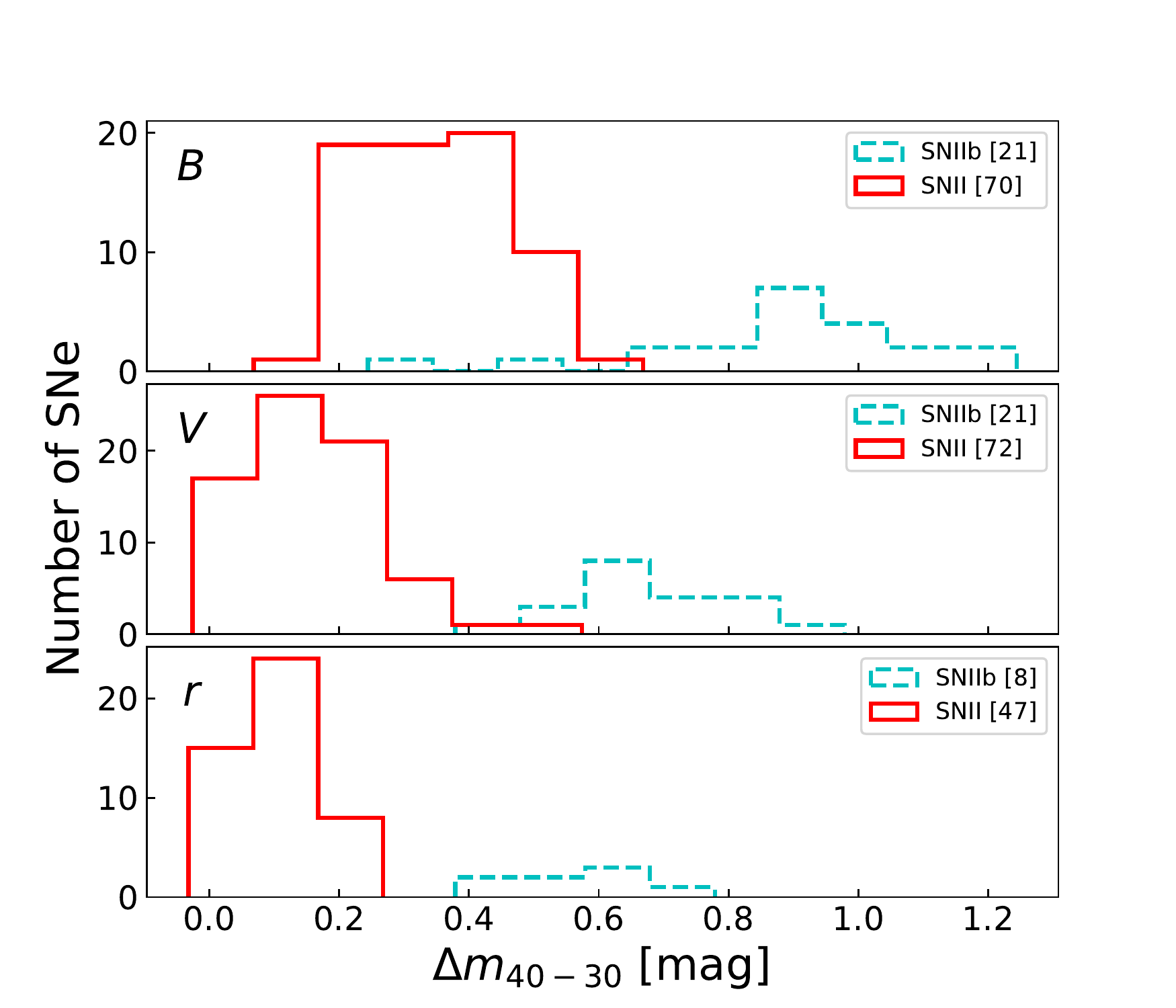}
    \caption{$\Delta m_{40-30}$ distributions. $B$-, $V$- and $r$-band from top to bottom. The total number of objects is specified in brackets for each band. SNe~II is plotted in red solid lines and SNe~IIb in cyan dashed lines.}
    \label{fig:deltamag}
\end{figure}

In Fig. \ref{fig:deltamag} we present histograms of $\Delta m_{40-30}$ for the three bands considered in this work. It can be seen in this figure that both SN populations are separated in the sense that SNe~IIb have larger $\Delta m_{40-30}$ values, which reflects the fact that SNe~IIb decline faster than SNe II after maximum light at these epochs. The Ashman's D values for this parameter are  $D_{B} \, = \, 3.0$, $D_{V} \, = \, 4.7$ and $D_{r} \, = \, 5.6$. The Silverman test also indicates the presence of two modes. These results confirm the visual conclusion that the distributions are bimodal.

\subsection{Derivatives}

\begin{figure}
  \includegraphics[width=\columnwidth]{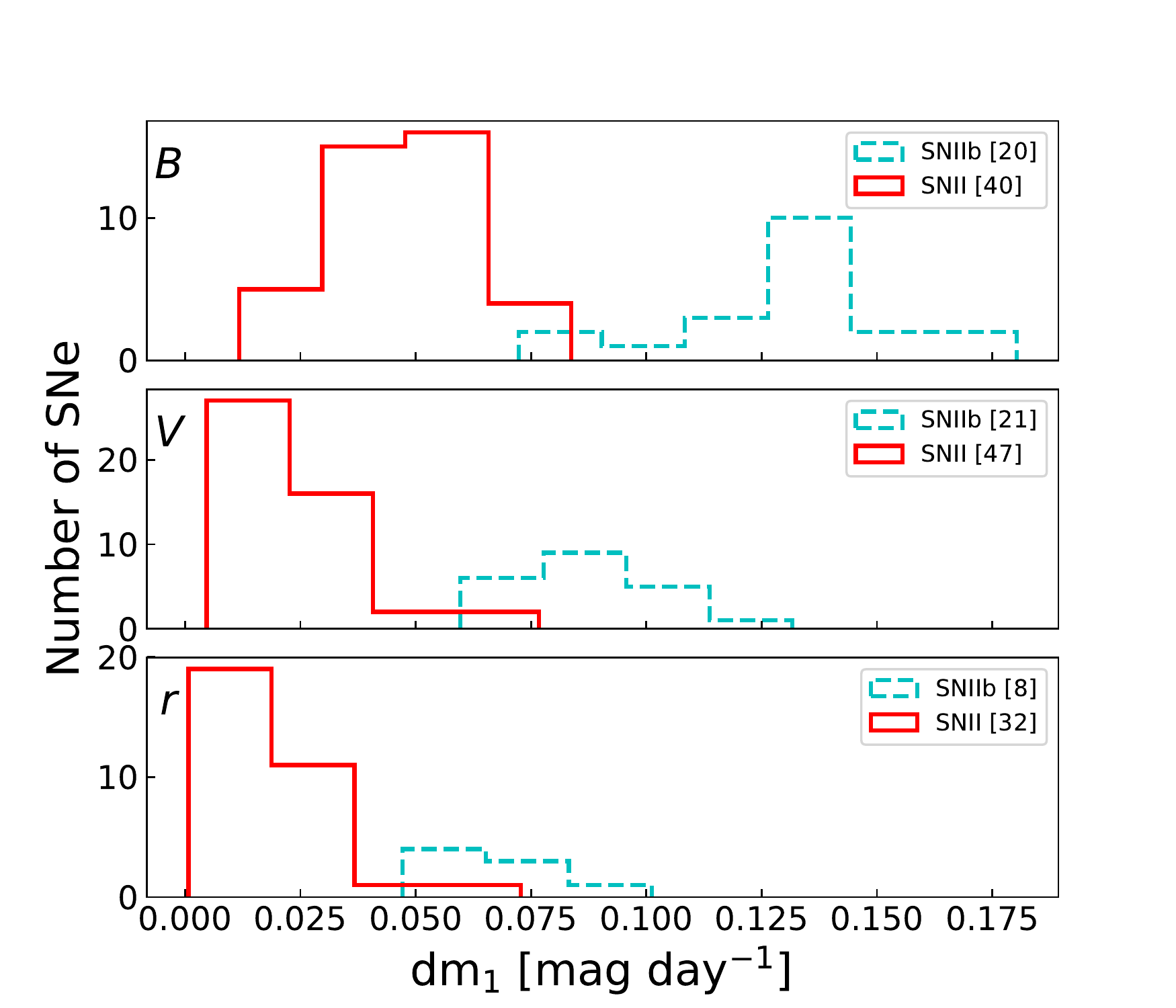}
  \includegraphics[width=\columnwidth]{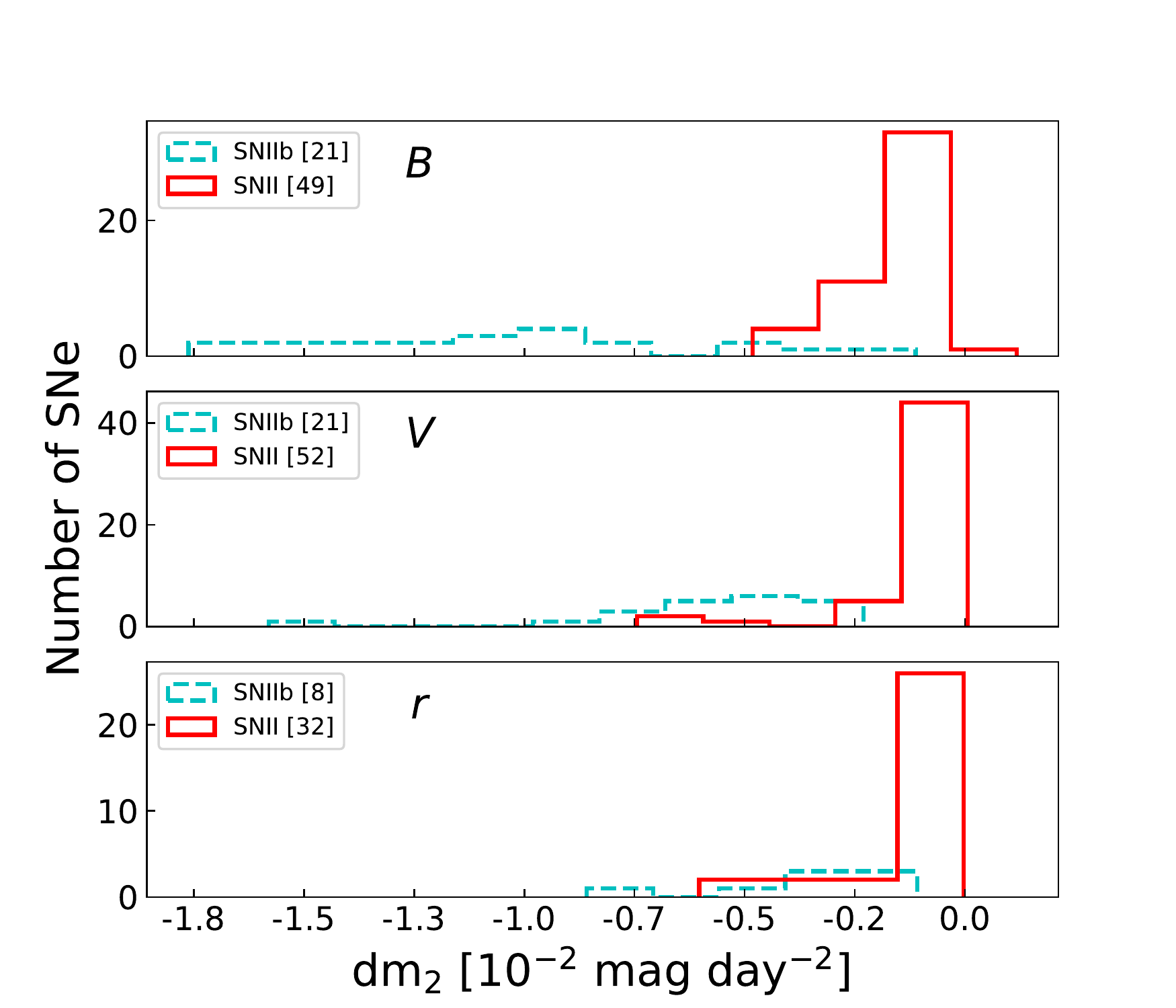}
    \caption{Top: $\mathrm{dm}_{1}$ distributions for all ($B$, $V$ and $r$ from top to bottom) bands. Bottom: $\mathrm{dm}_{2}$ distributions for all ($B$, $V$ and $r$ from top to bottom) bands. The total number of objects is specified in brackets for each band. SNe~II are plotted in red solid lines and SNe~IIb in cyan dashed lines.}
    \label{fig:derdist}
\end{figure}

In Fig. \ref{fig:derdist} we present the distributions of the first maximum of the first derivative ($\mathrm{dm}_{1}$, top panel) and of the first minimum of the second derivative ($\mathrm{dm}_{2}$, bottom panel) both after light curve peak for $B$-, $V$- and $r$-band. It can be seen that generally $\mathrm{dm}_{1}$ is larger for SNe~IIb meaning that they decline faster than SNe~II. The Ashman's D calculated for $\mathrm{dm}_{1}$ are: $D_{B} \, = \, 4.3$, $D_{V} \, = \, 4.2$ and $D_{r} \, = \, 4.1$. The Silverman test also indicates the presence of two modes, evidencing two different distributions for this parameter.
At the same time it can be observed that $\mathrm{dm}_{2}$ values for SNe~IIb are greater in absolute value (i.e., more negative) than those of SNe~II. This indicates that the change in curvature after the initial decline is greater for SNe~IIb than for SNe~II. We obtain $D_{B} \, = \, 2.8$, $D_{V} \, = \, 2.1$, and $D_{r} \, = \, 1.3$. These results imply that there might be only one distribution for $\mathrm{dm}_{2}$ in the $r$-band. However the Silverman test shows two modes for $B$- and $r$-band while showing only one mode for $V$-band, implying that this last band might be the one showing unimodal behaviour. Despite these results, the general shape of the light curves is different between both types of SNe in $V$- and $r$-band.

In general the slope of the decline after the first peak ($\mathrm{dm}_{1}$) of SNe~IIb seems to be smaller for redder bands, but this is not so clear for SNe~II between $V$- and $r$-band. 

\subsection{Comparison between parameters}

\begin{figure}
  \includegraphics[width=\columnwidth]{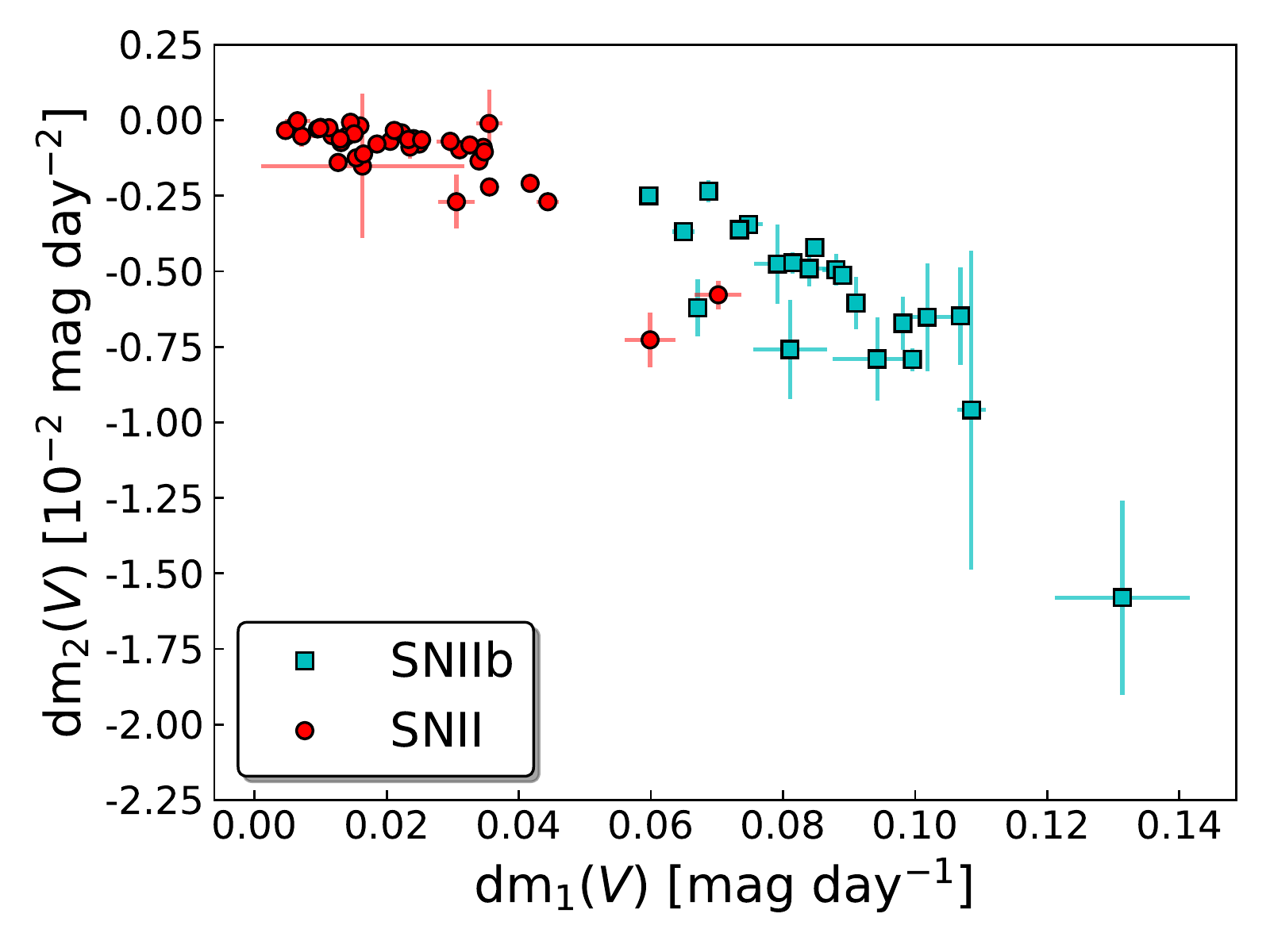}
    \caption{Comparison between $\mathrm{dm}_{1}$ and $\mathrm{dm}_{2}$ in the $V$-band. SNe~II are represented by red circles while SNe~IIb by cyan squares. Some error bars are smaller than the symbol size.}
    \label{fig:compder1der2}
\end{figure}

In Fig. \ref{fig:compder1der2} we present $\mathrm{dm}_{1}$ vs. $\mathrm{dm}_{2}$ in the $V$-band. There seems to be a correlation between these two parameters with a clear trend of increasing $\mathrm{dm}_{1}$  with decreasing $\mathrm{dm}_{2}$. However, the overlap in this parameter plane is very small between both SN types, indicating their distinct light-curve properties. We do not therefore conclude that this trend implies an underlying connection between SN~II and SN~IIb.

\begin{figure}
  \includegraphics[width=\columnwidth]{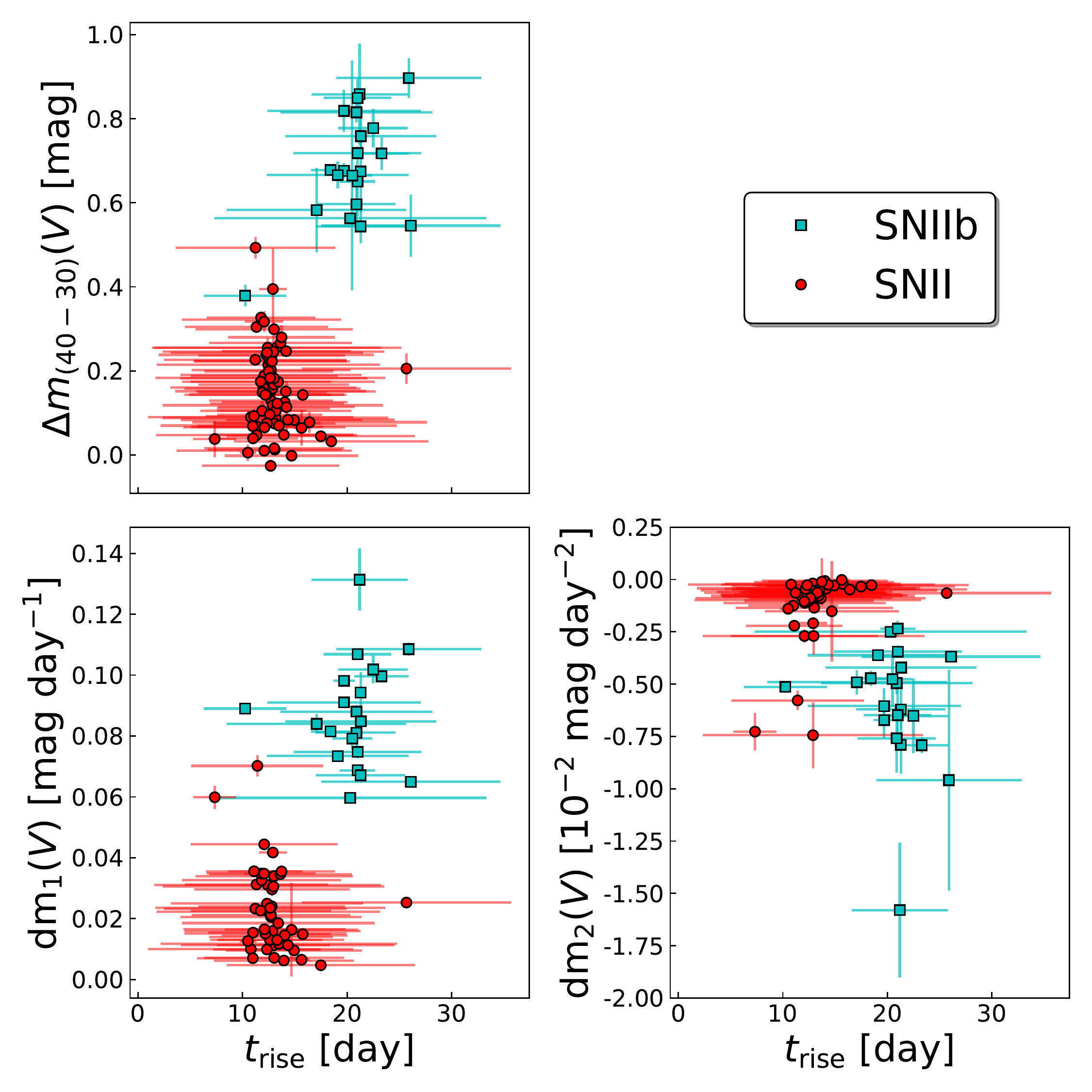}
    \caption{Comparison between parameters in the $V$-band. SNe~II are plotted as red circles while SNe~IIb are plotted as cyan squares. Top left panel: $\Delta m_{40-30}$ vs. $t_{\mathrm{rise}}$. Bottom left panel: $\mathrm{dm}_{1}$ vs. $t_{\mathrm{rise}}$. Bottom right panel: $\mathrm{dm}_{2}$ vs. $t_{\mathrm{rise}}$.}
    \label{fig:risevsall}
\end{figure}

In Fig. \ref{fig:risevsall} we plot each of the parameters described above against $t_{\mathrm{rise}}$ in $V$-band (see Table \ref{tab:resultsvbr} for the numeric values of each parameter). In all the panels it can be seen that SN~II and SN~IIb separate in two groups, with a small number of objects bridging them. These objects are discussed in the next section. We used Gaussian mixture models to performed clustering{\footnote{To perform clustering we used the scikit-learn (\citealt{scikit-learn}) GaussianMixture package.} over these plots in the three studied bands. When plotted out, we see that the ellipses that represent the Gaussian component do not touch. Nevertheless, when studying the $\mathrm{dm}_{2}$ parameter we see that some type II objects fall in the type IIb cluster, these objects are not the same nor the same amount for the different bands, so they are not considered as transitional events. The only systematically  ``misclassified'' object when studying our set of parameters versus $t_{\mathrm{rise}}$ is SN~2013ai, which always appears in the type IIb cluster. SN~2004ff is also ``misclassified'' for most parameters with the exception of $\mathrm{dm}_{1}$ versus $t_{\mathrm{rise}}$ in the $B$-band.}

\section{Discussion}
\label{sec:discussion}

We have studied a large sample of hydrogen-rich SN~II and SN~IIb light curves to test the hypothesis of a continuum of morphological properties among them.
Both qualitatively (visually) and quantitatively, SNe~II are found to have light curves that are distinct from those of SNe~IIb with no suggestion of a continuum of observational properties.
Here we discuss the interpretation of our results in terms of the nature of progenitors properties.

\subsection{Progenitors}

As described in the introduction of this work, it is generally accepted that slow-declining SNe~II (IIP) arise from stars that retain a significant fraction ($\gtrsim$ 2\msun, \citealt{2003ApJ...591..288H}) of their hydrogen rich envelopes and become red supergiants before exploding. Also, there are suggestions that faster-declining SNe~II (IIL)  arise from stars that lose a greater fraction of their hydrogen envelopes than slower-declining SNe~II. The observed continuum of light-curve decline rates among SNe~II may indicate that there is a continuum of progenitor properties, possibly a sequence of envelope masses. At the same time, the spectral transition of SNe~IIb from hydrogen dominated to helium dominated in turn suggests that their progenitors retained significantly less hydrogen than the majority of SNe~II. Therefore SNe~IIb could constitute a continuation of the envelope-mass sequence that is suggested for SNe~II. The dominant mass loss mechanism producing these degree of envelope stripping is still a matter of debate.

One explanation could be mass loss via strong winds or eruptions. In this scenario Wolf-Rayet (WR) stars have been considered probable progenitors of SE SNe.
However, single stars are only predicted to lose sufficient mass to explode as SE SNe when their initial masses are higher than 25\msun\ (see e.g. \citealt{2003ApJ...591..288H}).
Nevertheless, the rate of SE SNe is significantly higher than that predicted assuming single-star progenitors and a standard initial mass function (IMF, \citealt{2011MNRAS.412.1522S}). Also, the ejecta masses of SE SNe are lower than those expected from $>$ 20\msun\ progenitors, presenting a problem for single-star progenitors (e.g. \citealt{2011ApJ...741...97D}, \citealt{2016MNRAS.457..328L}, \citealt{2018A&A...609A.136T}). Nonetheless, recent stellar models using enhanced mass loss suggest that SN~IIb (and Ib) could come from stars $\sim$ 20-25\msun\ (\citealt{2013ApJ...764...21C}). In this case, a single-star progenitor continuum could be considered between slow-declining SNe~II (IIP, with the most hydrogen retained by their progenitors), fast-declining SNe~II (IIL, with less hydrogen) and SNe~IIb (with the least amount of hydrogen left in their progenitors).

In the second and more popular explanation for SNe~IIb progenitor's mass loss, hydrogen envelope stripping is achieved through the presence of a companion. This is supported by the strong evidence that the majority of massive stars reside in interacting binary systems (\citealt{2012Sci...337..444S}) and by the observed fractions of supernova types (\citealt{2011MNRAS.412.1522S}). While the confirmed progenitors of SNe~II are compatible with single-star evolutionary models (\citealt{2012IAUS..279..110V}, \citealt{2015PASA...32...16S} and references therein), those of SNe~IIb do not comply with this picture and appear to indicate a binary origin. Furthermore, for three SNe~IIb there is evidence of a binary companion (SN 1993J \citealt{2004Natur.427..129M}, SN 2001ig \citealt{2018ApJ...856...83R} and SN 2011dh \citealt{2011ApJ...739L..37M}, \citealt{2014ApJ...793L..22F}).

Progenitor properties can be indirectly studied via the observed properties of the SN light curves. The continuum observed between slow and fast declining SN~II light curves has been attributed to a possible continuum of progenitor properties.
In this work we do not find evidence for a continuum between the observed properties of hydrogen-rich SN~II and SN~IIb light curves. This may indicate that these events arise from different progenitor scenarios.

\subsubsection{Constraints from hydrodynamical modelling}

Using hydrodynamical models one can test whether the hydrogen rich envelope mass is the driving factor that shapes the light curves of SNe~II and IIb. Such a test has been carried out by simulating the explosion of stars with varying envelope masses using our own hydrodynamic code (\citealt{2011ApJ...729...61B}). A similar study was done by \citet{2015ApJ...814...63M}. We have calculated SN progenitors using the stellar evolution code MESA version 10398 (\citealt{2011ApJS..192....3P}, \citealt{2013ApJS..208....4P}). An  initial mass of 18 and 30\msun\ has been adopted, and produced pre-SN objects of different envelope masses by artificially varying the wind efficiency parameter, $\eta$, within the ``Dutch'' wind scheme (\citealt{1988A&AS...72..259D}, \citealt{2000A&A...360..227N}, \citealt{2001A&A...369..574V}, \citealt{2009A&A...497..255G}). We evolved the stellar models from the pre-main-sequence to the time of core collapse, defined as the moment when any part of the core exceeds an infall velocity of 1000 km/s. All pre-SN objects were exploded assuming a fixed energy of 10$^{51}$ erg and a radioactive $^{56}$Ni mass of 0.02\msun. A complete analysis based on these tests is deferred to a future publication, here we briefly describe these results to aid in our discussion and conclusions. 

The resulting light curves (Fig. \ref{fig:model}) are similar to those in  \citealt{2015ApJ...814...63M} (see their Fig. 7). However, our conclusions are distinct. We note that as the envelope mass decreases, the light curves decline faster during the recombination phase. This is in line with the range of SN~II light-curve morphologies from ``plateau'' to ``linear''. When the envelope mass falls below 0.5\msun\, the light curves show a bump that \citet{2015ApJ...814...63M} interpret as a signature of SNe~IIb. However, the properties of the computed light curves differ substantially from those of observed SNe~IIb. The bump luminosity is too low, and its timing and duration are too short to account for the observed SNe~IIb light curves. Also, the duration of the cooling phase preceding the bump is not compatible with observations. Our interpretation is that the envelope mass regulates on the zeroth order the shape of SN~II light curves, but it cannot account for SNe~IIb. Although, for the latter, there is evidence that a difference of $\sim$0.1\msun\ of hydrogen in the envelope divides SN~IIb into two categories: those with a compact progenitor (SN~cIIb) and those with an extended one (SN~eIIb, \citealt{2010ApJ...711L..40C}). Further investigation of this phenomenon is beyond the scope of this work. 

A recent population synthesis study of binary systems combined with explosion models presented by \citet{2018PASA...35...49E} claims that the variety of progenitor systems can produce a continuum of light-curve shapes including SNe~II (plateau and linear) and SNe~IIb. However, the synthetic light curves that they classify as of Type IIb do not comply with the observed properties (similarly to above). As seen in their Fig. 4, the SN~IIb models are too faint in general, and while some light curves are too wide, others peak too early or show a peculiar initial peak. Previous studies have shown that in order to reproduce the light-curve shapes of SNe~IIb, the outermost density structure of the progenitor needs to be modified relative to those of standard evolutionary models (see Fig. 1 in \citealt{2012ApJ...757...31B}). 

We conclude that the observed discontinuity in light-curve morphology between SNe~II and SNe~IIb must involve progenitor properties in addition to the mass of the hydrogen rich envelope. Further analysis is required to elucidate what such properties are and what the role of binarity is in determining them.

\subsection{SNe~II light-curve morphology: a clear continuum}

As mentioned in Sec. \ref{sec:intro}, most contemporary work on samples of SNe~II (this refers to all those events historically classified as IIP and IIL) conclude that a continuum exists between the slowest declining SNe~II and the faster declining events. While here our aim was not to further explore the IIP-IIL classification, it is clear that in all of our measured parameters - that are distinct from those defined in previous works, see e.g. \cite{2014ApJ...786...67A} - there is no evidence for bimodality. This work therefore further supports the existence of a continuum in the properties of slow and fast declining SNe~II.

\subsection{Possible transitional/outlying events} 
\label{out}

All of the qualitative and quantitative results and discussion we have presented so far have concluded that there is no clear continuum in light-curve properties between hydrogen-rich SNe~II and SNe~IIb. However, in several figures (see Fig. \ref{fig:compder1der2} and \ref{fig:risevsall}) there are a small number of SNe that appear to bridge the two populations. In this subsection we investigate whether any of these events could be considered transitional by studying their photometric and spectroscopic properties in more detail.

\subsubsection{Type II}
\begin{itemize}
  \item {\bf SN~2013ai:} This is the most noticeable SN~II outlier since it has an exceptionally large $t_{\mathrm{rise}}$ when compared to the rest of our SNe~II sample, it is also the only systematically ``misclassified'' object when performing a clustering analysis. We have checked that none of the other parameters considered indicated this object belonged to the SN~IIb group, meaning that the behaviour of the light curve (steepness and decline rate after peak) is otherwise typical of SNe~II. Nevertheless, we have inspected its spectra for similarities to SNe~IIb. None are found, and SN~2013ai also shows typical spectral features and spectral evolution to those of SNe~II. Therefore we consider it to be a spectroscopically normal SN~II with a slow rise. 
  \citet{2016ApJ...828..111R} study 59 SNe~II in the $R$-band. They find three different clusters of objects, one only composed of slowly rising events (II-S cluster). However, their II-S sample have a median $t_{\mathrm{rise}}$ of ~10 days while SN~2013ai has a $t_{\mathrm{rise}}$ of 25.5$\pm$10.5 days in the $r$-band. Therefore (even considering the error) the $t_{\mathrm{rise}}$ of SN~2013ai is still significantly longer than that of the II-S. This object is further analyzed by Davis et al. (in preparation).
  \item {\bf SN~2006Y:} This SN has large $\mathrm{dm}_{1}$ and $\mathrm{dm}_{2}$ values when compared to the rest of the SNe~II sample. SN~2006Y is an outlier on the observed trends of the \citet{2014ApJ...786...67A} sample: they find it to be the fastest decliner of their sample; this SN is again an outlier in the sample of \citet{2014ApJ...786L..15G}, showing the smallest H$\alpha$ absorption of their sample. However, its $t_{\mathrm{rise}}$ is typical of SNe~II and its light curve has a small ``plateau'' a few days after peak that is not observed in any SNe~IIb. Also it shows strong hydrogen lines throughout its spectral evolution. Therefore, while SN~2006Y shows abnormal properties for a SN~II, these do not position this event close to the parameter space of SNe~IIb.
    \item {\bf SN~2013fs:} This object also has large $\mathrm{dm}_{1}$ and $\mathrm{dm}_{2}$, when compared to the rest of the SNe~II sample. In all other parameters ($t_{\mathrm{rise}}$, $\Delta m_{40-30}$, overall shape of the light curve, spectral evolution) it behaves as a normal SN~II. 
\item {\bf SN~2008M:} This SN does not present any peculiarities except for a large $\mathrm{dm}_{2}$. While visual inspection of its light curve shows an early fast decline, later it has a normal ``plateau'' lasting for approximately 80 days. In no other parameter does this event appear similar to SNe~IIb.
\end{itemize}

\subsubsection{Type IIb}
\begin{itemize}
\item {\bf SN~2004ff:} This SN has an unusually small $t_{\mathrm{rise}}$ value. Checking our estimated time of maximum light we find that it is in agreement with that from \citet{2018A&A...609A.136T}. However, investigating the pre-discovery non-detection values this SN may have a poorly constrained explosion epoch. This conclusion arises from the fact that the non-detection limiting magnitude is 18~mag (Oct. 21.4) while the discovery magnitude is 17.8~mag (Oct. 28.39, \citealt{2004IAUC.8425....1P}). Such a shallow non-detection may not give an accurate constraint on the explosion epoch. If we consider the explosion epoch obtained with SNID - which agrees within uncertainties with that obtained by \citet{2018A&A...609A.136T}  via a fit to the photospheric radius before $r_\mathrm{max}$ - the $t_{\mathrm{rise}}$ agrees with that of the rest of the SNe~IIb sample. Examining the other parameters defined in this work we notice that $\Delta m_{40-30}$ seems to be positioning this object among SNe~II but not $\mathrm{dm}_{1}$. These two parameters account for steepness of the light curve but $\mathrm{dm}_{1}$ is independent of the explosion epoch and thus it is not affected by any possible error in the determination of the explosion epoch. The value of $\mathrm{dm}_{2}$ for SN~2004ff is consistent with the rest of the SN~IIb population. Finally, if the explosion epoch is wrong, the light curve would look like those of typical SNe IIb (with typical $t_{\mathrm{rise}}$ and $\Delta m_{40-30}$), but shifted in time. Therefore we conclude that it is not a transitional event.
  \end{itemize}

\subsubsection{The Type II SN~2001fa and Type II(?) SN~2007fz}

\begin{figure}
  \includegraphics[width=\columnwidth]{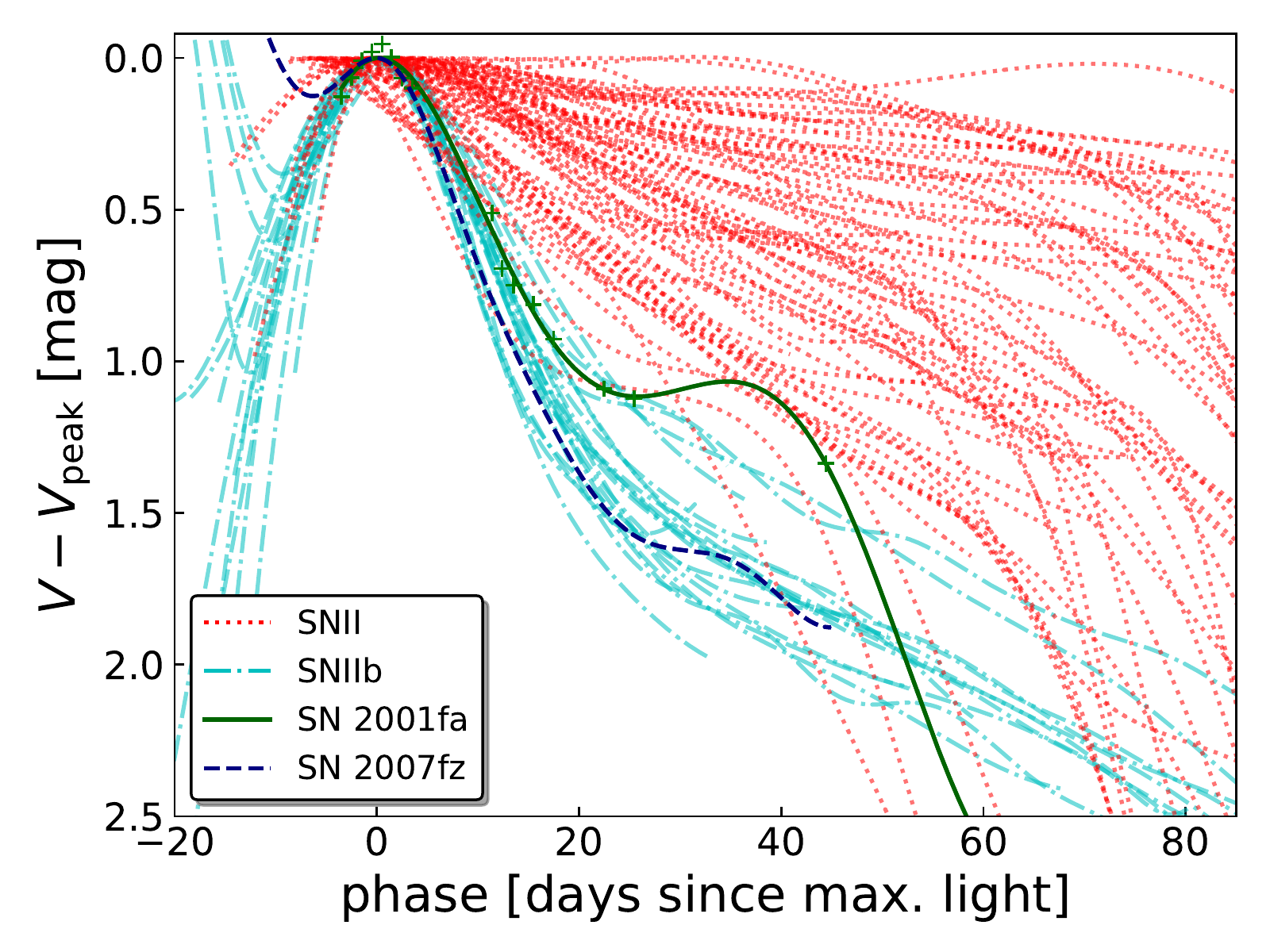}
    \caption{Resampled light curves normalized with respect to the $V$-band peak magnitude, SNe~II are plotted in red dotted lines while SNe~IIb are plotted in cyan dash-dot lines. In green solid lines is SN~2001fa (the photometric data points were added as green crosses to show the sparse sampling after 20 days) and in blue dashed lines SN~2007fz.}
    \label{fig:faran}
\end{figure}

These two objects are not included in our sample but they are in the \citet{2014MNRAS.445..554F} sample. Those authors studied 11 fast declining hydrogen-rich SNe~II and argued that SN~2001fa and SN~2007fz (shown in their Figure 8) could be transitional events between SNe~II and SNe~IIb. Given that within our sample we find a lack of transitional events, we decided to examine these two SNe in more detail. In Fig. \ref{fig:faran} the light curves of SN~2001fa and SN~2007fz are plotted in comparison to the SN~II and SN~IIb samples analyzed in our work.

The resampled light curve of SN~2001fa appears to be more consistent with the overall morphology of SN~II than SN~IIb light curves, showing a small ``plateau'' (however the cadence of the photometry at these epochs is not optimal). In addition, according to SNID  SN~2001fa is consistent with it being a SN~II, and therefore we conclude that this is not a transitional event. 

The resampled shape of the light curve of SN~2007fz is compatible with those of SNe~IIb. The $V$-band rise time (18.2 days) is also consistent with SNe~IIb. Regarding the spectra, \citet{2014MNRAS.445..554F} argued that they were similar to those of SNe~II. To test this we decided to run the four available spectra through SNID (after removing the narrow emission lines of host galaxy contamination). The SNID classification output is not conclusive as for the first three spectra the best match is a SN~IIb, while for the last one the best match is a SN~II. We further examined this last SNID classification output and notice that the SNe~II templates match quite well the H$\alpha$ emission for the studied spectra but they do not match the H$\alpha$ absorption nor the He lines. On the contrary, the SNe~IIb templates show an additional dip in the H$\alpha$ emission that is not seen on the studied spectra (as described by \citealt{2014MNRAS.445..554F}) but, match well both the H$\alpha$ absorption and the He lines. We conclude that SN~2007fz is most probably a Type IIb SN and not a SN~II (as claimed by \citealt{2014MNRAS.445..554F}).

\subsection{SNe~IIb without observed maxima}

In Section \ref{sec:intro} it is stated that only SNe~IIb with data points around maximum are considered. This could introduce a sample bias because there may be SNe~IIb with shorter rise times where the rise was not observed because it occurred sooner after explosion. Therefore, we searched the Open Supernova Catalog (\citealt{2017ApJ...835...64G}) for SNe~IIb that met the same criteria we use to select our sample but without observed photometric data points around maximum light. We found two objects (see Table \ref{tab:snIIbNOMAX}) that satisfied these requirements. The explosion epoch and date of maximum light were obtained by spectral matching using SNID as described in Sections \ref{exp} and \ref{sec:ml}. The mean $t_{\mathrm{rise}}$ for these objects $V$-band is 19.6 days, which is consistent with our results. Even though these are only two examples, the result suggests that our photometric selection criteria are not driving our results on the significant differences in rise times between SNe~II and SNe~IIb.

\section{Conclusions}
\label{sec:conc}
We present a morphological analysis of $B$-, $V$- and $r$-band light curves for 95 SNe (73 SNe~II and 22 SNe~IIb). Overall, we conclude that the two samples show distinct light-curve shapes and seem to form a bimodal distribution in all the defined parameters. In the $V$-band a $t_{\mathrm{rise}}$ of 17d, $\Delta m_{40-30}$ of 0.4mag, $\mathrm{dm}_{1}$ of 0.05mag$\cdot$day$^{-1}$ and $|\mathrm{dm}_{2}|$ of 0.003mag$\cdot$day$^{-2}$ serve as an approximate threshold between both types of SNe. In the $B$-band these thresholds are $t_{\mathrm{rise}}$ $=$ 13.7d, $\Delta m_{40-30}$ $=$ 0.6mag, $\mathrm{dm}_{1}$ $=$ 0.09mag$\cdot$day$^{-1}$ and $|\mathrm{dm}_{2}|$ $=$ 0.006mag$\cdot$day$^{-2}$. Finally in the $r$-band, $t_{\mathrm{rise}}$ $=$ 18.8d, $\Delta m_{40-30}$ $=$ 0.3mag, $\mathrm{dm}_{1}$ $=$ 0.04mag$\cdot$day$^{-1}$ and $|\mathrm{dm}_{2}|$ $=$ 0.002mag$\cdot$day$^{-2}$ can be used as thresholds, with lower values corresponding to SNe~II and higher (absolute) values corresponding to SNe~IIb.

These results argue against a continuum of observed properties between SN~II an SN~IIb.
Our main conclusions are:
\begin{itemize}
\item The clear separation seen between both SN types suggests that they may arise from distinct progenitor channels. Alternatively, even if they share a common origin, there must be a physical property (or a combination thereof) that under a continuum variation produces a sudden change in the resulting SN properties.
\item All the studied parameters are on average larger for SNe~IIb. Larger $\mathrm{dm}_{1}$ indicates that their post peak decline is steeper than that of SNe~II. Larger $|\mathrm{dm}_{2}|$ indicates that their change after the first decline post maximum is more pronounced than that of SNe~II.
\item Our findings agree with the recent claims that SNe~II form a single continuous family (\citealt{2014ApJ...786...67A}, \citealt{2015ApJ...799..208S}, \citealt{2016AJ....151...33G}, \citealt{2016ApJ...828..111R}). They also agree with the observed separation between SNe~II and IIb previously suggested by \citet{2012ApJ...756L..30A} and hinted by \citet{2016ApJ...828..111R} when they mention that the outliers in their SNe~II study are consistent with type IIb objects.
\end{itemize}

%
\section*{Acknowledgements}

This research has made use of the NASA/IPAC Extragalactic Database (NED), which is operated by the Jet Propulsion Laboratory, California Institute of Technology, under contract with the National Aeronautics and Space Administration. This material is based upon work supported by the US National Science Foundation (NSF) under grants AST-0306969, AST-0607438, AST-0908886, AST-1008343, AST-1211916, AST-1613426, AST-1613455, and AST-1613472, as well as in part by a  grant from the Danish Agency for Science and Technology and Innovation through a Sapere Aude Level 2 grant (PI M.D. Stritzinger). E.Y.H. and S.D acknowledges the support provided by the National Science Foundation under Grant No. AST-1613472. M. D. Stritzinger acknowledges support from a research grant (13261) from VILLUM FONDEN and a project grant  (8021-00170B) from IRFD (Independent Research Fund Denmark) grant. P.J.P acknowledges Santiago Gonz\'alez-Gait\'an and Claudia Guti\'errez for their help at the beginning of this work. G.F. acknowledge support from  ANPCyT grant PICT-2015-2734 ``Nacimiento y Muerte de Estrellas Masivas: Su relaci\'on con el Medio Interestelar.'' M.B., G.F., and P.J.P. acknowledge support from grant PIP-2015-2017-11220150100746CO of CONICET ``Estrellas Binarias y Supernovas''. 





\bibliographystyle{mnras}
\bibliography{biblio} 




\appendix

\section{Additional results}
\label{apend}

\begin{landscape}
  \begin{table}
    \caption{SNe~II sample (same in other tables). References: (1$^\mathrm{p}$) \citet{2002AJ....124.2490L}; (2$^\mathrm{p}$) \citet{2016AJ....151...33G}; (3$^\mathrm{p}$) \citet{2014ApJ...786...67A} and Contreras et al. in prep.; (4$^\mathrm{p}$) \citet{2015MNRAS.450.3137T}; (5$^\mathrm{p}$) \citet{2014ApJ...787..139D}; (6$^\mathrm{p}$) \citet{2016MNRAS.459.3939V}; (7$^\mathrm{p}$) \citet{2015MNRAS.448.2608V}; (8$^\mathrm{p}$) \citet{2016MNRAS.461.2003Y}; (9$^\mathrm{p}$) \citet{2016ApJ...832..139H}; (10$^\mathrm{p}$) \citet{2016MNRAS.462..137T}. (1$^\mathrm{s}$) WISeREP (\citealt{2012PASP..124..668Y}); (2$^\mathrm{s}$) \citet{2017ApJ...850...89G};(3$^\mathrm{s}$) \citet{2002AJ....124.2490L};(4$^\mathrm{s}$) \citet{2017PASP..129e4201S}; (5$^\mathrm{s}$) \citet{2015MNRAS.450.3137T}; (6$^\mathrm{s}$) \citet{2014ApJ...787..139D}; (7$^\mathrm{s}$) \citet{2016PASA...33...55C}; (8$^\mathrm{s}$) \citet{2015MNRAS.448.2608V}; (9$^\mathrm{s}$) \citet{2017ApJ...848....5B}; (10$^\mathrm{s}$) \citet{2016MNRAS.461.2003Y}; (11$^\mathrm{s}$) \citet{2014MNRAS.438L.101V}; (12$^\mathrm{s}$) \citet{2016ApJ...822....6D}; (13$^\mathrm{s}$) \citet{2015ApJ...806..160B}; (14$^\mathrm{s}$) \citet{2017NatPh..13..510Y};(15$^\mathrm{s}$) \citet{2016MNRAS.462..137T}.}
    \label{tab:snII}
    \begin{threeparttable}
    \begin{tabular}{lcccccccr} 
      \hline
      SN &   z\_host & Exp[MJD]  & V\_Max[MJD]  &  B\_Max[MJD] & r\_Max[MJD] & Phot. Ref. & Spec. Ref & Circular\\
      \hline 
      1999gi &   0.0020 &  51518.2(3.7)            &   51529.2(2.2)           & 51526.3(2.1)           &  51532.2\tnote{d} (3.4) & (1$^\mathrm{p}$) & (1$^\mathrm{s}$);(3$^\mathrm{s}$);(4$^\mathrm{s}$) & IAUC 7329\\  
      2002fa &   0.0600 &  52502.5\tnote{a} (8.0)  &   52515.9\tnote{c} (4.6) & 52510.9\tnote{d} (5.0) &  52518.9\tnote{d} (5.3) & (2$^\mathrm{p}$) & (2$^\mathrm{s}$)  & IAUC 7967\\
      2002gd &   0.0089 &  52551.5\tnote{a} (4.0)  &   52562.6(2.3)           & 52558.9(2.7)           &  52565.6\tnote{d} (3.5) & (2$^\mathrm{p}$) & (2$^\mathrm{s}$)  & IAUC 7986\\
      2002gw &   0.0103 &  52553.5\tnote{a} (8.0)  &   52572.0\tnote{c} (4.7) & 52567.0\tnote{d} (5.2) &  52575.0\tnote{d} (5.4) & (2$^\mathrm{p}$) & (2$^\mathrm{s}$)  & IAUC 7995\\
      2002hj &   0.0236 &  52562.5\tnote{a} (7.0)  &   52575.5\tnote{c} (5.6) & 52570.5\tnote{d} (6.0) &  52578.5\tnote{d} (6.2) & (2$^\mathrm{p}$) & (2$^\mathrm{s}$)  & IAUC 8006\\
      2002hx &   0.0310 &  52582.5\tnote{a} (9.0)  &   52595.4\tnote{c} (5.6) & 52590.4\tnote{d} (6.0) &  52598.5\tnote{d} (6.2) & (2$^\mathrm{p}$) & (2$^\mathrm{s}$)  & IAUC 8015\\
      2002ig &   0.0770 &  52570.5\tnote{a} (5.0)  &   52584.7\tnote{c} (3.6) & 52579.6\tnote{d} (4.2) &  52587.7\tnote{d} (4.5) & (2$^\mathrm{p}$) & (2$^\mathrm{s}$)  & IAUC 8020\\
      2003bl &   0.0146 &  52696.5\tnote{a} (4.0)  &   52711.2\tnote{c} (5.0) & 52706.2\tnote{d} (5.4) &  52714.2\tnote{d} (5.6) & (2$^\mathrm{p}$) & (2$^\mathrm{s}$)  & IAUC 8086\\   
      2003bn &   0.0128 &  52694.5\tnote{a} (3.0)  &   52707.1\tnote{c} (4.0) & 52702.1\tnote{d} (4.5) &  52710.1\tnote{d} (4.8) & (2$^\mathrm{p}$) & (2$^\mathrm{s}$)  & IAUC 8088\\   
      2003ci &   0.0304 &  52711.5\tnote{a} (8.0)  &   52723.8\tnote{c} (4.5) & 52718.8\tnote{d} (5.0) &  52726.9\tnote{d} (5.3) & (2$^\mathrm{p}$) & (2$^\mathrm{s}$)  & IAUC 8097\\
      2003cn &   0.0181 &  52717.5\tnote{a} (4.0)  &   52731.5\tnote{c} (4.5) & 52726.5\tnote{d} (5.0) &  52734.6\tnote{d} (5.2) & (2$^\mathrm{p}$) & (2$^\mathrm{s}$)  & IAUC 8101\\
      2003cx &   0.0370 &  52725.5\tnote{a} (5.0)  &   52740.4\tnote{c} (4.1) & 52735.4\tnote{d} (4.7) &  52743.5\tnote{d} (4.9) & (2$^\mathrm{p}$) & (2$^\mathrm{s}$)  & IAUC 8105\\
      2003ef &   0.0139 &  52757.5\tnote{a} (9.0)  &   52771.8\tnote{c} (4.8) & 52766.8\tnote{d} (5.3) &  52774.9\tnote{d} (5.5) & (2$^\mathrm{p}$) & (2$^\mathrm{s}$)  & IAUC 8131\\
      2003fb &   0.0175 &  52772.5\tnote{a} (10.0) &   52788.9\tnote{c} (5.1) & 52783.9\tnote{d} (5.5) &  52791.9\tnote{d} (5.8) & (2$^\mathrm{p}$) & (2$^\mathrm{s}$)  & IAUC 8143\\
      2003hd &   0.0395 &  52855.9\tnote{a} (5.0)  &   52870.1\tnote{c} (4.2) & 52865.1\tnote{d} (4.7) &  52873.1\tnote{d} (5.0) & (2$^\mathrm{p}$) & (2$^\mathrm{s}$)  & IAUC 8179\\
      2003hg &   0.0143 &  52865.5\tnote{a} (5.0)  &   52877.7\tnote{c} (5.9) & 52872.6\tnote{d} (6.3) &  52880.7\tnote{d} (6.5) & (2$^\mathrm{p}$) & (2$^\mathrm{s}$)  & IAUC 8184\\
      2003hl &   0.0082 &  52868.5\tnote{a} (5.0)  &   52881.8\tnote{c} (4.0) & 52876.8\tnote{d} (4.5) &  52884.8\tnote{d} (4.8) & (2$^\mathrm{p}$) & (2$^\mathrm{s}$)  & IAUC 8184\\
      2003hn &   0.0039 &  52866.5\tnote{a} (10.0) &   52879.1\tnote{c} (4.6) & 52874.1\tnote{d} (5.1) &  52882.2\tnote{d} (5.3) & (2$^\mathrm{p}$) & (2$^\mathrm{s}$)  & IAUC 8186\\
      2003ib &   0.0248 &  52891.5\tnote{a} (8.0)  &   52902.7\tnote{c} (3.5) & 52897.7\tnote{d} (4.1) &  52905.8\tnote{d} (4.4) & (2$^\mathrm{p}$) & (2$^\mathrm{s}$)  & IAUC 8201\\
      2003ip &   0.0180 &  52896.5\tnote{a} (4.0)  &   52908.2\tnote{c} (5.4) & 52903.2\tnote{d} (5.8) &  52911.3\tnote{d} (6.0) & (2$^\mathrm{p}$) & (2$^\mathrm{s}$)  & IAUC 8214\\
      2003iq &   0.0082 &  52919.5\tnote{a} (2.0)  &   52931.8\tnote{c} (4.7) & 52926.8\tnote{d} (5.2) &  52934.9\tnote{d} (5.4) & (2$^\mathrm{p}$) & (2$^\mathrm{s}$)  & CBET 48  \\
      2003E  &   0.0147 &  52629.5\tnote{a} (8.0)  &   52647.0\tnote{c} (4.1) & 52642.0\tnote{d} (4.6) &  52650.0\tnote{d} (4.9) & (2$^\mathrm{p}$) & (2$^\mathrm{s}$)  & IAUC 8044\\ 
      2003T  &   0.0279 &  52654.5\tnote{a} (10.0) &   52668.0\tnote{c} (5.3) & 52662.9\tnote{d} (5.7) &  52671.0\tnote{d} (5.9) & (2$^\mathrm{p}$) & (2$^\mathrm{s}$)  & IAUC 8058\\   
      2004er &   0.0147 &  53271.8\tnote{a} (2.0)  &   53287.6(1.8)           & 53282.6(1.9)           &  53292.9(1.6)           & (3$^\mathrm{p}$) & (2$^\mathrm{s}$)  & CBET 93  \\ 
      2004fc &   0.0061 &  53293.5\tnote{a} (1.0)  &   53306.5\tnote{c} (5.3) & 53301.4\tnote{d} (5.7) &  53309.5\tnote{d} (5.9) & (3$^\mathrm{p}$) & (2$^\mathrm{s}$)  & IAUC 8422\\   
      2004fx &   0.0089 &  53303.5\tnote{a} (4.0)  &   53316.2\tnote{c} (5.2) & 53311.2\tnote{d} (5.6) &  53319.2\tnote{d} (5.8) & (3$^\mathrm{p}$) & (2$^\mathrm{s}$)  & IAUC 8431\\
      2005an &   0.0107 &  53431.8\tnote{a} (6.0)  &   53444.8\tnote{c} (4.5) & 53439.8\tnote{d} (5.0) &  53447.8\tnote{d} (5.2) & (3$^\mathrm{p}$) & (2$^\mathrm{s}$)  & CBET 113 \\
      2005dt &   0.0257 &  53605.6\tnote{a} (9.0)  &   53616.4\tnote{c} (3.9) & 53611.4\tnote{d} (4.5) &  53619.4\tnote{d} (4.7) & (3$^\mathrm{p}$) & (2$^\mathrm{s}$)  & CBET 213 \\
      2005dx &   0.0267 &  53611.8\tnote{a} (7.0)  &   53625.9\tnote{c} (5.0) & 53620.9\tnote{d} (5.4) &  53629.0\tnote{d} (5.7) & (3$^\mathrm{p}$) & (2$^\mathrm{s}$)  & CBET 220 \\ 
      2005dz &   0.0190 &  53619.5\tnote{a} (4.0)  &   53632.2\tnote{c} (4.4) & 53627.2\tnote{d} (4.9) &  53635.3\tnote{d} (5.1) & (3$^\mathrm{p}$) & (2$^\mathrm{s}$)  & CBET 222 \\
      2005lw &   0.0257 &  53716.8\tnote{a} (10.0) &   53729.2\tnote{c} (3.7) & 53724.2\tnote{d} (4.3) &  53732.3\tnote{d} (4.6) & (3$^\mathrm{p}$) & (2$^\mathrm{s}$)  & CBET 318 \\
      2005me &   0.0224 &  53717.9\tnote{a} (10.0) &   53730.3\tnote{c} (4.2) & 53725.3\tnote{d} (4.7) &  53733.3\tnote{d} (5.0) & (3$^\mathrm{p}$) & (2$^\mathrm{s}$)  & CBET 333 \\
      2005J  &   0.0139 &  53379.8\tnote{a} (7.0)  &   53392.5\tnote{c} (5.1) & 53387.5\tnote{d} (5.5) &  53395.6\tnote{d} (5.8) & (3$^\mathrm{p}$) & (2$^\mathrm{s}$)  & IAUC 8467\\
      2005K  &   0.0274 &  53369.8\tnote{a} (8.0)  &   53382.5\tnote{c} (4.4) & 53377.4\tnote{d} (4.9) &  53385.5\tnote{d} (5.1) & (3$^\mathrm{p}$) & (2$^\mathrm{s}$)  & IAUC 8468\\
      2005Z  &   0.0192 &  53396.7\tnote{a} (6.0)  &   53409.4\tnote{c} (4.7) & 53404.4\tnote{d} (5.1) &  53412.5\tnote{d} (5.4) & (3$^\mathrm{p}$) & (2$^\mathrm{s}$)  & IAUC 8476\\
      2006ai &   0.0152 &  53781.6\tnote{a} (5.0)  &   53793.7\tnote{c} (4.9) & 53788.6\tnote{d} (5.4) &  53796.7\tnote{d} (5.6) & (3$^\mathrm{p}$) & (2$^\mathrm{s}$)  & CBET 406 \\  
      2006be &   0.0071 &  53802.8\tnote{a} (9.0)  &   53816.0\tnote{c} (5.9) & 53810.9\tnote{d} (6.3) &  53819.0\tnote{d} (6.5) & (3$^\mathrm{p}$) & (2$^\mathrm{s}$)  & CBET 449 \\ 
      2006bl &   0.0324 &  53822.7\tnote{a} (10.0) &   53835.9\tnote{c} (6.5) & 53830.9\tnote{d} (6.9) &  53839.0\tnote{d} (7.0) & (3$^\mathrm{p}$) & (2$^\mathrm{s}$)  & CBET 597 \\ 
      2006ee &   0.0154 &  53961.9\tnote{a} (4.0)  &   53975.0\tnote{c} (5.0) & 53969.9\tnote{d} (5.4) &  53978.0\tnote{d} (5.7) & (3$^\mathrm{p}$) & (2$^\mathrm{s}$)  & CBET 597 \\   
      2006iw &   0.0308 &  54010.7\tnote{a} (1.0)  &   54022.6\tnote{c} (4.2) & 54017.5\tnote{d} (4.7) &  54025.6\tnote{d} (4.9) & (3$^\mathrm{p}$) & (2$^\mathrm{s}$)  & CBET 663 \\
      2006qr &   0.0145 &  54062.8\tnote{a} (7.0)  &   54075.7\tnote{c} (4.7) & 54070.6\tnote{d} (5.2) &  54078.7\tnote{d} (5.4) & (3$^\mathrm{p}$) & (2$^\mathrm{s}$)  & CBET 766 \\
      2006Y  &   0.0336 &  53766.5\tnote{a} (4.0)  &   53777.9\tnote{c} (4.9) & 53772.9\tnote{d} (5.3) &  53781.0\tnote{d} (5.6) & (3$^\mathrm{p}$) & (2$^\mathrm{s}$)  & IAUC 8668\\  
      2007hm &   0.0251 &  54336.6\tnote{a} (6.0)  &   54348.9\tnote{c} (4.3) & 54343.9\tnote{d} (4.8) &  54351.9\tnote{d} (5.1) & (3$^\mathrm{p}$) & (2$^\mathrm{s}$)  & CBET 1050\\
      2007il &   0.0215 &  54349.8\tnote{a} (4.0)  &   54363.7\tnote{c} (5.3) & 54358.7\tnote{d} (5.8) &  54366.8\tnote{d} (6.0) & (3$^\mathrm{p}$) & (2$^\mathrm{s}$)  & CBET 1062\\
      2007ld &   0.0250 &  54376.5\tnote{a} (8.0)  &   54388.5\tnote{c} (3.9) & 54383.5\tnote{d} (4.4) &  54391.5\tnote{d} (4.7) & (3$^\mathrm{p}$) & (2$^\mathrm{s}$)  & CBET 1098\\
      \hline                                                                                                                       
    \end{tabular}
    \end{threeparttable}
  \end{table}
\end{landscape}

\begin{landscape}
  \begin{table}
    \contcaption{}
    \label{tab:snII_continued}
    \begin{threeparttable}
    \begin{tabular}{lcccccccr} 
      \hline
      SN &   z\_host & Exp[MJD]  & V\_Max[MJD]  &  B\_Max[MJD] & r\_Max[MJD] & Phot. Ref. & Spec. Ref & Circular\\
      \hline
      2007oc &   0.0048 &  54388.5\tnote{a} (3.0)  &   54401.0\tnote{c} (5.4) & 54396.0\tnote{d} (5.8) &  54404.0\tnote{d} (6.0) & (3$^\mathrm{p}$) & (2$^\mathrm{s}$)  & CBET 1114\\   
      2007od &   0.0058 &  54400.6\tnote{a} (5.0)  &   54412.5\tnote{c} (3.5) & 54407.5\tnote{d} (4.1) &  54415.5\tnote{d} (4.4) & (3$^\mathrm{p}$) & (2$^\mathrm{s}$)  & CBET 1116\\
      2007P  &   0.0408 &  54118.7\tnote{a} (5.0)  &   54131.3\tnote{c} (5.3) & 54126.3\tnote{d} (5.8) &  54134.4\tnote{d} (6.0) & (3$^\mathrm{p}$) & (2$^\mathrm{s}$)  & CBET 819 \\
      2007U  &   0.0260 &  54133.6\tnote{a} (6.0)  &   54144.9(3.3)           & 54140.7(3.3)           &  54150.1(4.8)           & (3$^\mathrm{p}$) & (2$^\mathrm{s}$)  & CBET 835 \\
      2007W  &   0.0097 &  54130.8\tnote{a} (7.0)  &   54142.9\tnote{c} (4.6) & 54137.8\tnote{d} (5.1) &  54145.9\tnote{d} (5.3) & (3$^\mathrm{p}$) & (2$^\mathrm{s}$)  & CBET 844 \\
      2007X  &   0.0095 &  54143.5\tnote{a} (5.0)  &   54156.3\tnote{c} (4.9) & 54151.3\tnote{d} (5.4) &  54159.3\tnote{d} (5.6) & (3$^\mathrm{p}$) & (2$^\mathrm{s}$)  & CBET 844 \\
      2008bh &   0.0145 &  54543.5\tnote{a} (5.0)  &   54556.5\tnote{c} (5.2) & 54551.4\tnote{d} (5.6) &  54559.5\tnote{d} (5.8) & (3$^\mathrm{p}$) & (2$^\mathrm{s}$)  & CBET 1311\\
      2008br &   0.0101 &  54555.7\tnote{a} (9.0)  &   54567.0\tnote{c} (3.4) & 54562.0\tnote{d} (4.0) &  54570.1\tnote{d} (4.3) & (3$^\mathrm{p}$) & (2$^\mathrm{s}$)  & CBET 1332\\ 
      2008bu &   0.0221 &  54566.8\tnote{a} (7.0)  &   54578.0\tnote{c} (3.0) & 54573.0\tnote{d} (3.7) &  54581.1\tnote{d} (4.0) & (3$^\mathrm{p}$) & (2$^\mathrm{s}$)  & CBET 1341\\ 
      2008gi &   0.0244 &  54742.7\tnote{a} (9.0)  &   54755.0\tnote{c} (5.0) & 54749.9\tnote{d} (5.4) &  54758.0\tnote{d} (5.6) & (3$^\mathrm{p}$) & (2$^\mathrm{s}$)  & CBET 1539\\ 
      2008ho &   0.0103 &  54792.7\tnote{a} (5.0)  &   54805.0\tnote{c} (3.9) & 54800.0\tnote{d} (4.5) &  54808.0\tnote{d} (4.8) & (3$^\mathrm{p}$) & (2$^\mathrm{s}$)  & CBET 1587\\   
      2008if &   0.0115 &  54807.8\tnote{a} (5.0)  &   54821.4\tnote{c} (4.7) & 54816.4\tnote{d} (5.1) &  54824.5\tnote{d} (5.4) & (3$^\mathrm{p}$) & (2$^\mathrm{s}$)  & CBET 1619\\
      2008in &   0.0052 &  54825.4\tnote{a} (2.0)  &   54838.5\tnote{c} (5.4) & 54833.4\tnote{d} (5.9) &  54841.5\tnote{d} (6.1) & (3$^\mathrm{p}$) & (2$^\mathrm{s}$)  & CBET 1636\\      
      2008K  &  0.0267 &  54475.5\tnote{a} (6.0)  &   54487.3\tnote{c} (4.7) & 54482.3\tnote{d} (5.2) &  54490.4\tnote{d} (5.4) & (3$^\mathrm{p}$) & (2$^\mathrm{s}$)  & CBET 1211\\
      2008M  &  0.0076 &  54471.7\tnote{a} (9.0)  &   54484.6\tnote{c} (5.5) & 54479.6\tnote{d} (5.9) &  54487.6\tnote{d} (6.1) & (3$^\mathrm{p}$) & (2$^\mathrm{s}$)  & CBET 1214\\
      2009aj &  0.0096 & 54880.5\tnote{a} (7.0)  & 54893.3\tnote{c} (2.5) & 54888.3\tnote{d} (3.3) & 54896.4\tnote{d} (3.7) & (3$^\mathrm{p}$)  & (2$^\mathrm{s}$)  & CBET 1704 \\   
      2009ao &  0.0111 & 54890.7\tnote{a} (4.0)  & 54903.9\tnote{c} (6.0) & 54898.9\tnote{d} (6.4) & 54906.9\tnote{d} (6.6) & (3$^\mathrm{p}$)  & (2$^\mathrm{s}$)  & CBET 1711 \\   
      2009au &  0.0094 & 54897.5\tnote{a} (4.0)  & 54909.3\tnote{c} (3.3) & 54904.2\tnote{d} (4.0) & 54912.3\tnote{d} (4.3) & (3$^\mathrm{p}$)  & (2$^\mathrm{s}$)  & CBET 1719 \\   
      2009bz &  0.0108 & 54915.8\tnote{a} (4.0)  & 54926.8\tnote{c} (3.5) & 54921.7\tnote{d} (4.1) & 54929.8\tnote{d} (4.4) & (3$^\mathrm{p}$)  & (2$^\mathrm{s}$)  & CBET 1748 \\   
      2009ib &  0.0043 & 55045.0\tnote{c} (7.6)  & 55057.1(1.6)           & 55052.1\tnote{d} (2.7) & 55060.1\tnote{d} (3.1) & (4$^\mathrm{p}$)  & (1$^\mathrm{s}$);(5$^\mathrm{s}$)     & CBET 1902 \\  
      2009N  &  0.0034 & 54846.8\tnote{a} (5.0)  & 54859.8\tnote{c} (4.4) & 54854.8\tnote{d} (4.9) & 54862.9\tnote{d} (5.2) & (3$^\mathrm{p}$)  & (2$^\mathrm{s}$)  & CBET 1670 \\  
      2012aw &  0.0026 & 56002.0(0.5)            & 56012.5(0.6)           & 56007.5(2.2)           & 56015.5(2.7)           & (5$^\mathrm{p}$)  & (1$^\mathrm{s}$);(6$^\mathrm{s}$)& CBET 3054 \\  
      2013ai &  0.0091 & 56339.7(10.0)           & 56365.4(1.0)           & 56359.9(0.9)           & 56365.2(1.4)           & (6$^\mathrm{p}$)  & (1$^\mathrm{s}$);(7$^\mathrm{s}$)& ATel 4849 \\  
      2013by &  0.0038 & 56401.0\tnote{b} (5.0)  & 56414.7(1.1)           & 56411.7(1.0)           & 56414.8(1.0)           & (7$^\mathrm{p}$)  & (1$^\mathrm{s}$);(7$^\mathrm{s}$);(8$^\mathrm{s}$);(9$^\mathrm{s}$)& CBET 3506 \\  
      2013ej &  0.0022 & 56497.5(0.9)            & 56510.4(0.9)           & 56508.7(0.9)           & 56515.9(1.0)           & (8$^\mathrm{p}$)  & (1$^\mathrm{s}$);(7$^\mathrm{s}$);(10$^\mathrm{s}$)& CBET 3606 \\
      &           &                         &                        &                        &                        &                 & (11$^\mathrm{s}$);(12$^\mathrm{s}$);(13$^\mathrm{s}$)& \\
      2013fs &  0.0118 & 56570.8(0.5)            & 56578.1(2.0)           & 56572.6(1.7)           & 56580.7(1.5)           & (6$^\mathrm{p}$)  & (1$^\mathrm{s}$);(7$^\mathrm{s}$);(14$^\mathrm{s}$)& CBET 3671 \\  
      2014cx &  0.0055 & 56901.9(0.5)            & 56917.5(0.6)           & 56910.3(0.2)           & 56925.2(2.3)           & (9$^\mathrm{p}$)  & (1$^\mathrm{s}$);(7$^\mathrm{s}$)& CBET 3963 \\  
      2014G  &  0.0045 & 56669.6(1.7)            & 56681.7(0.8)           & 56678.9(0.7)           & 56684.6(1.6)           & (10$^\mathrm{p}$) & (1$^\mathrm{s}$);(15$^\mathrm{s}$)& CBET 3787 \\ 
      \hline
    \end{tabular}
    \end{threeparttable}
    \begin{tablenotes}
      \small
      \item Notes: SN names (same in other tables) are listed in column 1. In column 2 the heliocentric redshift of the host obtained from NASA/IPAC Extragalactic Database (NED). In column 3 the explosion epoch of the SN and its error obtained from non-detections unless stated differently (see Sec. \ref{exp}). In columns 4, 5 and 6 the epoch of maximum light of the SN light curve and its error in $V$-, $B$- and $r$-band respectively obtained from the light curve unless stated differently (see Sec. \ref{sec:ml}). In column 7 photometry references. In column 8 spectral references. Finally in column 9 is the reference of the discovery circular.
    \begin{itemize}
    \item[a] Explosion epoch from \citet{2017ApJ...850...89G}.  
    \item[b] Explosion epoch estimation from spectral matching.
    \item[c] Maximum light epoch from spectral matching.
    \item[d] Maximum light epoch with respect to average in $V$-band from those SNe that have data points around maximum, see Table \ref{tab:difjdmax}).
    \end{itemize}
    \end{tablenotes}
  \end{table}
\end{landscape}

\begin{landscape}
  \begin{table}
    \caption{SNe~IIb sample. References:(1$^\mathrm{p}$) \citet{2018A&A...609A.134S}; (2$^\mathrm{p}$) The Open Supernova Catalog (\citealt{2017ApJ...835...64G}); (3$^\mathrm{p}$) \citet{2014MNRAS.445.1647M}, \citet{2015hsa8.conf..518M}; (4$^\mathrm{p}$) \citet{2018Natur.554..497B}; (5$^\mathrm{p}$) \citet{2014ApJS..213...19B}; (6$^\mathrm{p}$) \citet{2014Ap&SS.354...89B}; (7$^\mathrm{p}$) \citet{1993PASJ...45L..63O} ; (8$^\mathrm{p}$) \citet{1993PASJ...45L..59V}; (9$^\mathrm{p}$) \citet{1994AJ....107.1453B}; (10$^\mathrm{p}$) \citet{1994AJ....107.1022R}; (11$^\mathrm{p}$) \citet{1995AstL...21..598M}; (12$^\mathrm{p}$) \citet{1995A&AS..110..513B}; (13$^\mathrm{p}$) \citet{1996AJ....112..732R}; (14$^\mathrm{p}$) \citet{1999AJ....117..736Q};(15$^\mathrm{p}$) \citet{2016AJ....151...33G};(16$^\mathrm{p}$) \citet{2011ApJ...741...97D};(17$^\mathrm{p}$) \citet{2009PZ.....29....2T};(18$^\mathrm{p}$) \citet{2008MNRAS.389..955P};(19$^\mathrm{p}$) \citet{2014A&A...562A..17E};(20$^\mathrm{p}$) \citet{2013MNRAS.431..308K};(21$^\mathrm{p}$) \citet{2015MNRAS.454...95M};(22$^\mathrm{p}$) \citet{2014MNRAS.439.1807B}. (1$^\mathrm{s}$) \citet{2012PASP..124..668Y}; (2$^\mathrm{s}$) Holmbo et al. in prep.; (3$^\mathrm{s}$) \citet{2000AJ....120.1487M}; (4$^\mathrm{s}$) \citet{1995A&AS..110..513B}; (5$^\mathrm{s}$) \citet{2015A&A...573A..12J}; (6$^\mathrm{s}$) \citet{2014AJ....147...99M}; (7$^\mathrm{s}$) \citet{2009ApJ...703.1612H}; (8$^\mathrm{s}$) \citet{2008MNRAS.389..955P}; (9$^\mathrm{s}$) \citet{2011MNRAS.413.2140T}; (10$^\mathrm{s}$) \citet{2012MNRAS.424.1297O}; (11$^\mathrm{s}$) \citet{2011ApJ...742L..18A}; (12$^\mathrm{s}$) \citet{2014A&A...562A..17E}; (13$^\mathrm{s}$) \citet{2015A&A...580A.142E}; (14$^\mathrm{s}$) \citet{2013ApJ...767...71M};(15$^\mathrm{s}$) \citet{2013MNRAS.431..308K}; (16$^\mathrm{s}$) \citet{2015MNRAS.454...95M}; (17$^\mathrm{s}$) \citet{2014MNRAS.439.1807B}; (18$^\mathrm{s}$) \citet{2016PASA...33...55C}; (19$^\mathrm{s}$) \citet{2016MNRAS.460.1500S}; (20$^\mathrm{s}$) \citet{2017MNRAS.465.4650K}; (21$^\mathrm{s}$) \citet{2019MNRAS.482.1545S}}
    \label{tab:snIIb}
    \begin{threeparttable}
    \begin{tabular}{lcccccccr} 
      \hline
      SN &  z\_host & Exp[MJD]  & V\_Max[MJD]  &  B\_Max[MJD] & r\_Max[MJD] & Phot. Ref. & Spec. Ref & Circular\\
      \hline
      1993J           &  -0.0001 & 49071.7(2.6)           & 49095.0(0.1)  & 49093.7(0.1)          & 49095.6\tnote{b} (0.8)  & (2$^\mathrm{p}$);(7$^\mathrm{p}$);(8$^\mathrm{p}$); & (1$^\mathrm{s}$);(3$^\mathrm{s}$);& IAUC 5740\\
                      &            &                        &               &                       &                        & (9$^\mathrm{p}$);(10$^\mathrm{p}$);(11$^\mathrm{p}$);& (4$^\mathrm{s}$);(5$^\mathrm{s}$)& \\
                      &            &                        &               &                       &                        & (12$^\mathrm{p}$);(13$^\mathrm{p}$) & & \\
      1996cb          &  0.0024 & 50424.3(8.3)           & 50450.4(1.8)  & 50450.4(1.9)          & 50451.0\tnote{b} (2.0) & (2$^\mathrm{p}$);(14$^\mathrm{p}$) & (1$^\mathrm{s}$);(6$^\mathrm{s}$) & IAUC 6524\\    	 
      2003bg          &  0.0046 & 52697.3\tnote{a} (7.9) & 52717.6(10.3) & 52713.6(6.6)          & 52718.2\tnote{b} (10.4)& (2$^\mathrm{p}$);(15$^\mathrm{p}$) & (1$^\mathrm{s}$);(7$^\mathrm{s}$)& IAUC 8082\\   	 
      2004ex          &  0.0175  & 53288.8(0.5)           & 53307.2(1.8)  & 53306.1(1.5)          & 53307.8(1.6)           & (1$^\mathrm{p}$) & (2$^\mathrm{s}$) & IAUC 8418\\        
      2004ff          &  0.0226  & 53302.9(3.5)           & 53313.1(1.9)  & 53312.3(1.9)          & 53314.8(1.5)           & (1$^\mathrm{p}$) & (6$^\mathrm{s}$);(21$^\mathrm{s}$) & IAUC 8425\\        
      2006ba          &  0.0191  & 53802.0\tnote{a} (5.9) & 53823.1(1.6)  & 53821.8\tnote{b} (1.6)& 53823.7\tnote{b} (1.7) & (1$^\mathrm{p}$);(5$^\mathrm{p}$) & (1$^\mathrm{s}$);(2$^\mathrm{s}$) & CBET 443\\        
      2006el\tnote{*} &  0.0170 & 53962.8(0.5)           & 53983.8(1.6)  & 53982.5\tnote{b} (1.6)& 53984.4\tnote{b} (1.8) & (2$^\mathrm{p}$);(5$^\mathrm{p}$);(16$^\mathrm{p}$) & (1$^\mathrm{s}$);(6$^\mathrm{s}$)& CBET 605\\    	 
      2006T           &  0.0081  & 53759.0(7.0)           & 53780.3(1.7)  & 53779.2(1.5)          & 53781.6(1.9)           & (1$^\mathrm{p}$);(5$^\mathrm{p}$) & (2$^\mathrm{s}$);(6$^\mathrm{s}$) & IAUC 8666\\        
      2008aq          &  0.0080  & 54514.9(8.5)           & 54532.0(1.3)  & 54530.5(1.3)          & 54529.9(5.0)           & (1$^\mathrm{p}$);(5$^\mathrm{p}$);(6$^\mathrm{p}$)& (2$^\mathrm{s}$);(6$^\mathrm{s}$) & CBET 1271\\        
      2008ax          &  0.0019 & 54528.3(0.1)           & 54549.6(0.5)  & 54547.9(0.6)          & 54550.2\tnote{b} (0.9) & (2$^\mathrm{p}$);(6$^\mathrm{p}$); & (1$^\mathrm{s}$);(6$^\mathrm{s}$);& CBET 1286\\
                      &            &                        &               &                       &                        & (17$^\mathrm{p}$);(18$^\mathrm{p}$) & (8$^\mathrm{s}$);(9$^\mathrm{s}$) & \\
      2008bo          &  0.0049 & 54546.4\tnote{a} (4.4) & 54567.6(1.2)  & 54566.6(1.2)          & 54568.2\tnote{b} (1.4) & (2$^\mathrm{p}$);(5$^\mathrm{p}$);(6$^\mathrm{p}$) & (1$^\mathrm{s}$);(6$^\mathrm{s}$) & CBET 1324\\  
      2009mg          &  0.0076 & 55167.6\tnote{a} (7.0) & 55188.5(2.0)  & 55187.2(2.0)          & 55189.1\tnote{b} (2.2) & (2$^\mathrm{p}$);(6$^\mathrm{p}$) & (1$^\mathrm{s}$);(10$^\mathrm{s}$)& CBET 2071\\ 
      2009K           &  0.0117  & 54843.6(1.5)           & 54869.4(2.4)  & 54866.9(2.4)          & 54871.2(2.6)           & (1$^\mathrm{p}$) & (2$^\mathrm{s}$) & CBET 1663\\        
      2009Z           &  0.0251  & 54858.5\tnote{a} (6.3) & 54877.7(2.4)  & 54876.6(2.7)          & 54879.0(3.2)           & (1$^\mathrm{p}$) & (2$^\mathrm{s}$) & CBET 1685\\        
      2010jr          &  0.0124 & 55509.2(2.9)           & 55531.7(1.6)  & 55530.2(1.4)          & 55532.3\tnote{b} (1.8) & (2$^\mathrm{p}$);(6$^\mathrm{p}$) & - & CBET 2545\\     	 
      2011dh          &  0.0015 & 55712.4(0.5)           & 55732.1(0.9)  & 55729.8(0.8)          & 55732.8(1.2)           & (2$^\mathrm{p}$);(19$^\mathrm{p}$) & (1$^\mathrm{s}$);(11$^\mathrm{s}$)& CBET 2736\\
                      &            &                        &               &                       &                        &                                 & (12$^\mathrm{s}$);(13$^\mathrm{s}$)& \\
      2011ei          &  0.0093 & 55763.5\tnote{a} (3.6) & 55784.4(0.9)  & 55782.5(0.9)          & 55785.0\tnote{b} (1.2) & (2$^\mathrm{p}$);(6$^\mathrm{p}$) & (1$^\mathrm{s}$);(14$^\mathrm{s}$)& CBET 2777\\  
      2011fu          &  0.0185 & 55824.7\tnote{a} (4.2) & 55846.0(0.5)  & 55845.5(0.6)          & 55846.6\tnote{b} (0.9) & (2$^\mathrm{p}$);(20$^\mathrm{p}$);(21$^\mathrm{p}$) & (1$^\mathrm{s}$);(15$^\mathrm{s}$);(16$^\mathrm{s}$)& CBET 2827\\   	 
      2011hs          &  0.0057 & 55867.2\tnote{a} (3.1) & 55888.2(0.7)  & 55884.9(0.7)          & 55888.8\tnote{b} (1.1) & (2$^\mathrm{p}$);(6$^\mathrm{p}$);(22$^\mathrm{p}$) & (1$^\mathrm{s}$);(17$^\mathrm{s}$)& CBET 2902\\   	 
      2013ak          &  0.0034 & 56349.1\tnote{a} (7.1) & 56368.8(1.8)  & 56366.1(1.8)          & 56369.4\tnote{b} (1.9) & (2$^\mathrm{p}$);(6$^\mathrm{p}$) & (1$^\mathrm{s}$)& CBET 3437\\  	 
      2013df          &  0.0024 & 56443.9(6.9)           & 56469.8(0.4)  & 56468.8(0.4)          & 56470.5(1.3)           & (2$^\mathrm{p}$);(3$^\mathrm{p}$);(6$^\mathrm{p}$) & (1$^\mathrm{s}$);(18$^\mathrm{s}$);(19$^\mathrm{s}$)& CBET 3557\\
      2016gkg         &  0.0049 & 57651.2(0.1)           & 57671.7(1.9)  & 57670.5(1.9)          & 57672.3\tnote{b} (2.1) & (4$^\mathrm{p}$) & (1$^\mathrm{s}$);(20$^\mathrm{s}$)& vsnet 20188\\       
      \hline                                                                                                                                                        
    \end{tabular}                                                                                                       \end{threeparttable}                                            
    \begin{tablenotes}                                                                                                                                              
      \small                                                                                                                                                        
    \item Notes: SN names are listed in column 1. In column 2 the heliocentric redshift of the host obtained from NASA/IPAC Extragalactic Database (NED). In column 3 the explosion epoch of the SN and its error obtained from non-detections unless stated differently (see Sec. \ref{exp}). In columns 4, 5 and 6 the epoch of maximum light of the SN light curve and its error in $V$-, $B$- and $r$-band respectively obtained from the light curve unless stated differently (see Sec. \ref{sec:ml}). In column 7 photometry references. In column 8 spectral references from. Finally in column 9 is the reference of the discovery circular.
    \begin{itemize}
     \item[a] Explosion epoch estimation from spectral matching.
     \item[b] Maximum light epoch with respect to average in $V$-band from those SNe that have data points around maximum, see Table \ref{tab:difjdmax}).
     \item[*] This object last non-detection and first detection dates in \citet{2011ApJ...741...97D} differs from said dates in the discovery circular. We use the date that appear in \citet{2011ApJ...741...97D} since they are more constraining.
     \end{itemize}
    \end{tablenotes}
  \end{table}
\end{landscape}

\begin{landscape}
  \begin{table}
    \caption{SNe~IIb with no observed maximum sample. References:(1$^\mathrm{p}$) \citet{2018A&A...609A.134S}. (1$^\mathrm{s}$) Holmbo et al. in prep.}
    \label{tab:snIIbNOMAX}
    \begin{tabular}{lcccccr} 
      \hline
      SN &  z\_host & Exp[MJD] & V\_Max[MJD] & Phot. Ref. & Spec. Ref. & Ciruclar\\
      \hline
      2005bj & 0.022179 &  53451.1(10.18) & 53471.4(3.3) & (1$^\mathrm{p}$) & (1$^\mathrm{s}$) & IAUC 8511\\
      2006bf & 0.023867 &  53800.1(9.6)   & 53819.0(2.9) & (1$^\mathrm{p}$) & (1$^\mathrm{s}$) & IAUC 8693\\
      \hline
    \end{tabular}
    \begin{tablenotes}
      \small
    \item Notes: SN names are listed in column 1. In column 2 the heliocentric redshift of the host obtained from NASA/IPAC Extragalactic Database (NED). In column 3 the explosion epoch of the SN and its error (see Sec. \ref{exp}). In column 4 the epoch of maximum light of the SN light curve and its error in $V$-band. In column 5 the references from where the light curves were obtained. In column 6 the references from where the spectra were obtained. Finally in column 7 is the reference of the discovery circular.
    \end{tablenotes}
  \end{table}

  \begin{table}
    \caption{Parameters obtained with respect to explosion epoch.}
    \label{tab:resultsvbr}
    \begin{adjustbox}{max width=\linewidth}
      \begin{threeparttable}
        \begin{tabular}{lcccccccccccccccccr} 
          \hline
          SN       &  $t_{\mathrm{rise}}$  & $\mathrm{dm}_{1}$[$B$]  &  $t_{\mathrm{dm}_{1}[B]}$ & $\mathrm{dm}_{2}$[$B$] & $t_{\mathrm{dm}_{2}[B]}$ & $\Delta B_{40-30}$  & $t_{\mathrm{rise}}$      &   $\mathrm{dm}_{1}$[$V$]     & $t_{\mathrm{dm}_{1}[V]}$  &   $\mathrm{dm}_{2}$[$V$]  &  $t_{\mathrm{dm}_{2}[V]}$  & $\Delta V_{40-30}$ &  $t_{\mathrm{rise}}$  & $\mathrm{dm}_{1}$[$r$]  & $t_{\mathrm{dm}_{1}[r]}$  & $\mathrm{dm}_{2}$[$r$] & $t_{\mathrm{dm}_{2}[r]}$ & $\Delta r_{40-30}$\\
          \hline
          &       &       &     &      &   &  &    &   & SNe~IIb  &  &  &   &   &   &   &  &  & \\
          \hline
          & \multicolumn{6}{c}{$B$}& \multicolumn{6}{c}{$V$} & \multicolumn{6}{c}{$r$}\\
          \cmidrule(lr){2-7}\cmidrule(lr){8-13}\cmidrule(lr){14-19}
      1993J   &  22.0(2.6)  & 0.150(0.0001) & 30.2(2.6)  & -0.01355(0.00001) & 36.2(2.6)  & 1.08(0.077)   & 23.3(2.6)  & 0.099(0.0006) & 29.9(2.6)   & -0.00792(0.00038) & 32.7(2.6)  & 0.72(0.039)   &  23.8(2.7)   &  - ( - )      &  - ( - )   &  - ( - )          &  - ( - )   &  - ( - )	   \\
      1996cb  &  26.1(8.5)  & 0.134(0.0096) & 35.2(8.3)  & -0.01441(0.00276) & 40.6(8.3)  & 1.08(0.089)   & 26.1(8.5)  & 0.065(0.0017) & 37.7(8.3)   & -0.00369(0.00022) & 46.5(8.3)  & 0.54(0.074)   &  26.7(8.6)   &  - ( - )      &  - ( - )   &  - ( - )          &  - ( - )   &  - ( - )	   \\
      2003bg  &  16.3(10.3) & 0.090(0.0035) & 25.1(7.9)  & -0.00557(0.00136) & 30.4(7.9)  & 0.54(0.047)   & 20.3(13.0) & 0.060(0.0013) & 34.2(7.9)   & -0.00250(0.00010) & 44.6(7.9)  & 0.56(0.011)   &  20.9(13.0)  &  - ( - )      &  - ( - )   &  - ( - )          &  - ( - )   &  - ( - )	   \\
      2004ex  &  17.3(1.5)  & 0.124(0.0006) & 28.1(0.5)  & -0.00860(0.00023) & 36.1(0.5)  & 0.85(0.019)   & 18.4(1.9)  & 0.081(0.0009) & 29.9(0.5)   & -0.00472(0.00035) & 38.2(0.5)  & 0.68(0.014)   &  19.0(1.7)   & 0.065(0.0001) & 31.7(0.5)  & -0.00341(0.00002) & 40.7(0.5)  & 0.60(0.009) \\	  
      2004ff  &   9.4(4.0)  & 0.137(0.0027) & 17.9(3.5)  & -0.00977(0.00029) & 23.2(3.5)  & 0.24(0.046)   & 10.2(4.0)  & 0.089(0.0010) & 20.4(3.5)   & -0.00513(0.00017) & 26.7(3.5)  & 0.38(0.025)   &  11.9(3.8)   & 0.073(0.0009) & 20.7(3.5)  & -0.00440(0.00032) & 25.9(3.5)  & 0.38(0.007) \\	  
      2006ba  &  19.8(6.1)  &  - ( - )      &  - ( - )   & -0.00446(0.00025) & 50.3(5.9)  & 0.85(0.053)   & 21.0(6.1)  & 0.075(0.0021) & 34.5(5.9)   & -0.00345(0.00029) & 46.1(5.9)  & 0.72(0.008)   &  21.7(6.2)   & 0.061(0.0044) & 36.7(5.9)  & -0.00273(0.00112) & 47.6(5.9)  & 0.57(0.014) \\	  
      2006el  &  19.7(1.7)  &  - ( - )      &  - ( - )   & -0.00270(0.00007) & 52.2(2.5)  & 0.69(0.091)   & 21.0(1.7)  & 0.069(0.0013) & 32.4(1.6)   & -0.00235(0.00038) & 40.6(3.6)  & 0.65(0.083)   &  21.6(1.8)   &  - ( - )      &  - ( - )   &  - ( - )          &  - ( - )   &  - ( - )	   \\
      2006T   &  20.2(7.2)  & 0.132(0.0005) & 28.3(7.0)  & -0.00903(0.00025) & 34.6(7.0)  & 0.90(0.149)   & 21.3(7.2)  & 0.085(0.0001) & 31.7(7.0)   & -0.00421(0.00002) & 40.1(7.0)  & 0.76(0.044)   &  22.6(7.3)   & 0.064(0.0001) & 33.6(7.0)  & -0.00240(0.00003) & 42.6(7.0)  & 0.61(0.011) \\	  
      2008aq  &  15.6(8.6)  & 0.134(0.0027) & 24.9(8.5)  & -0.00953(0.00034) & 31.4(8.5)  & 0.66(0.175)   & 17.1(8.6)  & 0.084(0.0032) & 27.3(8.5)   & -0.00491(0.00058) & 34.6(8.5)  & 0.58(0.101)   &  15.0(9.8)   & 0.047(0.0021) & 36.5(8.5)  & -0.00143(0.00017) & 50.1(8.5)  & 0.45(0.010) \\	  
      2008ax  &  19.6(0.6)  & 0.163(0.0106) & 28.4(0.4)  & -0.01668(0.00235) & 33.0(1.4)  & 0.87(0.162)   & 21.3(0.5)  & 0.094(0.0067) & 29.9(0.5)   & -0.00790(0.00138) & 35.1(2.1)  & 0.67(0.098)   &  21.9(0.9)   &  - ( - )      &  - ( - )   &  - ( - )          &  - ( - )   &  - ( - )	   \\
      2008bo  &  20.2(4.6)  & 0.174(0.0104) & 30.3(4.4)  & -0.01762(0.00226) & 36.5(4.4)  & 1.21(0.163)   & 21.2(4.6)  & 0.131(0.0102) & 30.7(4.4)   & -0.01580(0.00322) & 35.2(4.4)  & 0.86(0.120)   &  21.8(4.6)   &  - ( - )      &  - ( - )   &  - ( - )          &  - ( - )   &  - ( - )	   \\
      2009mg  &  19.6(7.3)  & 0.130(0.0054) & 30.4(7.0)  & -0.01111(0.00691) & 38.3(7.0)  & 1.01(0.143)   & 20.9(7.2)  & 0.088(0.0021) & 33.5(7.0)   & -0.00495(0.00052) & 42.6(7.0)  & 0.81(0.023)   &  21.5(7.3)   &  - ( - )      &  - ( - )   &  - ( - )          &  - ( - )   &  - ( - )	   \\
      2009K   &  23.3(2.8)  & 0.072(0.0054) & 30.0(1.5)  &   - ( - )         &  - ( - )   &  - ( - )      & 25.8(2.8)  &  - ( - )      &  - ( - )    &  - ( - )          &  - ( - )   &  - ( - )      &  27.6(3.0)   &  - ( - )      &  - ( - )   &  - ( - )          &  - ( - )   &  - ( - )	   \\
      2009Z   &  18.0(6.9)  & 0.120(0.0021) & 29.4(6.3)  & -0.00944(0.00071) & 35.7(6.3)  & 0.86(0.125)   & 19.1(6.8)  & 0.073(0.0006) & 31.8(6.3)   & -0.00362(0.00006) & 40.5(6.3)  & 0.66(0.032)   &  20.4(7.1)   & 0.059(0.0003) & 34.7(6.3)  & -0.00248(0.00003) & 45.3(6.3)  & 0.56(0.006) \\	  
      2010jr  &  21.0(3.2)  & 0.136(0.0099) & 29.6(2.9)  & -0.01490(0.00694) & 38.7(4.2)  & 1.02(0.078)   & 22.5(3.3)  & 0.102(0.0045) & 29.8(2.9)   & -0.00652(0.00179) & 34.3(2.9)  & 0.78(0.046)   &  23.1(3.4)   &  - ( - )      &  - ( - )   &  - ( - )          &  - ( - )   &  - ( - )	   \\
      2011dh  &  17.4(1.0)  & 0.120(0.0160) & 28.3(0.5)  & -0.00809(0.00338) & 37.0(2.3)  & 0.87(0.038)   & 19.7(1.0)  & 0.098(0.0006) & 28.9(0.5)   & -0.00672(0.00087) & 33.1(0.5)  & 0.68(0.018)   &  20.4(1.3)   & 0.071(0.0003) & 29.0(0.8)  & -0.00327(0.00063) & 31.9(1.9)  & 0.58(0.014) \\	  
      2011ei  &  19.0(3.7)  & 0.136(0.0075) & 29.7(3.6)  & -0.01296(0.00844) & 36.3(4.9)  & 1.00(0.146)   & 20.9(3.7)  & 0.081(0.0056) & 30.2(3.6)   & -0.00759(0.00164) & 37.5(3.6)  & 0.60(0.057)   &  21.5(3.8)   &  - ( - )      &  - ( - )   &  - ( - )          &  - ( - )   &  - ( - )	   \\
      2011fu  &  20.8(4.3)  & 0.129(0.0009) & 28.7(4.2)  & -0.01262(0.00055) & 34.6(4.2)  & 0.75(0.125)   & 21.3(4.3)  & 0.067(0.0012) & 28.8(4.2)   & -0.00621(0.00094) & 40.4(4.2)  & 0.54(0.040)   &  21.9(4.3)   &  - ( - )      &  - ( - )   &  - ( - )          &  - ( - )   &  - ( - )	   \\
      2011hs  &  17.7(3.2)  & 0.105(0.0072) & 23.5(3.1)  & -0.00137(0.00339) & 24.6(4.3)  & 0.84(0.080)   & 21.0(3.2)  & 0.107(0.0012) & 29.8(3.1)   & -0.00648(0.00161) & 36.3(3.1)  & 0.85(0.045)   &  21.6(3.3)   &  - ( - )      &  - ( - )   &  - ( - )          &  - ( - )   &  - ( - )	   \\
      2013ak  &  17.0(7.3)  & 0.137(0.0018) & 28.3(7.1)  & -0.01034(0.00038) & 35.8(7.1)  & 0.95(0.167)   & 19.7(7.3)  & 0.091(0.0008) & 31.9(7.1)   & -0.00605(0.00086) & 43.0(7.1)  & 0.82(0.050)   &  20.3(7.4)   &  - ( - )      &  - ( - )   &  - ( - )          &  - ( - )   &  - ( - )	   \\
      2013df  &  24.9(6.9)  & 0.155(0.0007) & 32.7(6.9)  & -0.01463(0.00060) & 37.8(6.9)  & 1.24(0.111)   & 25.9(6.9)  & 0.108(0.0021) & 33.7(6.9)   & -0.00959(0.00528) & 38.7(6.9)  & 0.90(0.047)   &  26.6(7.0)   & 0.085(0.0116) & 34.8(6.9)  & -0.00858(0.00801) & 39.6(6.9)  & 0.70(0.035) \\	  
      2016gkg &  19.3(1.9)  & 0.131(0.0090) & 29.1(1.4)  & -0.01042(0.00190) & 35.8(2.6)  & 0.91(0.365)   & 20.5(1.9)  & 0.079(0.0035) & 30.7(1.5)   & -0.00476(0.00131) & 37.3(3.0)  & 0.66(0.273)   &  21.1(2.1)   &  - ( - )      &  - ( - )   &  - ( - )          &  - ( - )   &  - ( - )	   \\
      \hline
      \hline
          SN       &  $t_{\mathrm{rise}}$  & $\mathrm{dm}_{1}$[$B$]  &  $t_{\mathrm{dm}_{1}[B]}$ & $\mathrm{dm}_{2}$[$B$] & $t_{\mathrm{dm}_{2}[B]}$ & $\Delta B_{40-30}$  & $t_{\mathrm{rise}}$      &   $\mathrm{dm}_{1}$[$V$]     & $t_{\mathrm{dm}_{1}[V]}$  &   $\mathrm{dm}_{2}$[$V$]  &  $t_{\mathrm{dm}_{2}[V]}$  & $\Delta V_{40-30}$ &  $t_{\mathrm{rise}}$  & $\mathrm{dm}_{1}$[$r$]  & $t_{\mathrm{dm}_{1}[r]}$  & $\mathrm{dm}_{2}$[$r$] & $t_{\mathrm{dm}_{2}[r]}$ & $\Delta r_{40-30}$\\
          \hline
      \hline
            &       &       &     &      &   &  &    &   &  SNe~II &  &  &   &   &   &   &  &  & \\
      \hline
      & \multicolumn{6}{c}{$B$}& \multicolumn{6}{c}{$V$} & \multicolumn{6}{c}{$r$}\\
      \cmidrule(lr){2-7}\cmidrule(lr){8-13}\cmidrule(lr){14-19}
      1999gi  &   8.1(4.3)  & 0.047(0.0007) & 25.7(3.7)  & -0.00190(0.00010) & 38.2(3.7)  & 0.37(0.018)   & 11.0(4.3)  & 0.015(0.0008) & 22.1(3.7)   & -0.00125(0.00015) & 29.2(3.7)  & 0.04(0.009)   &  14.0(5.1)   &  - ( - )      &  - ( - )   &  - ( - )          &  - ( - )   &  - ( - )    \\	
      2002fa  &   8.4(9.5)  &  - ( - )      &  - ( - )   &  - ( - )          &  - ( - )   & 0.35(0.016)   & 13.4(9.2)  & 0.018(0.0009) & 33.7(8.0)   & -0.00079(0.00025) & 41.7(8.0)  & 0.17(0.011)   &  16.4(9.6)   &  - ( - )      &  - ( - )   &  - ( - )          &  - ( - )   &  - ( - )    \\	
      2002gd  &   7.4(4.8)  & 0.074(0.0023) & 20.6(4.0)  & -0.00412(0.00027) & 28.9(4.0)  & 0.31(0.033)   & 11.1(4.6)  & 0.035(0.0006) & 19.6(4.0)   & -0.00221(0.00017) & 25.2(4.0)  & 0.09(0.018)   &  14.1(5.3)   &  - ( - )      &  - ( - )   &  - ( - )          &  - ( - )   &  - ( - )    \\	
      2002gw  &  13.5(9.5)  &  - ( - )      &  - ( - )   & -0.00086(0.00003) & 40.7(8.0)  & 0.25(0.019)   & 18.5(9.3)  &  - ( - )      &  - ( - )    & -0.00027(0.00002) & 21.7(8.0)  & 0.03(0.005)   &  21.5(9.7)   &  - ( - )      &  - ( - )   &  - ( - )          &  - ( - )   &  - ( - )    \\	
      2002hj  &   8.0(9.2)  &  - ( - )      &  - ( - )   & -0.00108(0.00030) & 41.9(7.0)  & 0.35(0.018)   & 13.0(8.9)  &  - ( - )      &  - ( - )    & -0.00075(0.00024) & 37.2(7.0)  & 0.18(0.014)   &  16.0(9.3)   &  - ( - )      &  - ( - )   &  - ( - )          &  - ( - )   &  - ( - )    \\	
      2002hx  &   7.9(10.8) & 0.040(0.0006) & 52.2(9.0)  &  - ( - )          &  - ( - )   & 0.38(0.032)   & 12.9(10.6) & 0.030(0.0027) & 30.4(9.0)   & -0.00270(0.00089) & 40.9(9.0)  & 0.24(0.038)   &  16.0(10.9)  &  - ( - )      &  - ( - )   &  - ( - )          &  - ( - )   &  - ( - )    \\	
      2002ig  &   9.1(6.5)  &  - ( - )      &  - ( - )   &  - ( - )          &  - ( - )   & 0.44(0.030)   & 14.2(6.2)  &  - ( - )      &  - ( - )    &  - ( - )          & - ( - )   & 0.25(0.010)   &  17.2(6.7)   &  - ( - )      &  - ( - )   &  - ( - )          &  - ( - )   &  - ( - )    \\	
      2003bl  &   9.7(6.7)  &  - ( - )      &  - ( - )   &  - ( - )          &  - ( - )   & 0.26(0.018)   & 14.7(6.4)  & 0.016(0.0154) & 16.2(6.7)   & -0.00152(0.00239) & 23.4(5.1)  & -0.002(0.009) &  17.7(6.9)   &  - ( - )      &  - ( - )   &  - ( - )          &  - ( - )   &  - ( - )    \\	
      2003bn  &   7.6(5.4)  &  - ( - )      &  - ( - )   & -0.00111(0.00004) & 37.9(3.0)  & 0.29(0.012)   & 12.6(5.0)  & 0.013(0.0004) & 26.8(3.0)   & -0.00074(0.00006) & 38.4(3.0)  & 0.09(0.006)   &  15.6(5.7)   &  - ( - )      &  - ( - )   &  - ( - )          &  - ( - )   &  - ( - )    \\	
      \hline
      \end{tabular}
    \end{threeparttable}
    \end{adjustbox}
  \end{table}
\end{landscape}

 \begin{landscape}
  \begin{table}
  \contcaption{}
    \label{tab:resultsvbr_continued}
        \begin{adjustbox}{max width=\linewidth}
      \begin{threeparttable}
       \begin{tabular}{lcccccccccccccccccr} 
        \hline
          SN       &  $t_{\mathrm{rise}}$  & $\mathrm{dm}_{1}$[$B$]  &  $t_{\mathrm{dm}_{1}[B]}$ & $\mathrm{dm}_{2}$[$B$] & $t_{\mathrm{dm}_{2}[B]}$ & $\Delta B_{40-30}$  & $t_{\mathrm{rise}}$      &   $\mathrm{dm}_{1}$[$V$]     & $t_{\mathrm{dm}_{1}[V]}$  &   $\mathrm{dm}_{2}$[$V$]  &  $t_{\mathrm{dm}_{2}[V]}$  & $\Delta V_{40-30}$ &  $t_{\mathrm{rise}}$  & $\mathrm{dm}_{1}$[$r$]  & $t_{\mathrm{dm}_{1}[r]}$  & $\mathrm{dm}_{2}$[$r$] & $t_{\mathrm{dm}_{2}[r]}$ & $\Delta r_{40-30}$\\
          \hline
      \hline
            &       &       &     &      &   &  &    &   &  SNe~II &  &  &   &   &   &   &  &  & \\
      \hline
      & \multicolumn{6}{c}{$B$}& \multicolumn{6}{c}{$V$} & \multicolumn{6}{c}{$r$}\\
      \cmidrule(lr){2-7}\cmidrule(lr){8-13}\cmidrule(lr){14-19}
      2003ci  &   7.3(9.4)  &  - ( - )      &  - ( - )   &  - ( - )          &  - ( - )   & 0.44(0.019)   & 12.3(9.2)  & 0.025(0.0009) & 35.9(8.0)   & -0.00079(0.00016) & 51.9(8.0)  & 0.24(0.014)   &  15.4(9.6)   &  - ( - )      &  - ( - )   &  - ( - )          &  - ( - )   &  - ( - )    \\	
      2003cn  &   9.0(6.4)  &  - ( - )      &  - ( - )   & -0.00170(0.00048) & 30.1(4.9)  & 0.28(0.016)   & 14.0(6.0)  & 0.014(0.0004) & 47.8(4.0)   & -0.00007(0.00004) & 51.3(4.0)  & 0.12(0.003)   &  17.1(6.6)   &  - ( - )      &  - ( - )   &  - ( - )          &  - ( - )   &  - ( - )    \\	
      2003cx  &   9.9(6.8)  & 0.029(0.0021) & 27.0(8.9)  & -0.00099(0.00019) & 45.9(6.1)  & 0.26(0.019)   & 14.9(6.5)  & 0.009(0.0009) & 25.9(5.0)   & -0.00029(0.00015) & 41.8(5.0)  & 0.08(0.011)   &  18.0(7.0)   &  - ( - )      &  - ( - )   &  - ( - )          &  - ( - )   &  - ( - )    \\	
      2003ef  &   9.3(10.4) &  - ( - )      &  - ( - )   & -0.00140(0.00010) & 31.1(9.0)  & 0.31(0.036)   & 14.3(10.2) & 0.011(0.0006) & 19.5(9.0)   & -0.00025(0.00009) & 30.4(10.9) & 0.08(0.005)   &  17.4(10.5)  &  - ( - )      &  - ( - )   &  - ( - )          &  - ( - )   &  - ( - )    \\	
      2003fb  &  11.4(11.4) & 0.022(0.0029) & 45.0(10.0) &  - ( - )          &  - ( - )   & 0.20(0.010)   & 16.4(11.2) &  - ( - )      &  - ( - )    & -0.00048(0.00016) & 38.3(10.0) & 0.08(0.026)   &  19.4(11.5)  &  - ( - )      &  - ( - )   &  - ( - )          &  - ( - )   &  - ( - )    \\	
      2003hd  &   9.2(6.9)  & 0.050(0.0007) & 21.9(5.0)  & -0.00163(0.00022) & 38.2(5.0)  & 0.38(0.022)   & 14.2(6.5)  &  - ( - )      &  - ( - )    &  - ( - )          &  - ( - )   & 0.11(0.063)   &  17.2(7.1)   &  - ( - )      &  - ( - )   &  - ( - )          &  - ( - )   &  - ( - )    \\	
      2003hg  &   7.1(8.0)  & 0.047(0.0008) & 29.1(5.0)  & -0.00156(0.00004) & 51.9(5.0)  & 0.44(0.009)   & 12.2(7.8)  & 0.015(0.0006) & 28.6(5.0)   & -0.00045(0.00003) & 52.7(5.0)  & 0.14(0.009)   &  15.2(8.2)   &  - ( - )      &  - ( - )   &  - ( - )          &  - ( - )   &  - ( - )    \\	
      2003hl  &   8.3(6.8)  & 0.043(0.0024) & 27.6(5.0)  & -0.00157(0.00034) & 42.7(6.7)  & 0.38(0.016)   & 13.3(6.4)  & 0.013(0.0004) & 32.3(5.0)   & -0.00063(0.00006) & 44.6(5.0)  & 0.12(0.006)   &  16.3(6.9)   &  - ( - )      &  - ( - )   &  - ( - )          &  - ( - )   &  - ( - )    \\	
      2003hn  &   7.6(11.2) & 0.056(0.0052) & 21.3(10.0) & -0.00142(0.00040) & 27.4(10.0) & 0.41(0.030)   & 12.6(11.0) & 0.023(0.0008) & 26.6(10.0)  & -0.00089(0.00039) & 33.7(10.0) & 0.18(0.021)   &  15.7(11.3)  &  - ( - )      &  - ( - )   &  - ( - )          &  - ( - )   &  - ( - )    \\	
      2003ib  &   6.2(9.0)  &  - ( - )      &  - ( - )   &  - ( - )          &  - ( - )   & 0.31(0.010)   & 11.2(8.7)  & 0.023(0.0006) & 34.6(8.0)   & -0.00064(0.00007) & 46.9(8.0)  & 0.23(0.002)   &  14.3(9.1)   &  - ( - )      &  - ( - )   &  - ( - )          &  - ( - )   &  - ( - )    \\	
      2003ip  &   6.7(7.0)  & 0.041(0.0001) & 23.0(17.5) &  - ( - )          &  - ( - )   & 0.41(0.053)   & 11.7(6.7)  & 0.023(0.0016) & 52.9(4.0)   &  - ( - )          &  - ( - )   & 0.17(0.010)   &  14.8(7.2)   &  - ( - )      &  - ( - )   &  - ( - )          &  - ( - )   &  - ( - )    \\	
      2003iq  &   7.3(5.6)  &  - ( - )      &  - ( - )   & -0.00089(0.00008) & 33.0(5.5)  & 0.26(0.005)   & 12.3(5.1)  & 0.010(0.0003) & 20.5(2.0)   & -0.00027(0.00004) & 37.2(2.8)  & 0.07(0.008)   &  15.4(5.8)   &  - ( - )      &  - ( - )   &  - ( - )          &  - ( - )   &  - ( - )    \\	
      2003E   &  12.5(9.2)  &  - ( - )      &  - ( - )   & -0.00073(0.00002) & 41.6(8.0)  & 0.19(0.015)   & 17.5(9.0)  & 0.005(0.0009) & 32.6(8.0)   & -0.00034(0.00006) & 55.0(8.0)  & 0.04(0.014)   &  20.5(9.4)   &  - ( - )      &  - ( - )   &  - ( - )          &  - ( - )   &  - ( - )    \\	
      2003T   &   8.4(11.5) & 0.044(0.0062) & 20.1(10.0) & -0.00142(0.00021) & 43.0(10.0) & 0.34(0.034)   & 13.5(11.3) & 0.012(0.0011) & 20.9(13.6)  & -0.00051(0.00024) & 29.2(12.5) & 0.07(0.014)   &  16.5(11.6)  &  - ( - )      &  - ( - )   &  - ( - )          &  - ( - )   &  - ( - )    \\	
      2004er  &  10.8(2.8)  & 0.042(0.0008) & 31.5(2.0)  & -0.00131(0.00010) & 46.1(2.0)  & 0.40(0.018)   & 15.8(2.7)  & 0.015(0.0013) & 30.8(3.4)   & -0.00022(0.00012) & 37.3(8.2)  & 0.14(0.014)   &  21.1(2.6)   & 0.009(0.0007) & 38.6(2.0)  & -0.00045(0.00019) & 51.1(4.2)  & 0.07(0.003) \\
      2004fc  &   7.9(5.8)  & 0.043(0.0002) & 23.8(1.0)  & -0.00136(0.00002) & 39.6(1.0)  & 0.34(0.014)   & 13.0(5.4)  & 0.011(0.0003) & 31.1(1.4)   & -0.00032(0.00004) & 58.2(1.9)  & 0.11(0.023)   &  16.0(6.0)   & 0.009(0.0001) & 29.6(1.0)  & -0.00048(0.00003) & 43.3(1.3)  & 0.08(0.014) \\
      2004fx  &   7.7(6.9)  & 0.026(0.0037) & 20.8(10.6) & -0.00298(0.00096) & 26.3(4.1)  & 0.07(0.029)   & 12.7(6.6)  &  - ( - )      &  - ( - )    & -0.00060(0.00036) & 23.6(20.2) & -0.02(0.014)  &  15.7(7.1)   & 0.001(0.0003) & 20.9(4.0)  & -0.00102(0.00015) & 24.9(4.0)  & -0.03(0.010)\\
      2005an  &   8.0(7.8)  & 0.070(0.0023) & 25.1(6.0)  & -0.00296(0.00028) & 36.5(6.0)  & 0.52(0.046)   & 13.0(7.5)  & 0.034(0.0009) & 29.1(6.0)   & -0.00135(0.00012) & 39.0(6.0)  & 0.30(0.017)   &  16.0(8.0)   & 0.021(0.0002) & 32.6(6.0)  & -0.00053(0.00004) & 44.9(6.0)  & 0.20(0.003) \\
      2005dt  &   5.8(10.0) &  - ( - )      &  - ( - )   & -0.00293(0.00085) & 23.7(23.0) & 0.18(0.041)   & 10.8(9.8)  & 0.010(0.0003) & 24.2(9.0)   & -0.00024(0.00003) & 47.7(9.0)  & 0.09(0.005)   &  13.8(10.2)  &  - ( - )      &  - ( - )   &  - ( - )          &  - ( - )   & 0.03(0.004) \\	
      2005dx  &   9.1(8.9)  & 0.065(0.0034) & 26.7(7.0)  & -0.00390(0.00069) & 39.0(7.0)  & 0.49(0.064)   & 14.1(8.6)  &  - ( - )      &  - ( - )    & -0.00044(0.00034) & 34.4(7.0)  & 0.15(0.009)   &  17.2(9.0)   &  - ( - )      &  - ( - )   &  - ( - )          &  - ( - )   & 0.08(0.010) \\	
      2005dz  &   7.7(6.3)  & 0.035(0.0004) & 27.5(4.0)  & -0.00119(0.00003) & 49.8(4.0)  & 0.33(0.011)   & 12.7(5.9)  & 0.014(0.0002) & 29.3(4.0)   & -0.00053(0.00002) & 47.2(4.0)  & 0.13(0.008)   &  15.8(6.5)   & 0.010(0.0002) & 36.0(4.0)  & -0.00046(0.00007) & 51.9(4.7)  & 0.10(0.008) \\
      2005lw  &   7.4(10.9) & 0.048(0.0020) & 23.7(10.0) &  - ( - )          &  - ( - )   & 0.46(0.012)   & 12.4(10.7) & 0.022(0.0003) & 39.2(10.0)  & -0.00042(0.00004) & 56.4(10.0) & 0.21(0.006)   &  15.5(11.0)  & 0.016(0.0002) & 38.2(10.0) & -0.00050(0.00004) & 51.4(10.0) & 0.15(0.007) \\
      2005me  &   7.4(11.1) &  - ( - )      &  - ( - )   & -0.00173(0.00007) & 33.8(10.0) & 0.46(0.048)   & 12.4(10.9) & 0.031(0.0004) & 25.4(10.0)  & -0.00098(0.00012) & 38.0(10.0) & 0.25(0.024)   &  15.4(11.2)  &  - ( - )      &  - ( - )   &  - ( - )          &  - ( - )   & 0.16(0.008) \\	
      2005J   &   7.7(8.9)  & 0.051(0.0003) & 26.8(7.0)  & -0.00179(0.00006) & 40.1(7.0)  & 0.43(0.029)   & 12.7(8.7)  & 0.020(0.0003) & 30.6(7.0)   & -0.00070(0.00008) & 42.2(7.0)  & 0.19(0.008)   &  15.8(9.1)   & 0.011(0.0001) & 38.1(7.0)  & -0.00034(0.00001) & 54.3(7.0)  & 0.10(0.003) \\
      2005K   &   7.6(9.4)  &  - ( - )      &  - ( - )   &  - ( - )          &  - ( - )   & 0.29(0.010)   & 12.7(9.1)  &  - ( - )      &  - ( - )    &  - ( - )          &  - ( - )   & 0.15(0.010)   &  15.7(9.5)   &  - ( - )      &  - ( - )   & -0.00138(0.00027) & 20.7(8.0)  & 0.07(0.013) \\	
      2005Z   &   7.7(7.9)  &  - ( - )      &  - ( - )   &  - ( - )          &  - ( - )   & 0.39(0.011)   & 12.7(7.6)  & 0.021(0.0009) & 29.6(6.0)   & -0.00034(0.00014) & 40.6(6.0)  & 0.20(0.004)   &  15.8(8.0)   & 0.012(0.0003) & 40.2(12.1) & -0.00005(0.00005) & 49.0(6.0)  & 0.12(0.004) \\
      2006ai  &   7.0(7.3)  & 0.063(0.0003) & 22.0(5.0)  & -0.00228(0.00008) & 33.6(5.0)  & 0.43(0.030)   & 12.1(7.0)  & 0.044(0.0017) & 21.0(5.0)   & -0.00270(0.00031) & 32.8(5.0)  & 0.19(0.034)   &  15.1(7.5)   & 0.036(0.0027) & 20.6(5.0)  & -0.00262(0.00042) & 31.9(5.0)  & 0.09(0.032) \\
      2006be  &   8.1(11.0) &  - ( - )      &  - ( - )   & -0.00157(0.00007) & 21.2(9.0)  & 0.27(0.031)   & 13.2(10.8) &  - ( - )      &  - ( - )    &  - ( - )          & - ( - )   & 0.09(0.007)   &  16.2(11.1)  &  - ( - )      &  - ( - )   & -0.00027(0.00006) & 32.3(9.0)  & 0.05(0.006) \\	
      2006bl  &   8.2(12.1) &  - ( - )      &  - ( - )   &  - ( - )          &  - ( - )   & 0.48(0.011)   & 13.2(11.9) &  - ( - )      &  - ( - )    &  - ( - )          & - ( - )   & 0.25(0.005)   &  16.3(12.2)  &  - ( - )      &  - ( - )   &  - ( - )          &  - ( - )   & 0.16(0.004) \\	
      2006ee  &   8.0(6.8)  &  - ( - )      &  - ( - )   & -0.00105(0.00007) & 27.3(4.0)  & 0.21(0.011)   & 13.1(6.4)  &  - ( - )      &  - ( - )    & -0.00085(0.00044) & 24.6(15.2) & 0.01(0.004)   &  16.1(6.9)   & 0.004(0.0096) & 19.7(19.0) & -0.00072(0.00021) & 32.0(4.0)  & -0.03(0.015)\\	
      2006iw  &   6.8(4.8)  & 0.042(0.0015) & 24.9(1.0)  & -0.00145(0.00018) & 38.6(4.1)  & 0.34(0.008)   & 11.9(4.3)  &  - ( - )      &  - ( - )    &  - ( - )          & - ( - )   & 0.10(0.023)   &  14.9(5.0)   &  - ( - )      &  - ( - )   &  - ( - )          &  - ( - )   & 0.05(0.016) \\	
      2006qr  &   7.8(8.7)  &  - ( - )      &  - ( - )   & -0.00066(0.00002) & 35.1(7.0)  & 0.35(0.013)   & 12.9(8.4)  & 0.016(0.0002) & 35.6(7.0)   & -0.00019(0.00002) & 50.2(7.0)  & 0.16(0.009)   &  15.9(8.8)   & 0.011(0.0011) & 54.0(10.4) &  - ( - )          &  - ( - )   & 0.08(0.007) \\
      2006Y   &   6.4(6.7)  & 0.080(0.0004) & 16.4(4.0)  & -0.00482(0.00038) & 26.7(4.0)  & 0.21(0.015)   & 11.4(6.3)  & 0.070(0.0035) & 16.8(4.0)   & -0.00578(0.00047) & 24.2(4.0)  & 0.07(0.009)   &  14.5(6.9)   & 0.059(0.0005) & 16.5(4.0)  & -0.00560(0.00007) & 23.0(4.0)  & 0.03(0.003) \\
      2007hm  &   7.3(7.7)  &  - ( - )      &  - ( - )   &  - ( - )          &  - ( - )   & 0.24(0.013)   & 12.3(7.4)  &  - ( - )      &  - ( - )    &  - ( - )          & - ( - )   & 0.14(0.007)   &  15.3(7.9)   &  - ( - )      &  - ( - )   &  - ( - )          &  - ( - )   & 0.09(0.005) \\	
      2007il  &   8.9(7.0)  & 0.025(0.0006) & 22.1(4.0)  & -0.00046(0.00001) & 51.0(4.0)  & 0.23(0.010)   & 13.9(6.7)  & 0.006(0.0001) & 14.5(4.0)   & -0.00013(0.00001) & 41.4(5.5)  & 0.05(0.004)   &  17.0(7.2)   & 0.006(0.0009) & 33.5(4.0)  & -0.00063(0.00010) & 44.2(4.0)  & 0.05(0.003) \\
      2007ld  &   7.0(9.1)  &  - ( - )      &  - ( - )   &  - ( - )          &  - ( - )   & 0.33(0.023)   & 12.0(8.9)  &  - ( - )      &  - ( - )    &  - ( - )          & - ( - )   & 0.16(0.010)   &  15.0(9.3)   &  - ( - )      &  - ( - )   & -0.00046(0.00000) & 39.9(8.0)  & 0.08(0.011) \\	
      2007oc  &   7.5(6.5)  &  - ( - )      &  - ( - )   & -0.00175(0.00009) & 25.4(3.0)  & 0.35(0.012)   & 12.5(6.2)  &  - ( - )      &  - ( - )    & -0.00102(0.00012) & 30.5(3.0)  & 0.20(0.007)   &  15.5(6.7)   & 0.018(0.0001) & 20.7(3.0)  & -0.00036(0.00009) & 36.6(3.9)  & 0.15(0.003) \\
      2007od  &   6.9(6.5)  &  - ( - )      &  - ( - )   &  - ( - )          &  - ( - )   & 0.29(0.011)   & 11.9(6.1)  &  - ( - )      &  - ( - )    &  - ( - )          & - ( - )   & 0.15(0.003)   &  14.9(6.7)   &  - ( - )      &  - ( - )   &  - ( - )          &  - ( - )   & 0.07(0.002) \\	
      2007P   &   7.6(7.6)  & 0.056(0.0015) & 33.3(5.0)  &  - ( - )          &  - ( - )   & 0.55(0.014)   & 12.6(7.3)  &  - ( - )      &  - ( - )    &  - ( - )          & - ( - )   & 0.22(0.013)   &  15.7(7.8)   & 0.018(0.0020) & 33.3(5.0)  & -0.00035(0.00020) & 55.7(8.3)  & 0.17(0.010) \\
      2007U   &   7.1(6.9)  & 0.061(0.0005) & 28.2(6.0)  & -0.00198(0.00007) & 41.8(6.0)  & 0.54(0.024)   & 11.3(6.9)  & 0.031(0.0008) & 37.1(6.0)   &  - ( - )          & - ( - )   & 0.30(0.005)   &  16.5(7.7)   & 0.020(0.0004) & 39.8(6.0)  &  - ( - )          &  - ( - )   & 0.19(0.006) \\
      2007W   &   7.0(8.7)  &  - ( - )      &  - ( - )   & -0.00077(0.00009) & 17.7(7.0)  & 0.24(0.014)   & 12.1(8.4)  &  - ( - )      &  - ( - )    & -0.00060(0.00016) & 27.4(11.5) & 0.01(0.007)   &  15.1(8.8)   &  - ( - )      &  - ( - )   &  - ( - )          &  - ( - )   & -0.01(0.003)\\	
      2007X   &   7.8(7.3)  & 0.061(0.0006) & 24.0(5.0)  & -0.00218(0.00010) & 33.7(5.0)  & 0.45(0.029)   & 12.8(7.0)  & 0.024(0.0002) & 34.2(5.0)   & -0.00061(0.00003) & 49.2(5.0)  & 0.24(0.002)   &  15.8(7.5)   & 0.015(0.0002) & 39.2(5.0)  & -0.00046(0.00004) & 54.3(5.0)  & 0.14(0.006) \\
      2008bh  &   7.9(7.5)  & 0.044(0.0004) & 24.4(5.0)  &  - ( - )          &  - ( - )   & 0.41(0.010)   & 13.0(7.2)  &  - ( - )      &  - ( - )    &  - ( - )          & - ( - )   & 0.17(0.005)   &  16.0(7.7)   &  - ( - )      &  - ( - )   &  - ( - )          &  - ( - )   & 0.11(0.005) \\	
      2008br  &   6.3(9.8)  &  - ( - )      &  - ( - )   &  - ( - )          &  - ( - )   & 0.21(0.010)   & 11.3(9.6)  &  - ( - )      &  - ( - )    &  - ( - )          & - ( - )   & 0.05(0.004)   &  14.4(10.0)  &  - ( - )      &  - ( - )   &  - ( - )          &  - ( - )   & -0.01(0.003)\\	
      2008bu  &   6.2(7.9)  &  - ( - )      &  - ( - )   &  - ( - )          &  - ( - )   &  - ( - )      & 11.2(7.6)  &  - ( - )      &  - ( - )    &  - ( - )          & - ( - )   & 0.49(0.026)   &  14.3(8.1)   & 0.019(0.0195) & 24.6(11.6) & -0.00371(0.00125) & 30.7(10.4) & 0.04(0.034) \\
      2008gi  &   7.2(10.5) &  - ( - )      &  - ( - )   &  - ( - )          &  - ( - )   & 0.44(0.016)   & 12.3(10.3) &  - ( - )      &  - ( - )    &  - ( - )          & - ( - )   & 0.24(0.006)   &  15.3(10.6)  & 0.017(0.0003) & 33.0(9.0)  & -0.00030(0.00006) & 46.6(9.0)  & 0.16(0.002) \\
      2008ho  &   7.3(6.7)  & 0.035(0.0020) & 18.5(5.0)  & -0.00290(0.00016) & 31.2(5.0)  &  - ( - )      & 12.3(6.4)  &  - ( - )      &  - ( - )    &  - ( - )          & - ( - )   &  - ( - )      &  15.3(6.9)   &  - ( - )      &  - ( - )   &  - ( - )          &  - ( - )   &  - ( - )	   \\
      2008if  &   8.6(7.2)  & 0.061(0.0001) & 26.7(5.0)  & -0.00175(0.00008) & 44.4(5.0)  & 0.54(0.018)   & 13.6(6.8)  & 0.035(0.0001) & 17.5(5.0)   & -0.00090(0.00003) & 39.6(5.0)  & 0.27(0.012)   &  16.7(7.3)   & 0.027(0.0059) & 18.4(7.2)  & -0.00119(0.00115) & 25.4(5.0)  & 0.15(0.003) \\
      2008in  &   8.0(6.2)  & 0.046(0.0009) & 16.4(2.0)  & -0.00174(0.00018) & 28.8(2.0)  & 0.23(0.008)   & 13.1(5.8)  &  - ( - )      &  - ( - )    &  - ( - )          &  - ( - )   & 0.07(0.007)   &  16.1(6.4)   & 0.010(0.0009) & 23.8(2.0)  & -0.00072(0.00053) & 32.6(2.0)  & 0.04(0.003) \\
      2008K   &   6.8(7.9)  &  - ( - )      &  - ( - )   &  - ( - )          &  - ( - )   & 0.50(0.019)   & 11.8(7.6)  & 0.033(0.0006) & 35.3(6.0)   & -0.00082(0.00010) & 51.0(6.0)  & 0.32(0.012)   &  14.9(8.1)   & 0.023(0.0008) & 29.8(6.0)  & -0.00092(0.00020) & 38.1(6.0)  & 0.21(0.011) \\
      2008M   &   7.9(10.7) &  - ( - )      &  - ( - )   & -0.00073(0.00062) & 14.6(9.0)  & 0.36(0.043)   & 12.9(10.5) &  - ( - )      &  - ( - )    & -0.00744(0.00158) & 14.6(9.0)  & 0.12(0.030)   &  15.9(10.9)  &  - ( - )      &  - ( - )   & -0.00342(0.00094) & 16.1(11.3) & 0.09(0.010) \\	
      2009aj  &   7.8(7.7)  &  - ( - )      &  - ( - )   & -0.00064(0.00000) & 29.9(7.0)  & 0.31(0.013)   & 12.8(7.4)  & 0.030(0.0021) & 17.0(7.0)   & -0.00070(0.00033) & 39.3(7.0)  & 0.22(0.013)   &  15.9(7.9)   & 0.022(0.0002) & 19.7(7.0)  & -0.00056(0.00003) & 41.5(7.0)  & 0.18(0.010) \\
      2009ao  &   8.1(7.5)  & 0.012(0.0157) & 10.7(15.6) & -0.00011(0.00019) & 13.8(4.0)  & 0.35(0.024)   & 13.2(7.2)  &  - ( - )      &  - ( - )    &  - ( - )          &  - ( - )   & 0.10(0.010)   &  16.2(7.7)   &  - ( - )      &  - ( - )   &  - ( - )          &  - ( - )   & 0.04(0.009) \\	
      2009au  &   6.7(5.6)  & 0.056(0.0001) & 41.6(4.0)  &  - ( - )          &  - ( - )   & 0.54(0.006)   & 11.8(5.2)  & 0.035(0.0005) & 43.4(4.0)   &  - ( - )          &  - ( - )   & 0.33(0.005)   &  14.8(5.8)   & 0.028(0.0001) & 44.6(4.0)  &  - ( - )          &  - ( - )   & 0.26(0.006) \\
      2009bz  &   5.9(5.7)  &  - ( - )      &  - ( - )   &  - ( - )          &  - ( - )   & 0.21(0.006)   & 11.0(5.3)  & 0.007(0.0004) & 33.9(4.0)   &  - ( - )          &  - ( - )   & 0.07(0.005)   &  14.0(5.9)   & 0.006(0.0001) & 38.4(4.0)  &  - ( - )          &  - ( - )   & 0.05(0.004) \\
      2009ib  &   7.1(8.1)  & 0.053(0.0009) & 14.6(7.6)  & -0.00202(0.00018) & 27.5(7.6)  & 0.23(0.033)   & 12.1(7.8)  & 0.016(0.0011) & 21.8(7.6)   & -0.00112(0.00020) & 29.1(7.6)  & 0.06(0.012)   &  15.1(8.2)   &  - ( - )      &  - ( - )   &  - ( - )          &  - ( - )   &  - ( - )	   \\
      2009N   &   8.0(7.0)  & 0.045(0.0007) & 14.2(5.0)  & -0.00143(0.00032) & 28.8(5.0)  & 0.23(0.017)   & 13.0(6.7)  & 0.007(0.0020) & 20.6(5.0)   & -0.00053(0.00036) & 30.6(5.0)  & 0.01(0.006)   &  16.1(7.2)   & 0.002(0.0013) & 23.3(5.0)  & -0.00059(0.00022) & 32.8(5.0)  & -0.02(0.008)\\
      2012aw  &   5.5(2.3)  &  - ( - )      &  - ( - )   &  - ( - )          &  - ( - )   &  - ( - )      & 10.5(0.8)  & 0.013(0.0000) & 17.4(0.5)   & -0.00140(0.00001) & 23.3(0.5)  & 0.00(0.019)   &  13.5(2.8)   &  - ( - )      &  - ( - )   &  - ( - )          &  - ( - )   &  - ( - )	   \\
      2013ai  &  20.2(10.0) & 0.059(0.0015) & 37.6(10.0) & -0.00204(0.00023) & 47.4(10.0) & 0.54(0.035)   & 25.7(10.0) & 0.025(0.0004) & 39.6(10.0)  & -0.00065(0.00010) & 45.9(10.0) & 0.20(0.036)   &  25.5(10.1)  & 0.002(0.0074) & 28.1(10.0) & -0.00013(0.00035) & 29.2(10.0) & 0.08(0.038) \\
      2013by  &  10.7(5.1)  & 0.074(0.0012) & 29.7(5.0)  & -0.00452(0.00189) & 37.8(5.9)  & 0.62(0.048)   & 13.7(5.1)  & 0.035(0.0020) & 21.4(5.0)   & -0.00011(0.00112) & 22.5(5.6)  & 0.28(0.029)   &  13.8(5.1)   & 0.034(0.0005) & 21.0(5.0)  & -0.00256(0.00014) & 26.6(5.0)  & 0.12(0.009) \\
      2013ej  &  11.2(1.3)  & 0.052(0.0013) & 11.7(0.9)  & -0.00286(0.00004) & 13.5(0.9)  & 0.46(0.026)   & 12.9(1.3)  & 0.042(0.0006) & 32.0(0.9)   & -0.00209(0.00005) & 45.6(0.9)  & 0.39(0.096)   &  18.4(1.4)   & 0.023(0.0005) & 31.2(2.1)  & -0.00066(0.00017) & 40.6(3.0)  & 0.22(0.104) \\
      2013fs  &   1.8(1.8)  & 0.048(0.0015) & 23.9(0.5)  & -0.00161(0.00019) & 38.1(0.5)  & 0.38(0.052)   &  7.3(2.1)  & 0.060(0.0038) & 12.3(0.7)   & -0.00727(0.00090) & 15.9(1.1)  & 0.04(0.044)   &  9.9(1.5)    & 0.044(0.0030) & 14.0(0.7)  & -0.00603(0.00076) & 18.4(1.3)  & 0.08(0.038) \\
      2014cx  &   8.4(0.6)  & 0.033(0.0015) & 22.4(2.1)  & -0.00121(0.00017) & 31.7(4.2)  & 0.21(0.056)   & 15.6(0.8)  & 0.006(0.0019) & 27.7(4.5)   & -0.00002(0.00024) & 29.7(9.6)  & 0.06(0.041)   &  23.3(2.4)   & 0.005(0.0009) & 30.4(2.9)  & -0.00057(0.00015) & 38.1(5.0)  & 0.03(0.049) \\
      2014G   &   9.3(1.8)  & 0.060(0.0004) & 21.9(1.7)  & -0.00133(0.00022) & 27.3(2.0)  & 0.50(0.027)   & 12.1(1.9)  & 0.035(0.0007) & 20.7(1.7)   & -0.00105(0.00010) & 24.5(1.7)  & 0.32(0.025)   &  15.0(2.3)   & 0.029(0.0010) & 26.0(4.7)  & -0.00066(0.00046) & 30.5(4.0)  & 0.26(0.015) \\ 
      \hline
        \end{tabular}
        \begin{tablenotes}
        \item In column 1 we present the SN name. In the next six columns we present $t_{\mathrm{rise}}$, $\mathrm{dm}_{1}$($B$), $t_{\mathrm{dm}_{1}(B)}$ (which is the phase corresponding to $\mathrm{dm}_{1}$($B$)), $\mathrm{dm}_{2}$($B$), $t_{\mathrm{dm}_{2}(B)}$ (which is the phase corresponding to $\mathrm{dm}_{2}$($B$)) and $\Delta B_{40-30}$. The same scheme is repeated in the following twelve columns for $V$- and $r$-band respectively. The Monte Carlo method was used to obtain the errors for $\mathrm{dm}_{1}$ and $\mathrm{dm}_{2}$, the errors in the corresponding phases are considered to be the maximum between the error in the explosion epoch and the error obtained using the Monte Carlo method.
        \end{tablenotes}
      \end{threeparttable}
    \end{adjustbox}
  \end{table}
\end{landscape}

\begin{figure}
  \includegraphics[width=\columnwidth]{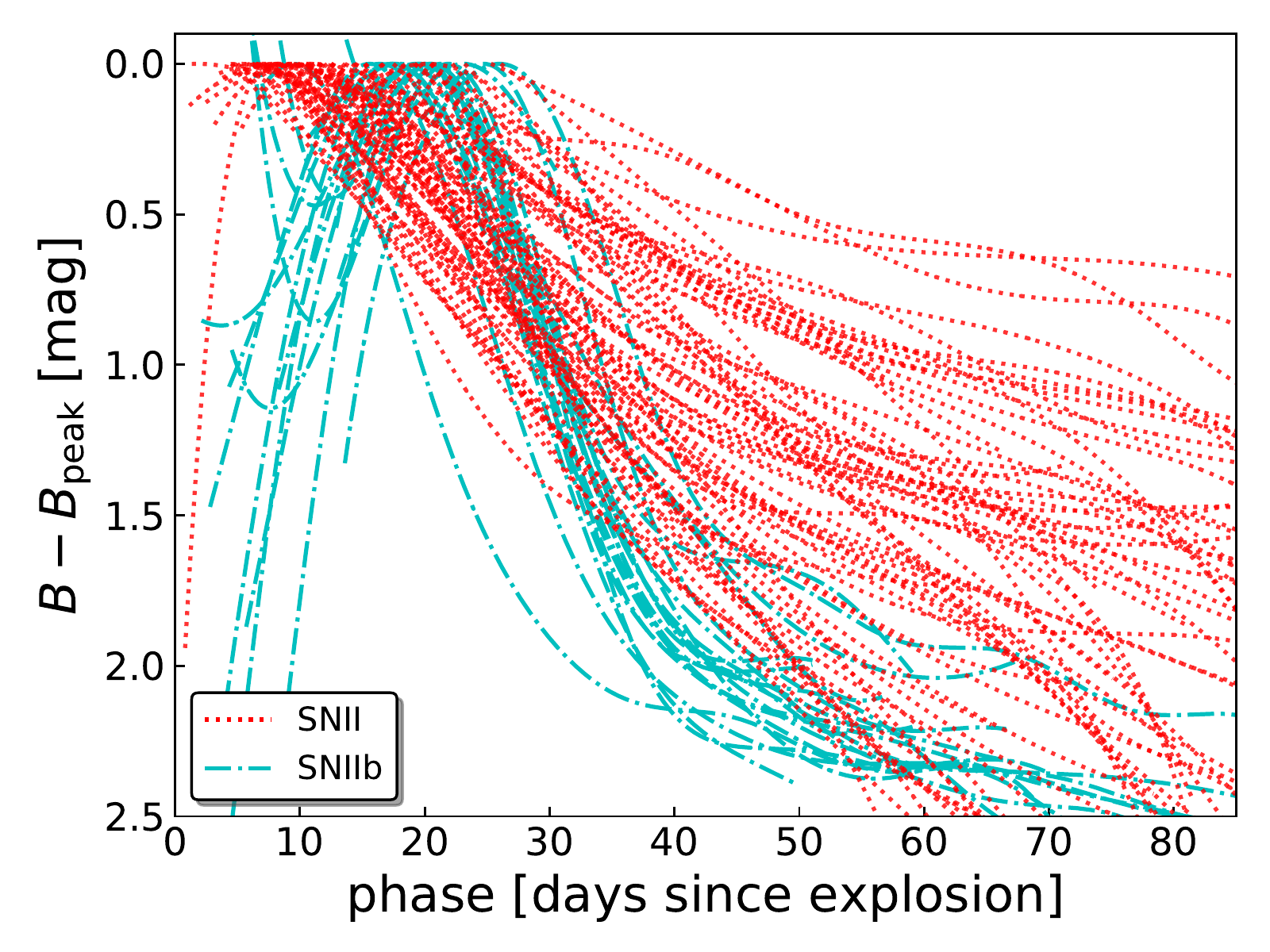}
  \includegraphics[width=\columnwidth]{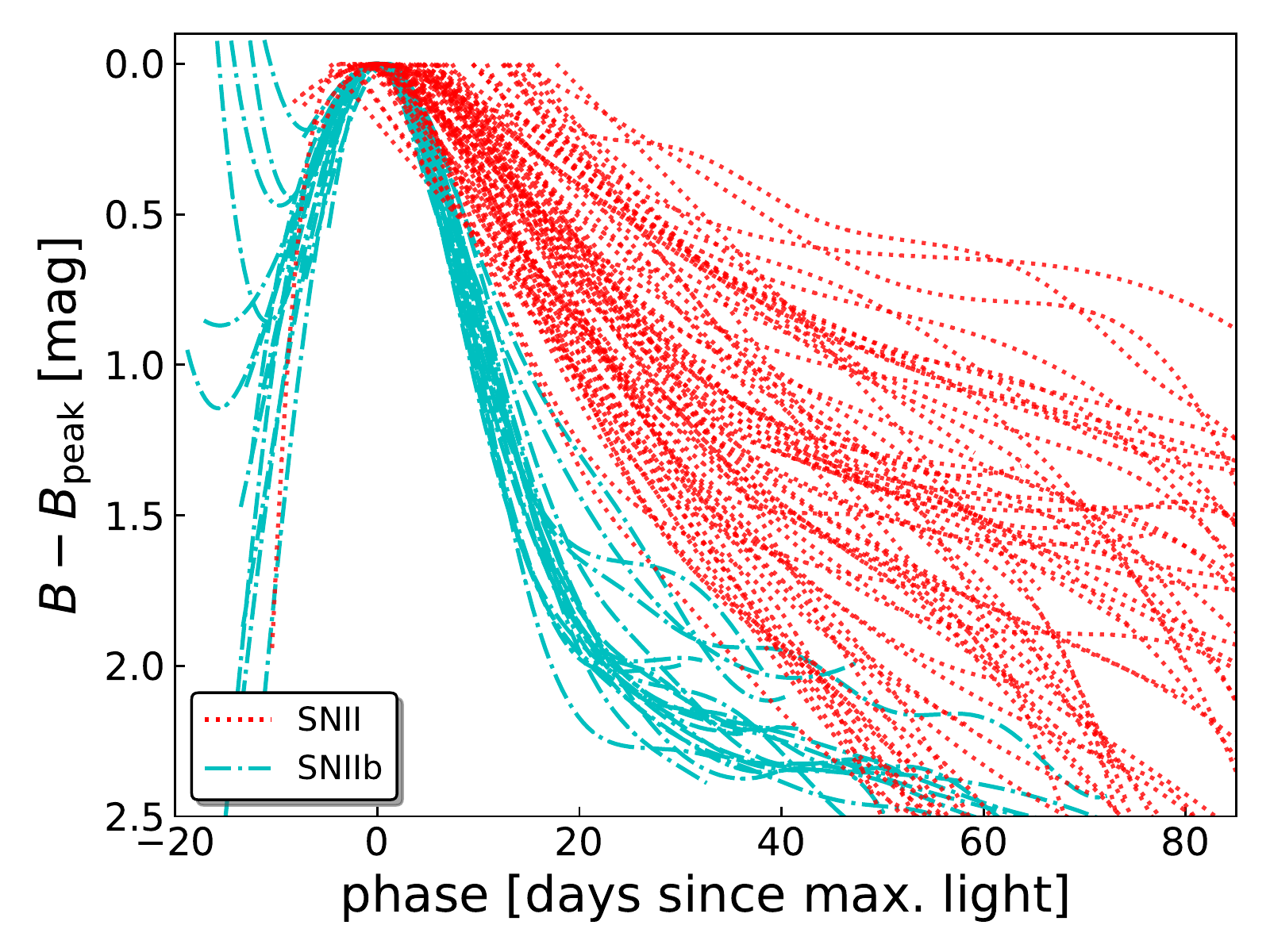}
  \caption{Resampled light curves normalized with respect to the $B$-band peak magnitude, SNe~II are plotted in red dotted lines while SNe~IIb are plotted in cyan dash-dot lines. Top panel: phase with respect to explosion epoch. Bottom panel: phase with respect to date of maximum light.}
  \label{fig:normB}
\end{figure}

\begin{figure}
  \includegraphics[width=\columnwidth]{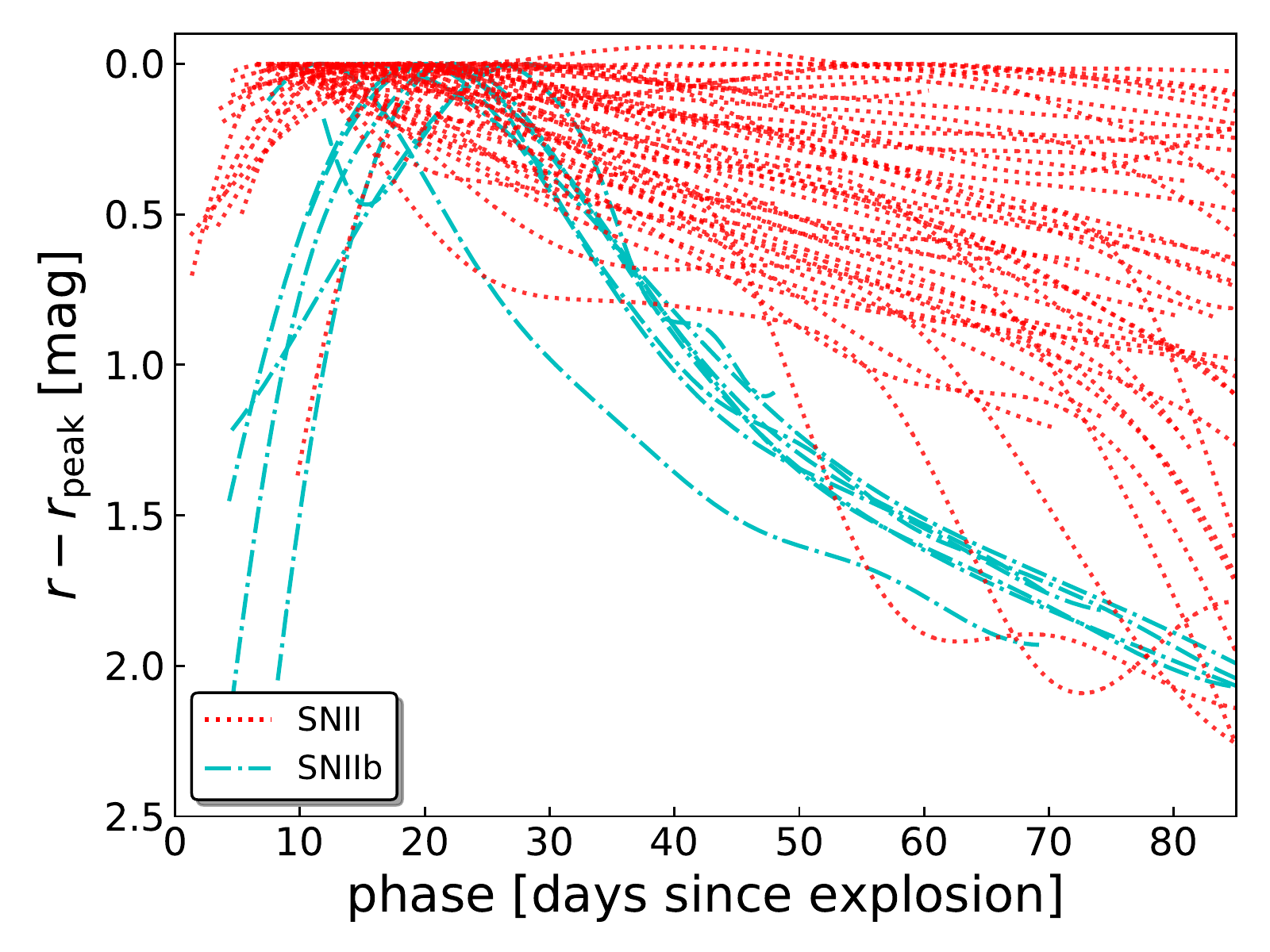}
  \includegraphics[width=\columnwidth]{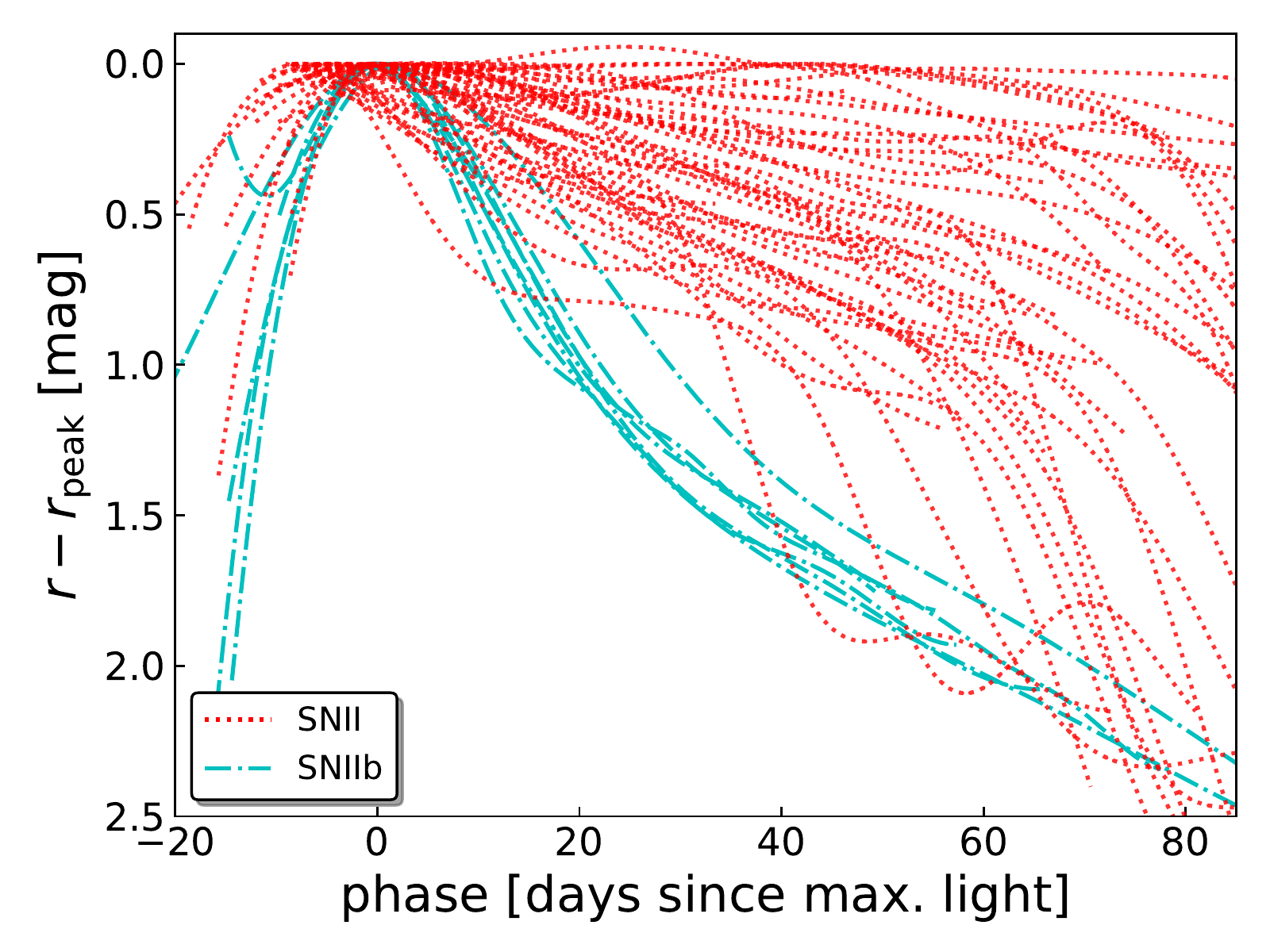}
  \caption{Resampled light curves normalized with respect to the $r$-band peak magnitude, SNe~II are plotted in red dotted lines while SNe~IIb are plotted in cyan dash-dot lines. Top panel: phase with respect to explosion epoch. Bottom panel: phase with respect to date of maximum light.}
  \label{fig:normr}
\end{figure}

\begin{figure}
  \includegraphics[width=\columnwidth]{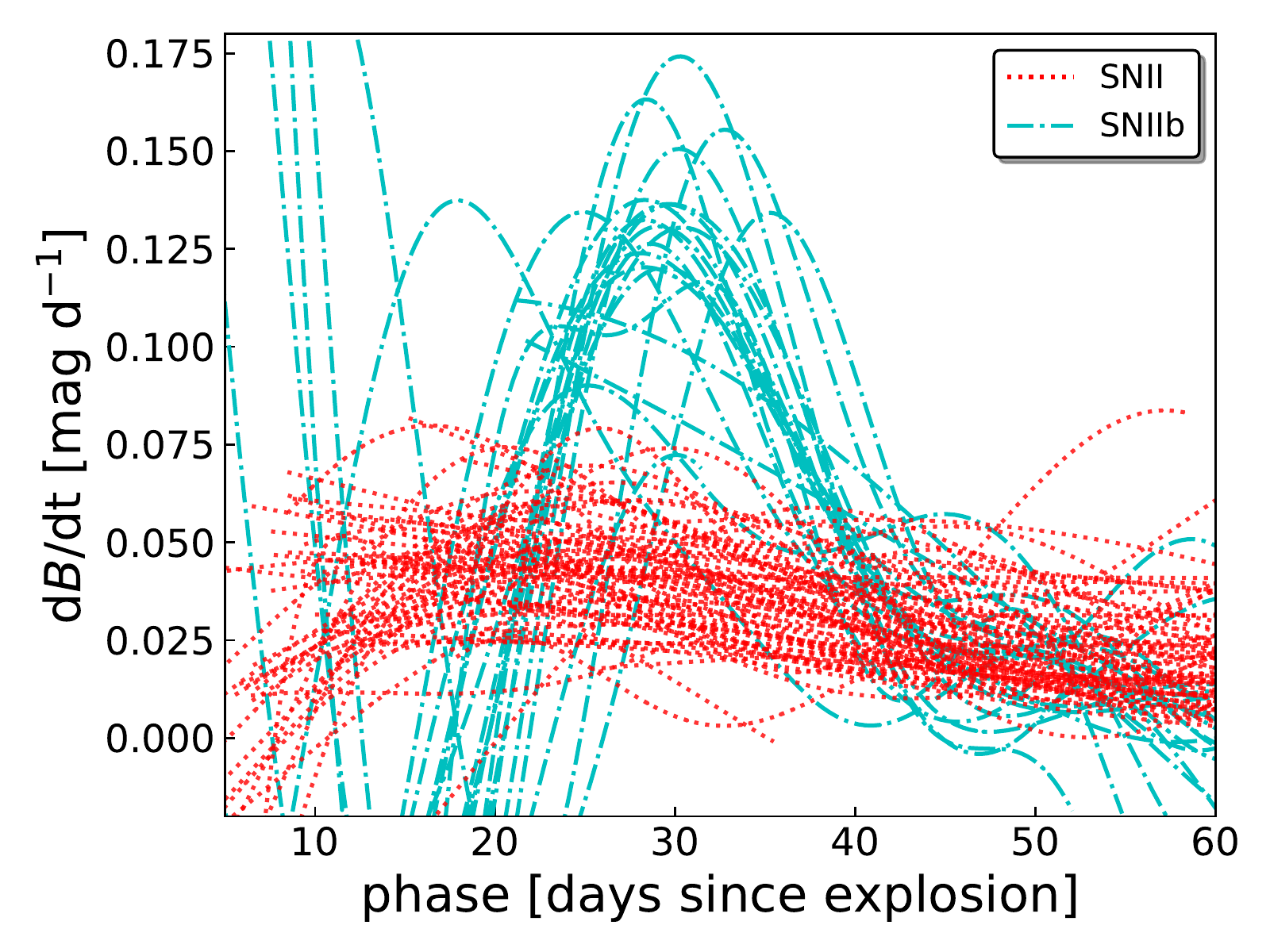}
  \includegraphics[width=\columnwidth]{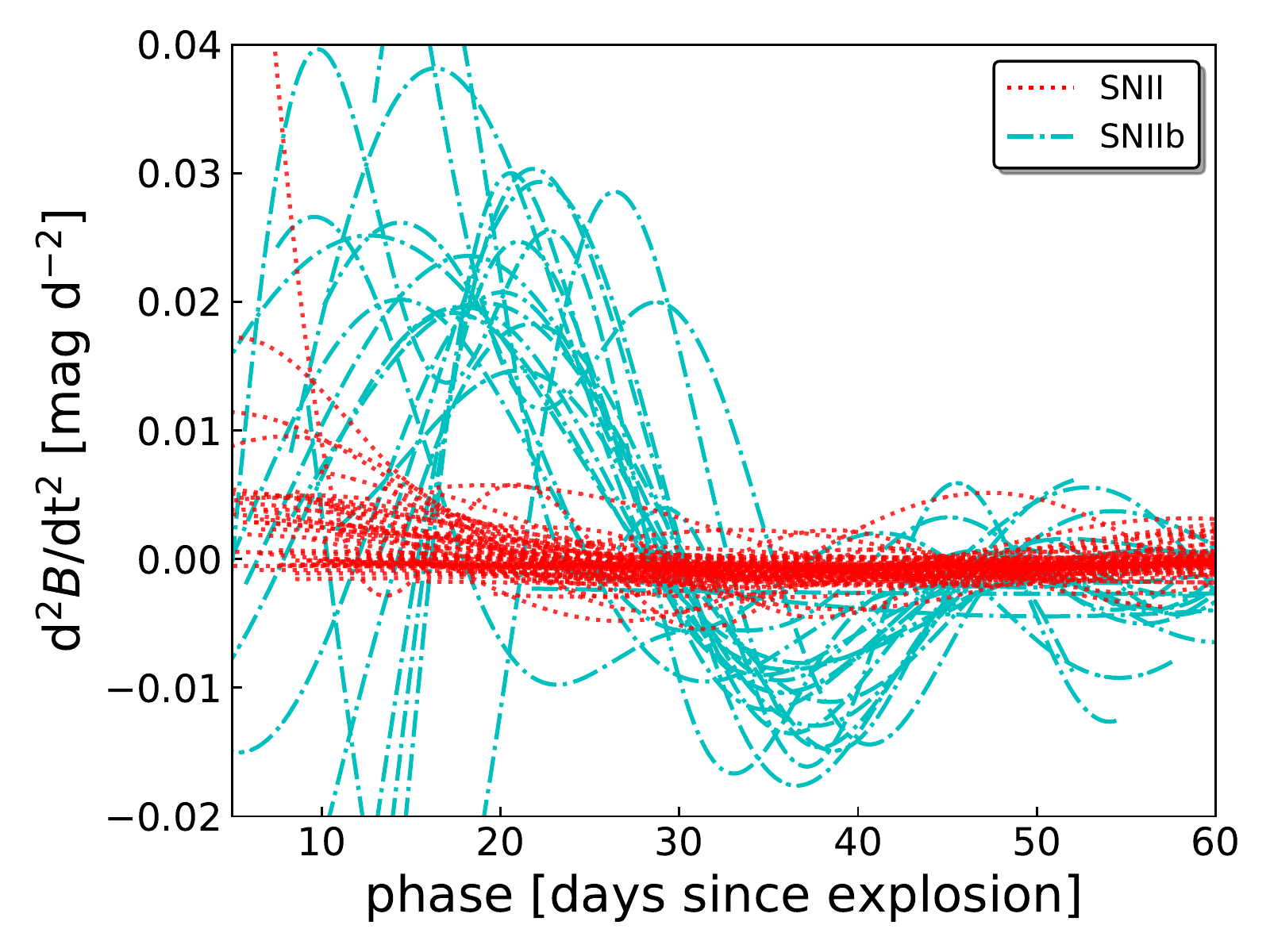}
  \caption{Derivative of the $B$-band light curve with respect to explosion. SNe~II are plotted in red dotted lines while SNe~IIb are plotted in cyan dash-dot lines. Top panel: first derivative. Bottom panel: second derivative. }
  \label{fig:derB}
\end{figure}

\begin{figure}
  \includegraphics[width=\columnwidth]{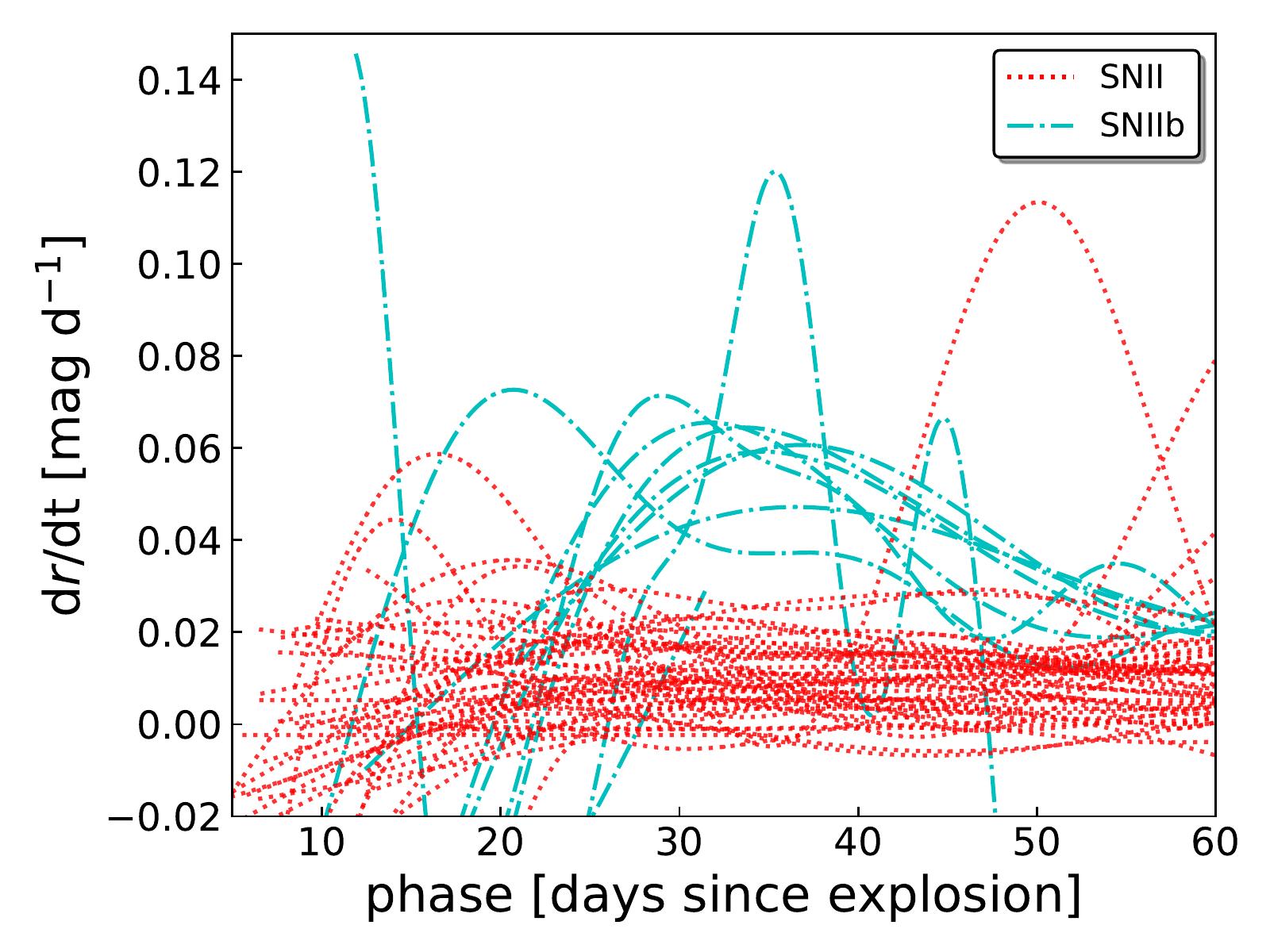}
  \includegraphics[width=\columnwidth]{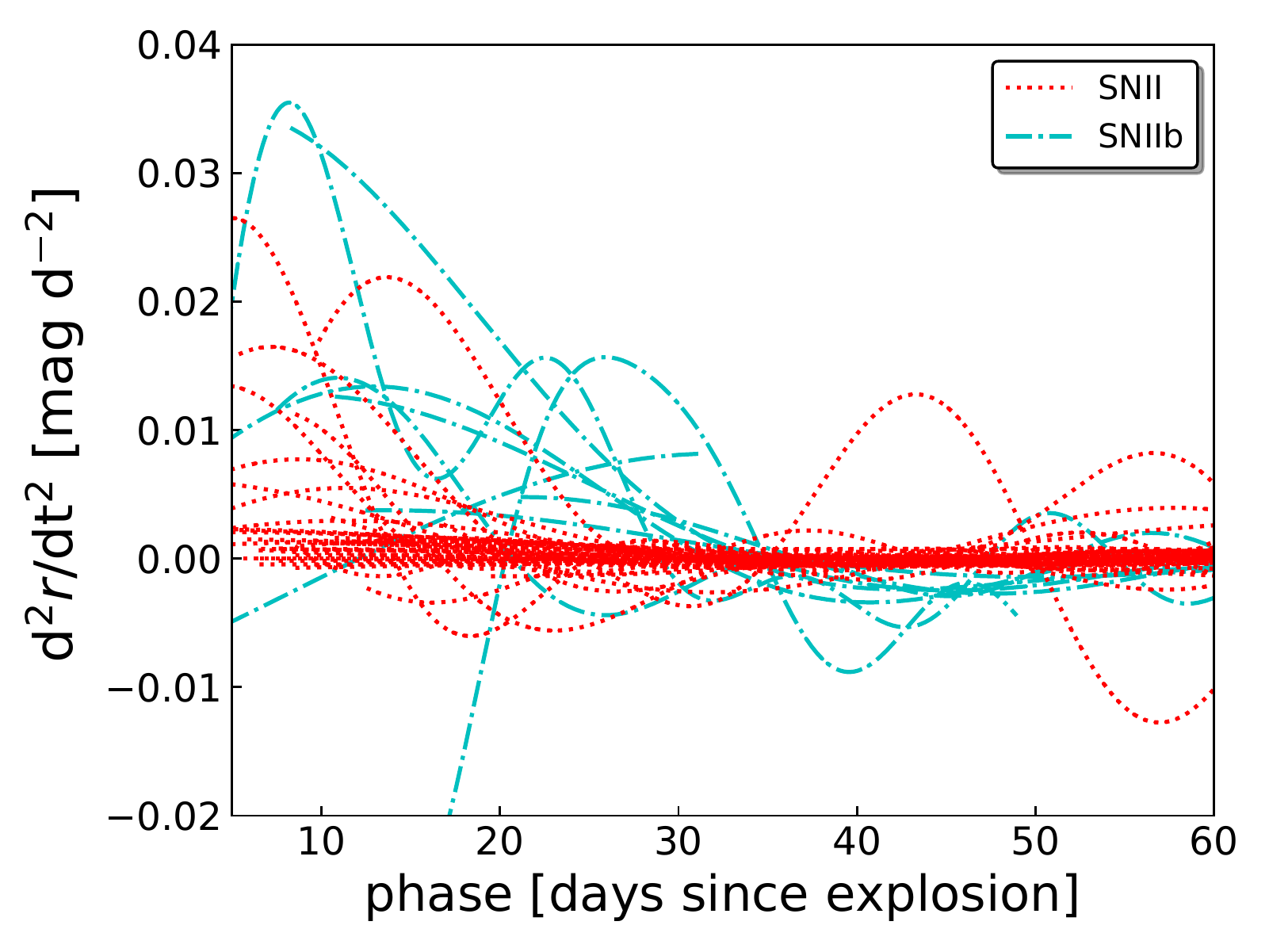}
  \caption{Derivative of the $r$-band light curve with respect to explosion. SNe~II are plotted in red dotted lines while SNe~IIb are plotted in cyan dash-dot lines. Top panel: first derivative. Bottom panel: second derivative. }
  \label{fig:derr}
\end{figure}

\begin{figure}
  \includegraphics[width=\columnwidth]{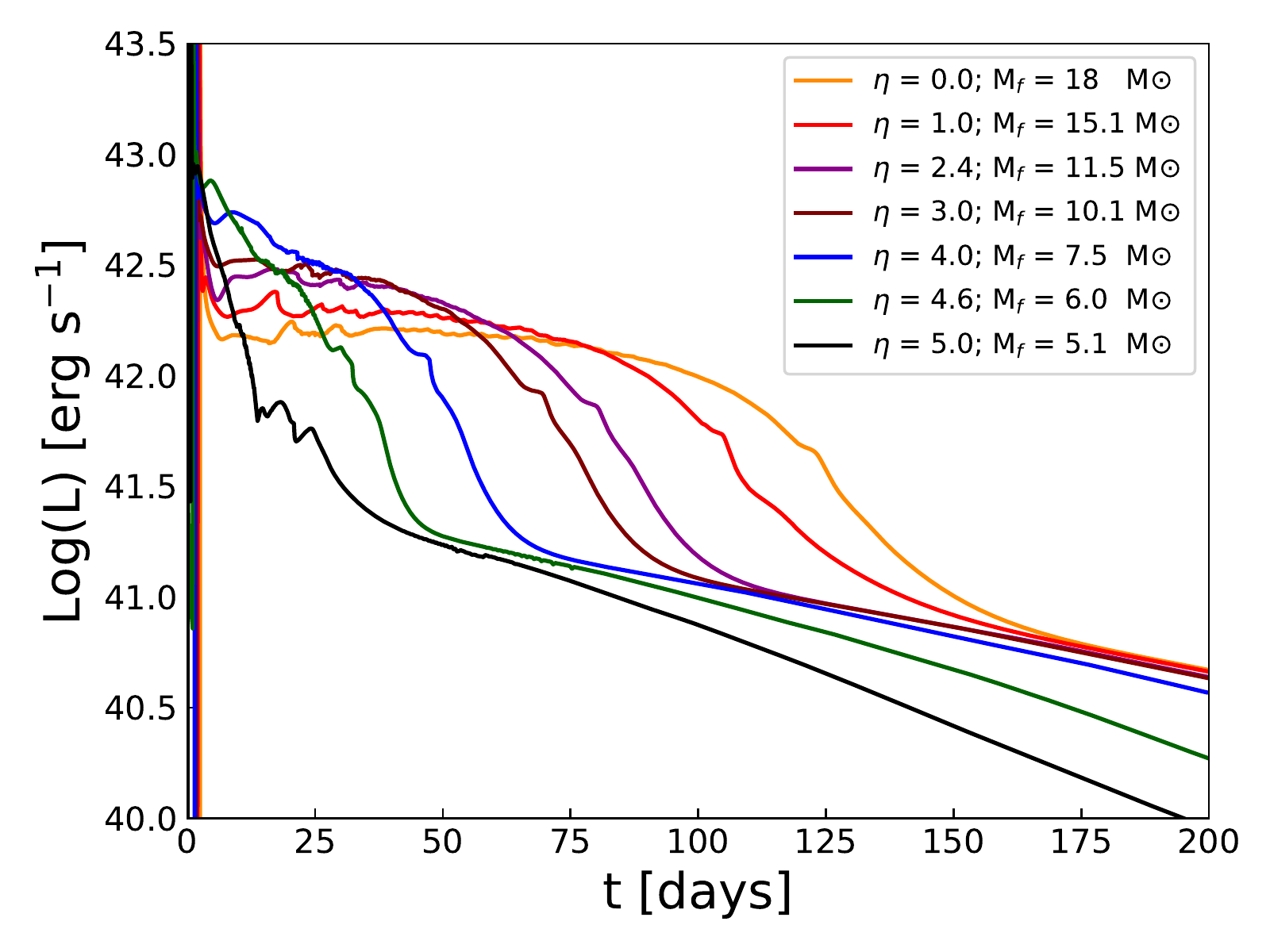}
  \includegraphics[width=\columnwidth]{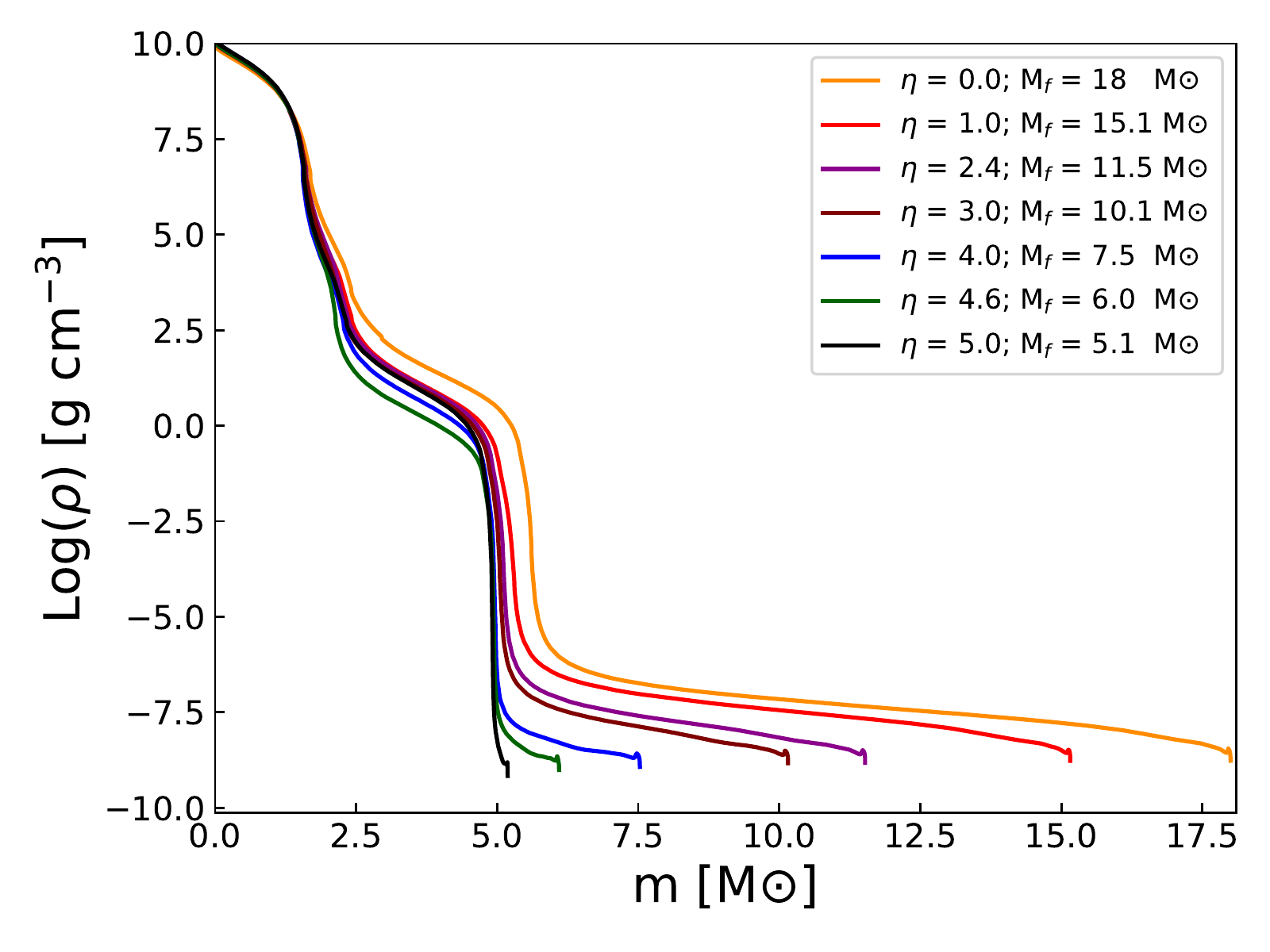}
  \caption{Models of a 18\msun\ progenitor with different values of the wind efficiency ($\eta$), considering an explosion energy of 1 foe and $^{56}$Ni mass of 0.02\msun. A model without mass loss by winds ($\eta$=0.0) is shown for comparison. M$_{f}$ denotes the final mass (pre-explosion) of the progenitor. Top panel: Bolometric light curves. Bottom panel: Initial density profiles.}
  \label{fig:model}
\end{figure}

\bsp	
\label{lastpage}
\end{document}